\documentclass[useAMS,usenatbib,usegraphicx]{mn2e}

\usepackage{times}
\usepackage{lscape}
\usepackage{subfigure}

\newcommand{\HST}{{\it HST}}
\newcommand{\Spitzer}{{\it Spitzer}}
\newcommand{\myr}{${\rm M_{\sun}yr^{-1}}$}
\newcommand{\um}{$\umu$m}
\newcommand{\zphot}{z_{\rm phot}}
\newcommand{\hyperz}{{\sc Hyperz}}

\title[A photometric redshift survey of SMGs] {The LABOCA survey of the
  Extended Chandra Deep Field South: \\
  A photometric redshift survey of submillimetre galaxies} 
\author[J. L. Wardlow et al.]{
\parbox[t]{\textwidth}{
J.\,L.\ Wardlow$^{1}$\thanks{E-mail: j.l.wardlow@durham.ac.uk},
Ian\ Smail$^{2}$, 
K.\,E.\,K.\ Coppin$^{2}$,
D.\,M.\ Alexander$^{1}$,
W.\,N.\ Brandt$^{3}$,
A.\,L.\,R.\ Danielson$^{1}$,
B.\ Luo$^{3}$,
A.\,M.\ Swinbank$^{2}$,
F.\ Walter$^{4}$,
A.\ Wei\ss$^{5}$,
Y.\,Q.\ Xue$^{3}$,
S.\ Zibetti$^{4}$,
F.\ Bertoldi$^{6}$,
A.\,D.\ Biggs$^{7}$,
S.\,C.\ Chapman$^{8}$,
H.\ Dannerbauer$^{4}$,
J.\,S.\ Dunlop$^{9}$,
E.\ Gawiser$^{10}$,
R.\,J.\ Ivison$^{9,11}$,
K.\,K.\ Knudsen$^{6}$,
A.\ Kov{\'a}cs$^{5}$,
C.\,G.\ Lacey$^{2}$,
K.\,M. Menten$^{5}$,
N.\ Padilla$^{12}$,
H.-W.\ Rix$^{4}$,
 and P.\,P.\ van der Werf$^{13}$ }\\\\
$^{1}$Department of Physics, Durham University, South Road, Durham, DH1 3LE, UK \\
$^{2}$Institute for Computational Cosmology, Durham University, South Road, Durham, DH1 3LE, UK\\
$^{3}$Department of Astronomy and Astrophysics, 525 Davey Lab, Pennsylvania State University, University Park, PA 16802, USA\\
$^{4}$Max-Planck-Institut f\"ur Astronomie, K\"onigstuhl 17, D-69117 Heidelberg, Germany\\
$^{5}$Max-Planck-Institut f\"ur Radioastronomie, Auf dem H\"ugel 69, D-53121 Bonn, Germany\\
$^{6}$Argelander Institut f\"ur Astronomie, Auf dem H\"ugel 71, D-53121 Bonn, Germany\\
$^{7}$European Southern Observatory, Karl-Schwarzschild Strasse 2, D-85748 Garching, Germany\\
$^{8}$Institute of Astronomy, Madingley Road, Cambridge, CB3 0HA, UK\\
$^{9}$SUPA (Scottish University Physics Alliance), Institute for Astronomy, University of Edinburgh, Royal Observatory, Edinburgh EH9 3HJ, UK\\
$^{10}$Physics and Astronomy Department, Rutgers University, Piscataway, NJ 08854, USA\\
$^{11}$UK Astronomy Technology Centre, Royal Observatory, Blackford Hill, Edinburgh EH9 3HJ, UK\\
$^{12}$Departmento de Astronomia y Astrofisica, Pontificia Universidad Catolica de Chile, Santiago, Chile \\
$^{13}$Leiden Observatory, Leiden University, P.O. Box 9513, 2300 RA Leiden, The Netherlands
}

\begin{document}
%\date{Accepted year month day. Received year month day; in original form year month day}

\pagerange{\pageref{firstpage}--\pageref{lastpage}} \pubyear{2010}

\maketitle \label{firstpage}

\begin{abstract}
We derive photometric redshifts from 17-band optical to mid-infrared
photometry of 74 robust radio, $24$\,\um\ and \Spitzer\ IRAC
counterparts to 68 of the 126 submillimetre galaxies (SMGs) selected at
$870$\,\um\ by LABOCA observations in the Extended \textit{Chandra}
Deep Field South (ECDFS).  We test the photometric redshifts of the
SMGs against the extensive archival spectroscopy in the ECDFS. The
median photometric redshift of identified SMGs is $z=2.2\pm0.1$, the
interquartile range is $z=1.8$--2.7 and we identify 10 ($\sim15\%$)
high-redshift ($z\ge 3$) SMGs.  We derive a simple redshift estimator
for SMGs based on the IRAC 3.6 and 8\,\um\ fluxes which is accurate to
$\Delta z\sim0.4$ for SMGs at $z<4$.  A statistical analysis of sources
around unidentified SMGs identifies a population of likely counterparts
with a redshift distribution peaking at $z=2.5\pm0.3$, which likely
comprises $\sim60\%$ of the unidentified SMGs. This confirms that the
bulk of the undetected SMGs are co-eval with those detected in the
radio/mid-infrared. We conclude that at most $\sim15\%$ of all the SMGs
are below the flux limits of our IRAC observations and lie at $z\ga3$
and hence around $\sim30\%$ of all SMGs have $z\ga3$. We estimate
that the full $S_{870\mu m}>4$~mJy SMG population has a median redshift of
$2.5\pm0.6$.  In contrast to previous suggestions we find no
significant correlation between submillimetre flux and redshift.  The
median stellar mass of the SMGs derived from SED fitting is
$(9.2\pm0.9)\times10^{10}\,M_{\sun}$ and the interquartile range is
(4.7--14)$\times10^{10}\,M_{\sun}$, although we caution that the
uncertainty in the star-formation histories results in a factor of
$\sim5$ uncertainty in these stellar masses.  Using a single
temperature modified blackbody fit with $\beta=1.5$ the median
characteristic dust temperature of SMGs is $35.9\pm 1.4$\,K and the
interquartile range is $28.5$--43.3\,K.  The infrared luminosity
function shows that SMGs at $z=2$--3 typically have higher far-infrared
luminosities and luminosity density than those at $z=1$--2. This is
mirrored in the evolution of the star-formation rate density (SFRD) for
SMGs which peaks at $z\sim 2$.  The maximum contribution of bright SMGs
to the global SFRD ($\sim5\%$ for SMGs with $S_{870\mu m}\ga4$\,mJy;
$\sim50\%$ for SMGs with $S_{870\mu m}>1$\,mJy) also occurs at
$z\sim2$.
\end{abstract}


\section{Introduction}
\label{sec:intro}

Observations in the millimetre and submillimetre wavebands provide a
uniquely powerful route to survey the distant Universe for intense
dust-obscured starbursts \citep{Blain93}.  This is due to the negative
K-correction arising from the shape of the spectral energy
distribution (SED) of the dust emission in the rest-frame
far-infrared, which results in an almost constant apparent flux for
sources with a fixed luminosity at $z\sim 1$--8.

Over the past decade, a series of ever larger surveys in the
submillimetre and millimetre wavebands have mapped out a population of
sources at mJy-flux limits with a surprisingly high surface density
\citep[e.g.][]{Smail97, Barger98, Hughes98, Eales99, Bertoldi00,
Bertoldi07, Coppin06, Knudsen08, Weiss09, Austermann10}. The mJy
fluxes of these sources imply far-infrared luminosities of $\ga
10^{12}$ \,L$_\odot$, if the sources are at cosmological distances,
$z\ga 1$, classing them as ultraluminous infrared galaxies
\cite[ULIRGs;][]{Sanders96}. Their high surface density is far in
excess of that expected from a ``no evolution'' model, suggesting very
strong evolution of the population: $\propto (1+z)^4$ \citep{Smail97,
Blain99}.  If this results from strong luminosity evolution of
starburst galaxies (as opposed to obscured AGN; \citealt{Alexander05}
then a significant fraction of the massive star formation (and metal
production) at high redshift may be occurring in this population.

To confirm this evolution and understand the  physical
processes driving it requires redshifts for the submillimetre galaxies (SMGs).
Due to the coarse spatial resolution of the submillimetre and
millimetre maps from which the SMGs can be identified, combined with
their optical faintness (in part due to their high dust obscuration),
it has proved challenging to measure their spectroscopic redshift
distribution \citep[e.g.][]{Barger99, Chapman03a, Chapman05}.  

In fact, spectroscopic redshifts are not necessary to map the broad
evolution of the SMG population and cruder photometric redshifts can
be sufficient, if they are shown to be reliable.
Various photometric redshift techniques have therefore been applied in
an attempt to trace the evolution of SMGs, using their
optical/near-/mid-infrared or far-infrared/radio SEDs
\citep[e.g.][]{Carilli99, Smail00, Ivison04, Pope05, Pope06, Ivison07, Aretxaga07,
Clements08, Dye08, Biggs10}.  

Both spectroscopic and photometric analyses suggest that the bulk of
the SMG population lies at $z\ga 1$, with an apparent peak at $z\sim
2.2$ for the subset of SMGs which can be located through their $\mu$Jy
radio emission \citep{Chapman05}.  Nevertheless, there are significant
disagreements between the different studies \citep[see
e.g.][]{Chapman05, Clements08, Dye08}, which may arise in part due to
differing levels and types of incompleteness in the identifications
and biases in the redshift measurements.  The most serious of these is
the incompleteness due to challenges in reliably locating the correct
SMG counterpart.  They are typically identified through statistical
arguments and physical correlations based on radio, mid- or
near-infrared emission \citep[e.g.][]{Ivison98, Ivison00, Ivison05,
Smail99, Pope05, Bertoldi07, Hainline09, Biggs10}, but these locate
only $\sim 60$--80\% of SMGs.  The expectation is that the SMGs whose
counterparts are missed could potentially include the highest redshift
(and thus the faintest in the radio and mid-infrared) examples,
biasing the derived evolution \citep{Ivison05}.  Attempts to address
this incompleteness through time-intensive submillimetre
interferometry have located a small fraction of previously
unidentified SMGs \citep[e.g.][]{Dannerbauer02, Dannerbauer08,
Younger07, Younger09, Wang07} but the nature and redshifts of this
unidentified subset of SMGs remains a critical issue for studies of
the population as a whole.

In this paper we use optical, near- and mid-infrared photometry to
study SMGs detected in the Extended {\it Chandra} Deep Field South
(ECDFS) by the Large APEX BOlometer CAmera
\citep[LABOCA;][]{Siringo09} on the Atacama Pathfinder EXperiment
\citep[APEX;][]{Gusten06} 12-m telescope in the LABOCA ECDFS
Submillimetre Survey \citep[LESS;][]{Weiss09}. LESS mapped the full
$30'\times30'$ ECDFS at $870$-\um\ to a noise level of $\sigma_{870\mu
m}\approx1.2$ mJy beam$^{-1}$, for a beam with angular resolution of
$19\farcs2$. 126 SMGs were detected at $>3.7\sigma$ significance
(equivalent to a false-detection rate of $\sim4\%$, \citealt{Weiss09})
and robust or tentative radio, $24$\,\um\ or IRAC mid-infrared
counterparts are identified to 93 (71 robust and 22 tentative) SMGs
\citep{Biggs10}. Here we determine photometric redshifts for the 91
(68 robust and 23 tentative) of these SMGs with detectable optical and
near-infrared counterparts in new and archival multiband photometry of
the ECDFS (described in \S\ref{sec:dr}).  LESS is an ideal survey for
this purpose because of its panoramic, deep and uniform submillimetre
coverage and extensive auxiliary data, including spectroscopy of
sufficient SMG counterparts to adequately test our photometric
redshifts.  In addition, the large size of the survey allows us to
statistically measure the redshift distribution of the SMGs that we
are unable to locate directly, in order to test if their redshift
distribution differs significantly from the identified population.

The plan of the paper is as follows: in \S\ref{sec:dr} we derive
multiband photometry from new and archival observations; while in
\S\ref{sec:analysis} we describe our the photometric redshift
estimates and tests of their reliability. The photometric redshifts,
SED fits, absolute $H$-band magnitudes, infrared luminosities, dust
temperatures and star-formation rates of SMGs are presented and
discussed in \S\ref{sec:results} and we present our conclusions in
\S\ref{sec:conc}. Throughout this paper we use deboosted submillimetre
fluxes from \citet{Weiss09}, J2000 coordinates and $\Lambda$CDM
cosmology with $\Omega_{M}=0.3$, $\Omega_{\Lambda}=0.7$ and
$H_{0}=70{\rm kms^{-1}Mpc^ {-1}}$. All photometry is on the AB
magnitude system, in which $23.9\ {\rm m_ {AB}}=1\umu{\rm Jy}$, unless
otherwise stated.

\section{Observations and Data Reduction}
\label{sec:dr}

\begin{table}
\begin{minipage}{84mm} 
\caption{Summary of photometry employed in this paper.}
\setlength{\tabcolsep}{0.8 mm}
\begin{tabular}{lccl}
\hline
Filter & $\lambda_{\rm effective}$ & Detection limit  & Reference \\
 & (\um) &($3\sigma$; mag) &  \\
\hline
MUSYC WFI $U$   & 0.35 & 26.9 & \citet{Taylor09} \\
MUSYC WFI $U38$ & 0.37 & 25.4 & \citet{Taylor09} \\
VIMOS $U$   & 0.38 & ~28.4\footnote{The listed depth of the VIMOS $ U$ band is that of the central region. The typical depth in the shallower outskirts is 28.0 mag} & \citet{Nonino09} \\
MUSYC WFI $B$   & 0.46 & 26.8 & \citet{Taylor09} \\
MUSYC WFI $V$   & 0.54 & 26.7 & \citet{Taylor09} \\
MUSYC WFI $R$   & 0.66 & 25.8 & \citet{Taylor09} \\
MUSYC WFI $I$   & 0.87 & 24.9 & \citet{Taylor09} \\
MUSYC Mosaic II $z$   & 0.91 & 24.5 & \citet{Taylor09} \\
MUSYC ISPI $J$   & 1.25 & 23.6 & \citet{Taylor09} \\
HAWK-I $J$  & 1.26 & 25.7 & Zibetti et al. (in prep.) \\  
MUSYC SofI $H$   & 1.66 & 23.0 & \citet{Taylor09} \\
MUSYC ISPI $K$   & 2.13 & 22.7 & \citet{Taylor09} \\
HAWK-I $K_s$  & 2.15 & 25.3 & Zibetti et al. (in prep.) \\
SIMPLE IRAC $3.6$\um & 3.58 & 24.6 & \citet{Damen10} \\
SIMPLE IRAC $4.5$\um & 4.53 & 24.4 & \citet{Damen10} \\
SIMPLE IRAC $5.8$\um & 5.79 & 22.8 & \citet{Damen10} \\
SIMPLE IRAC $8.0$\um & 8.05 & 23.5 & \citet{Damen10} \\
\hline
\vspace{-6mm}
\end{tabular}
\label{tab:photsum}
\end{minipage} 
\end{table}

In this paper we consider the optical and infrared counterparts to 126
SMGs in the ECDFS detected at $\ge3.7\sigma$ \citep{Weiss09} and
identified by VLA radio, MIPS \citep{Rieke04} 24\,\um\ and IRAC
\citep{Fazio04} emission \citep{Biggs10}.  Following convention and
\citet{Biggs10} we consider robust counterparts as those with a
corrected Poissonian probability of being unassociated with the
submillimetre source \citep[{\it p};][]{Downes86} of $p\le0.05$ in one
or more of the radio, 24\um, or IRAC datasets, or $p=0.05$ -- 0.10 in
two or more; tentative counterparts are those with $p=0.05$ -- 0.10 in
only one of the three bands.

Six of the SMGs have multiple robust counterparts; of these four SMGs
(LESS\,2, LESS\,27, LESS\,49 and LESS\,74) have two counterparts with
photometric redshifts (\S\ref{sec:calcphotz}) consistent with them
being at the same distance and possibly physically associated. Two
SMGs (LESS\,10 and LESS\,49) each have two robust counterparts with
photometric redshifts and SEDs that suggest they are not physically
associated. In these cases, from the information currently available,
it is not possible to determine which of the two counterparts is the
source of the submillimetre flux, or whether the LABOCA detection is a
blend of emission from two galaxies.  To avoid bias we have included
all of the multiple counterparts in our analysis, but we note that
their small number means that their inclusion does not significantly
affect our results.

\subsection{Optical and infrared photometry}
\label{sec:dr_opt}

SMGs typically have faint optical and near-infrared counterparts
\citep[e.g.][]{Ivison02} so we require deep photometry for accurate
photometric redshift estimates. The ECDFS was chosen for this survey
because it is an exceptionally well-studied field, and as such we are
able to utilise data from extensive archival imaging and spectroscopic
surveys.  For completeness and uniformity we only consider surveys
that cover a large fraction of the ECDFS rather than the smaller and
deeper central CDFS region.  Therefore, we utilise the MUltiwavelength
Survey by Yale-Chile \citep[MUSYC;][]{Gawiser06} near-infrared survey
for $U$ to $K$-band imaging \citep{Taylor09}, and the
\Spitzer\ IRAC/MUSYC Public Legacy in ECDFS (SIMPLE) imaging for
\Spitzer\ IRAC data \citep{Damen10}.  We also include $U$-band
data from the deep GOODS/VIMOS imaging survey of the CDFS
\citep{Nonino09}; although this covers only $\sim60\%$ of LESS SMGs it
is valuable for galaxies that are undetected at short wavelengths in
the shallower MUSYC survey.

In addition, we have carried out deep near-infrared observations in the $J$
and $K_s$ bands with HAWK-I \citep{Pirard04, Casali06,
Kissler-Patig08} at the ESO-VLT (ID: 082.A-0890, P.I.\ N.\ Padilla). The
ECDFS was covered with a mosaic of 16 pointings in each band,
with a total exposure time of 0.75 and 1.1 hours per pointing, in the $J$
and $K_s$ bands respectively.
The median seeing is 0.7$^{\prime\prime}$ in $J$ and
0.5$^{\prime\prime}$ in $K_s$. Data reduction has been performed using
an upgraded version of the official ESO pipeline for HAWK-I, customized
calibration has been obtained from observations of photometric standard
stars. More details and catalogues will be published in Zibetti et
al.\ (in preparation).

For accurate photometric redshifts we require consistent photometry in
apertures which sample the same emitting area in each of the 17
filters.  For consistency between surveys and to ensure that all
detected SMG counterparts are included in this study we extract
photometry from the available survey imaging rather than relying on
the catalogued sources.  SMGs are typically brighter at mid-infrared
than optical wavelengths due to their high redshifts and extreme dust
obscuration. Therefore, we use {\sc SExtractor} \citep{Bertin96} to
create a source list from a combined image of the four IRAC channels,
which is weighted such that a given magnitude receives equal
contributions from all of the input images.  Real sources are required
to have at least 4 contiguous $0.6''\times 0.6''$ pixels with fluxes at
least 1.5 times the background noise.  In addition, we visually check
the area within $15^{\prime\prime}$ of each LABOCA source to ensure that no
potential SMG counterparts are missed.  

We next use {\sc apphot} in {\sc iraf} to measure the fluxes in
$3.8^{\prime\prime}$ diameter apertures for each of the four IRAC
bands. We then cut the catalogues to $\ge3\sigma$ based on the
background noise, and finally apply aperture corrections as derived by
the SWIRE team \citep{Surace05} to obtain total source magnitudes.  The
resolution in the $U$- to $K$-band imaging is better than IRAC
(FWHM $\le 1.5^{\prime\prime}$ compared to $\sim2^{\prime\prime}$ for
IRAC) and so we convolve each $U$- to $ K$-band image to match
the $1.5^{\prime\prime}$ seeing of the worst band. We next use {\sc
apphot} to measure photometry in $3^{\prime\prime}$ diameter apertures
at the positions of the IRAC-selected sources.  In all cases, we only
allow {\sc apphot} to re-centroid the aperture if centroiding does not
cause the extraction region to be moved to a nearby source, as flagged
by {\sc iraf}'s CIER parameter when the centroid shift is $>0.5^{\prime\prime}$.
We have not performed any deblending of the photometry but examination
of the images suggests fewer than $\sim10\%$ of the SMG counterparts
are affected.  We note here that the photometric extraction process is
not restricted to SMGs and yields photometry (which allows us to
calculate consistent photometric redshifts) for IRAC-selected sources
throughout the ECDFS.

Finally, to ensure equivalent photometry between the IRAC and
optical-to-near-infrared filters we create simulated IRAC images of
point sources. Using these images we calculate that the correction between
the measured IRAC total magnitudes and the photometry extracted from
$3^{\prime\prime}$ diameter apertures on $1.5^{\prime\prime}$ seeing images is
$-0.014\pm 0.017$ magnitudes, and as such we do not apply any
systematic corrections to the IRAC magnitudes at this stage.  In
\S\ref{sec:calcphotz} we calibrate the photometry prior to photometric
redshift calculation in a process which corrects for small residual
offsets. A summary of our photometry is presented in Table
\ref{tab:photsum}.

The median number of photometric filters per
SMG counterpart is 15 and we require detections in at least three
photometric filters in order to calculate photometric redshifts. Our
final sample therefore contains 74 optical counterparts to the 68
robustly identified SMGs with sufficient detectable
optical-to-infrared emission.  In \S\ref{sec:nz_cf} we show that the
exclusion of the additional 23, tentatively identified, counterparts does not bias our
results.

\subsection{Spectroscopy}

We employ spectroscopy of the ECDFS to calibrate our
photometry with the SED templates (\S\ref{sec:calcphotz}) and test our
photometric redshifts (\S\ref{sec:testphotz}). We have examined the
spectroscopic redshift catalogues from many archival surveys
\citep[][Koposov et al. in prep.]{Cristiani00, Croom01, Bunker03,
Dickinson04, LeFevre04, Stanway04, Strolger04, Szokoly04, vanderWel04,
Zheng04, Daddi05, Doherty05, Mignoli05, Grazian06, Ravikumar07,
Kriek08, Vanzella08, Popesso09, Treister09, Balestra10} and also our
own on-going spectroscopic survey of LESS sources with the VLT (PID:
183.A-0666, P.I.\ I.\ Smail), which will be published in full in
Danielson et al.\ (in prep.).

\section{Analysis}
\label{sec:analysis}

\subsection{Photometric redshift calculation}
\label{sec:calcphotz}

We use \hyperz\footnote{We use \hyperz\ version 10.0
 (http://www.ast.obs-mip.fr/users/roser/hyperz/)} \citep{Bolzonella00}
 to calculate the photometric redshifts of counterparts to LESS SMGs
 \citep{Biggs10}. \hyperz\ compares a model SED to observed magnitudes
 and computes $\chi^2$ for each combination of spectral type, age,
 reddening and redshift and thus statistically determines the most
 likely redshift of the galaxy.  We use the elliptical (E), Sb, single
 burst (Burst) and constant star-formation (Im) spectral templates
 from \citet{Bruzual93} which are provided with \hyperz, and allow
 reddening \citep{Calzetti00} of $A_V=0$--5 in steps of 0.2.  This
 combination of templates and $A_V$ was shown by \citet{Wardlow10} to
 be sufficient for calculating photometric redshifts of SMGs. Redshifts
 between 0 and 7 are considered and galaxy ages are required to be
 less than the age of the Universe at the appropriate redshift.  In
 \S\ref{sec:params} we show that the \hyperz-derived galaxy ages
 cannot be reliably determined, but we note here that the requirement
 for SMGs to be younger than the Universe does not significantly
 affect the derived redshifts.  Galaxies are assigned zero flux in any
 filter in which they are not detected, with an error equal to the
 $1\sigma$ detection limit of that filter.  To ensure that galaxies at
 $z\sim2$--3 do not have their redshifts systematically underestimated
 we have modified the handling of the Lyman-$\alpha$ forest in
 \hyperz, such that intragalactic absorption in the models is
 increased and three different levels of absorption are considered in
 the fitting process.  The reliability of the calculated redshifts and
 the validity of these settings is tested in full in
 \S\ref{sec:testphotz}.

We test for small systematic discrepancies between the photometry and
model SEDs prior to using \hyperz\ to calculate photometric redshifts
of SMGs.  This is done by running \hyperz\ on 1796 galaxies and AGN
with spectroscopic redshifts in the ECDFS and requiring a fit at the
observed redshift. We then compare the model and measured magnitudes
for each galaxy, and iteratively adjust the zeropoints of the filters
with the largest systematic offsets.  This yields significant offsets
for the following filters: VIMOS $ U$ ($0.083$ mag), MUSYC $U$
($-0.091$ mag), $U38$ ($-0.074$ mag), $R$ ($0.049$ mag), $
I$ ($0.048$ mag), $z$ ($0.095$ mag), HAWK-I $J$ ($0.043$ mag),
IRAC 3.6 ($0.043$ mag) and IRAC 8.0\,\um\ ($0.110$ mag). The typical uncertainties in these corrections are
$\pm 0.02$ and the remaining
eight filters have no significant corrections. 

The calibrated photometry of the robust LESS SMG counterparts is listed
in Table~\ref{tab:photo} and in Table~\ref{tab:info} we provide the
coordinates, photometric redshifts, absolute rest-frame $H$-band magnitudes,
far-infrared luminosities and characteristic
dust temperatures of the SMGs (\S\ref{sec:tdlirsfrd}). We also provide
the reduced $\chi^2$ of the best fit SED at the derived photometric
redshift and the number of filters in which the SMG was
detected and undetected (but observed). We caution
that the reduced $\chi^2$ for galaxies with only a few photometric
detections is typically low ($\la0.5$) but the error on the photometric
redshift is typically large, since there are only weak limits on the
SED from the photometry.  Therefore, the values of the reduced $\chi^2$
should be considered in conjunction with the number of photometric
detections when considering the reliability of the photometric
redshifts. 

 The median reduced $\chi^2$ of the SMG counterparts is 2.3
(2.1 if only the galaxies with reduced $\chi^2\le10$ are considered).
This suggests that our photometric errors are slightly overestimated
and lead to apparently overly-precise photometric redshift limits.
Indeed, we find that the \hyperz\ 99\% confidence intervals more
reliably represent the $1\sigma$ errors, yielding $\sim68\%$ of SMGs
with photometric redshifts consistent with the spectroscopic redshifts.
Therefore, throughout this paper we use the \hyperz\ 99\% confidence
intervals on the photometric redshift estimates to represent
the 1-$\sigma$ uncertainty.
Of the 74 SMG counterparts examined there are eight with
poor fits of the SED to the photometry (indicated with reduced
$\chi^2>10$). Of these, one (LESS\,39) is blended in the optical
imaging and two (LESS\,66 and LESS\,81) lie in stellar halos.  LESS\,66
is also likely to be a quasar, as is LESS\,96, and another four SMGs
with reduced $\chi^2>10$ (LESS\,19, LESS\,57, LESS\,75 and LESS\,111)
have excess 8\,\um\ flux compared to the best-fit SED, which is
indicative of an AGN component (see \S\ref{sec:agncontamination} for a
full discussion). Since we did not include any quasar or AGN templates
in the fitting procedure it is unsurprising that these sources are not
well represented by the employed SEDs. We note here that, as we show in
\S\ref{sec:agncontamination}, the exclusion of AGN templates does not
bias our photometric redshift estimates.

\begin{table*}
\begin{minipage}{175mm} 
\caption{The catalogue of 74 robust counterparts to LESS SMGs, their
  photometric redshift estimates, reduced $\chi^2$ of the best-fit SED
  and the number of photometric filters in which the galaxy is
  observed.  We also present the absolute rest-frame $H$-band
  magnitudes, the derived far-infrared luminosities and characteristic
  dust temperatures of the SMGs. }  
\setlength{\tabcolsep}{1.3 mm}
\begin{tabular}{llccccccccl}
\hline
SMG$^{a}$ & Short name & RA$^{b}$ & Dec$^{b}$ & $z_{\rm phot}$$^{c}$ & $\chi^{2}_{\rm red}$$^{d}$ & Filters$^{e}$ & $M_H$$^{f}$ & $L_{FIR}$$^{g}$ & $T_D$$^{h}$ & ID type$^{i}$ \\
 &  &  &  &  &  &  & (mag) & ($10^{12}L_{\sun}$) & (K) & \\
\hline
LESSJ033302.5-275643 & LESS\,2a & 03$^{\rm h}$33$^{\rm m}$02$\fs$55&$-27\degr$56$\arcmin$44$\farcs$7 & $1.80^{+0.35}_{-0.14}$  & $2.8$  & 16 [1] & $-23.42$ & $<1.5$ & $<19.9$ & M\\
LESSJ033302.5-275643 & LESS\,2b & 03$^{\rm h}$33$^{\rm m}$02$\fs$68&$-27\degr$56$\arcmin$42$\farcs$6 & $2.27^{+0.16}_{-0.55}$  & $1.1$  & 8 [9]  & $-23.15$ & $30.9^{+6.4}_{-15.3}$ & $44.2^{+2.5}_{-7.6}$ & R\\
LESSJ033321.5-275520 & LESS\,3 & 03$^{\rm h}$33$^{\rm m}$21$\fs$50&$-27\degr$55$\arcmin$20$\farcs$1 & $3.92^{+0.54}_{-0.72}$   & $0.5$  & 5 [10] & $-24.66$ & $<8.9$ & $<35.2$ & M\\
LESSJ033257.1-280102 & LESS\,6 & 03$^{\rm h}$32$^{\rm m}$57$\fs$15&$-28\degr$01$\arcmin$01$\farcs$5 & $0.40^{+0.09}_{-0.03}$   & $4.3$  & 16 [1] & $-20.29$ & $0.09^{+0.08}_{-0.03}$ & $12.8^{+1.1}_{-0.8}$ & RM\\
LESSJ033315.6-274523 & LESS\,7 & 03$^{\rm h}$33$^{\rm m}$15$\fs$41&$-27\degr$45$\arcmin$24$\farcs$0 & $2.81^{+0.18}_{-0.07}$   & $6.9$  & 16 [1] & $-25.50$ & $16.2^{+4.3}_{-2.4}$ & $41.1^{+2.6}_{-1.9}$ & RM\\
LESSJ033211.3-275210 & LESS\,9 & 03$^{\rm h}$32$^{\rm m}$11$\fs$35&$-27\degr$52$\arcmin$12$\farcs$9 & $4.63^{+0.10}_{-1.10}$   & $2.4$  & 6 [9]  & $-25.29$ & $20.3^{+5.9}_{-11.7}$ & $48.3^{+3.4}_{-10.0}$ & RM\\
LESSJ033219.0-275219 & LESS\,10a & 03$^{\rm h}$32$^{\rm m}$19$\fs$04&$-27\degr$52$\arcmin$14$\farcs$3 & $2.46^{+0.15}_{-0.15}$ & $6.0$  & 12 [5] & $-23.46$ & $8.7^{+2.4}_{-2.0}$ & $34.5^{+2.2}_{-2.2}$ & R\\
LESSJ033219.0-275219 & LESS\,10b & 03$^{\rm h}$32$^{\rm m}$19$\fs$30&$-27\degr$52$\arcmin$19$\farcs$1 & $0.91^{+0.07}_{-0.05}$ & $4.5$  & 15 [2] & $-23.32$ & $0.8^{+0.3}_{-0.2}$ & $18.6^{+1.1}_{-1.0}$ & R\\
LESSJ033213.6-275602 & LESS\,11 & 03$^{\rm h}$32$^{\rm m}$13$\fs$84&$-27\degr$55$\arcmin$59$\farcs$8 & $2.60^{+0.30}_{-0.36}$  & $3.2$  & 7 [9]  & $-24.04$ & $9.9^{+4.4}_{-3.8}$ & $35.9^{+3.5}_{-4.0}$ & R\\
LESSJ033248.1-275414 & LESS\,12 & 03$^{\rm h}$32$^{\rm m}$47$\fs$96&$-27\degr$54$\arcmin$16$\farcs$1 & $3.92^{+1.02}_{-2.11}$  & $0.1$  & 6 [11] & $-24.06$ & $18.2^{+17.3}_{-15.6}$ & $45.6^{+9.8}_{-19.7}$ & RM\\
LESSJ033152.6-280320 & LESS\,14 & 03$^{\rm h}$31$^{\rm m}$52$\fs$47&$-28\degr$03$\arcmin$18$\farcs$6 & $3.56^{+0.92}_{-0.56}$  & $0.8$  & 7 [9]  & $-24.74$ & $32.6^{+26.6}_{-12.5}$ & $51.3^{+10.6}_{-6.7}$ & RM\\
LESSJ033333.4-275930 & LESS\,15 & 03$^{\rm h}$33$^{\rm m}$33$\fs$35&$-27\degr$59$\arcmin$29$\farcs$4 & $1.95^{+3.05}_{-0.39}$  & $0.2$  & 4 [8]  & $-23.59$ & $<1.8$ & $<22.6$ & M\\
LESSJ033218.9-273738 & LESS\,16 & 03$^{\rm h}$32$^{\rm m}$18$\fs$70&$-27\degr$37$\arcmin$43$\farcs$5 & $1.09^{+0.08}_{-0.09}$  & $4.1$  & 17 [0] & $-24.05$ & $1.2^{+0.4}_{-0.4}$ & $20.8^{+1.4}_{-1.5}$ & R\\
LESSJ033207.6-275123 & LESS\,17 & 03$^{\rm h}$32$^{\rm m}$07$\fs$26&$-27\degr$51$\arcmin$20$\farcs$1 & $1.55^{+0.11}_{-0.11}$  & $1.0$  & 17 [0] & $-24.11$ & $6.6^{+1.5}_{-1.4}$ & $32.7^{+2.0}_{-2.1}$ & RM\\
LESSJ033205.1-274652 & LESS\,18 & 03$^{\rm h}$32$^{\rm m}$04$\fs$87&$-27\degr$46$\arcmin$47$\farcs$4 & $2.07^{+0.08}_{-0.09}$  & $2.4$  & 16 [1] & $-24.88$ & $13.8^{+2.1}_{-2.0}$ & $40.2^{+2.1}_{-2.1}$ & RM\\
LESSJ033208.1-275818 & LESS\,19 & 03$^{\rm h}$32$^{\rm m}$08$\fs$23&$-27\degr$58$\arcmin$13$\farcs$7 & $2.11^{+0.11}_{-0.10}$  & $10.3$ & 10 [6] & $-22.80$ & $3.4^{+1.3}_{-1.1}$ & $28.3^{+2.3}_{-2.2}$ & RI\\
LESSJ033316.6-280018$^{j}$ & LESS\,20 & 03$^{\rm h}$33$^{\rm m}$16$\fs$77&$-28\degr$00$\arcmin$15$\farcs$8 & $2.80^{+0.17}_{-0.27}$  & $2.2$  & 9 [7] & $-24.28$ & $903^{+132}_{-190}$ & $124.6^{+7.8}_{-10.5}$ & RM\\
LESSJ033147.0-273243 & LESS\,22 & 03$^{\rm h}$31$^{\rm m}$46$\fs$90&$-27\degr$32$\arcmin$38$\farcs$8 & $1.95^{+0.34}_{-0.38}$  & $2.4$  & 6 [4]  & $-24.67$ & $10.4^{+5.8}_{-4.6}$ & $36.6^{+4.6}_{-5.1}$ & RM\\
LESSJ033336.8-274401 & LESS\,24 & 03$^{\rm h}$33$^{\rm m}$36$\fs$97&$-27\degr$43$\arcmin$58$\farcs$1 & $1.72^{+0.29}_{-0.36}$  & $2.6$  & 11 [2] & $-24.06$ & $4.1^{+2.6}_{-2.1}$ & $29.2^{+3.6}_{-4.3}$ & RM\\
LESSJ033157.1-275940 & LESS\,25 & 03$^{\rm h}$31$^{\rm m}$56$\fs$85&$-27\degr$59$\arcmin$38$\farcs$9 & $2.28^{+0.09}_{-0.15}$  & $3.0$  & 13 [2] & $-24.47$ & $8.1^{+1.8}_{-2.1}$ & $36.4^{+2.4}_{-2.8}$ & RM\\
LESSJ033149.7-273432 & LESS\,27a & 03$^{\rm h}$31$^{\rm m}$49$\fs$88&$-27\degr$34$\arcmin$30$\farcs$4 & $2.10^{+1.00}_{-0.88}$ & $0.1$  & 4 [11] & $-22.83$ & $<2.1$ & $<25.1$ & I\\
LESSJ033149.7-273432 & LESS\,27b & 03$^{\rm h}$31$^{\rm m}$49$\fs$92&$-27\degr$34$\arcmin$36$\farcs$7 & $2.46^{+0.42}_{-0.72}$ & $1.9$  & 7 [6]  & $-23.73$ & $<3.1$ & $<28.1$ & MI\\
LESSJ033336.9-275813 & LESS\,29 & 03$^{\rm h}$33$^{\rm m}$36$\fs$88&$-27\degr$58$\arcmin$08$\farcs$8 & $2.64^{+4.36}_{-0.87}$  & $0.1$  & 4 [8]  & $-24.13$ & $8.4^{+78.5}_{-5.8}$ & $36.8^{+44.0}_{-9.2}$ & R\\
LESSJ033150.0-275743 & LESS\,31 & 03$^{\rm h}$31$^{\rm m}$49$\fs$77&$-27\degr$57$\arcmin$40$\farcs$4 & $3.63^{+0.84}_{-0.70}$  & $0.3$  & 6 [9]  & $-24.30$ & $9.9^{+10.0}_{-5.4}$ & $41.8^{+8.5}_{-7.3}$ & RI\\
LESSJ033217.6-275230 & LESS\,34 & 03$^{\rm h}$32$^{\rm m}$17$\fs$60&$-27\degr$52$\arcmin$28$\farcs$1 & $0.86^{+0.11}_{-0.05}$  & $3.8$  & 17 [0] & $-23.53$ & $<0.3$ & $<15.6$ & M\\
LESSJ033149.2-280208 & LESS\,36 & 03$^{\rm h}$31$^{\rm m}$48$\fs$94&$-28\degr$02$\arcmin$13$\farcs$6 & $2.49^{+0.53}_{-0.31}$  & $0.3$  & 7 [7]  & $-24.58$ & $7.8^{+6.4}_{-3.0}$ & $36.7^{+6.2}_{-4.2}$ & RM\\
LESSJ033336.0-275347 & LESS\,37 & 03$^{\rm h}$33$^{\rm m}$36$\fs$01&$-27\degr$53$\arcmin$49$\farcs$4 & $3.52^{+0.26}_{-0.36}$  & $4.0$  & 11 [1] & $-24.95$ & $<6.9$ & $<37.3$ & M\\
LESSJ033144.9-273435 & LESS\,39 & 03$^{\rm h}$31$^{\rm m}$45$\fs$00&$-27\degr$34$\arcmin$36$\farcs$3 & $2.59^{+0.16}_{-0.06}$  & $12.6$ & 13 [1] & $-24.25$ & $8.2^{+2.8}_{-1.8}$ & $37.7^{+3.3}_{-2.9}$ & RM\\
LESSJ033246.7-275120 & LESS\,40 & 03$^{\rm h}$32$^{\rm m}$46$\fs$77&$-27\degr$51$\arcmin$20$\farcs$7 & $1.90^{+0.10}_{-0.11}$  & $3.1$  & 17 [0] & $-23.61$ & $10.5^{+2.0}_{-1.8}$ & $39.7^{+2.7}_{-2.8}$ & RM\\
LESSJ033110.5-275233 & LESS\,41 & 03$^{\rm h}$31$^{\rm m}$10$\fs$09&$-27\degr$52$\arcmin$36$\farcs$3 & $2.74^{+4.26}_{-0.91}$  & $0.0$  & 4 [0]  & $-25.56$ & $<4.0$ & $<29.9$ & I\\
LESSJ033307.0-274801 & LESS\,43 & 03$^{\rm h}$33$^{\rm m}$06$\fs$63&$-27\degr$48$\arcmin$01$\farcs$9 & $1.67^{+0.23}_{-0.14}$  & $2.0$  & 8 [9]  & $-23.35$ & $<1.3$ & $<22.8$ & MI\\
LESSJ033131.0-273238 & LESS\,44 & 03$^{\rm h}$31$^{\rm m}$31$\fs$19&$-27\degr$32$\arcmin$38$\farcs$6 & $2.49^{+0.00}_{-0.08}$  & $2.8$  & 11 [0] & $-24.82$ & $14.8^{+1.7}_{-2.6}$ & $43.0^{+2.9}_{-3.1}$ & RM\\
LESSJ033256.0-273317 & LESS\,47 & 03$^{\rm h}$32$^{\rm m}$55$\fs$99&$-27\degr$33$\arcmin$18$\farcs$9 & $2.90^{+0.14}_{-0.42}$  & $1.5$  & 8 [6]  & $-23.77$ & $<4.5$ & $<32.7$ & MI\\
LESSJ033237.8-273202 & LESS\,48 & 03$^{\rm h}$32$^{\rm m}$38$\fs$00&$-27\degr$31$\arcmin$59$\farcs$4 & $1.91^{+0.36}_{-0.43}$  & $0.2$  & 4 [1]  & $-24.57$ & $7.5^{+4.8}_{-3.8}$ & $35.0^{+4.9}_{-5.8}$ & RM\\
LESSJ033124.5-275040 & LESS\,49a & 03$^{\rm h}$31$^{\rm m}$24$\fs$45&$-27\degr$50$\arcmin$37$\farcs$5 & $1.50^{+0.15}_{-0.10}$ & $5.0$  & 12 [1] & $-23.22$ & $1.8^{+1.0}_{-0.6}$ & $25.0^{+2.6}_{-2.4}$ & RM\\
LESSJ033124.5-275040 & LESS\,49b & 03$^{\rm h}$31$^{\rm m}$24$\fs$69&$-27\degr$50$\arcmin$46$\farcs$4 & $3.31^{+0.22}_{-0.38}$ & $0.7$  & 11 [2] & $-24.13$ & $35.9^{+8.5}_{-10.6}$ & $58.4^{+4.8}_{-6.4}$ & R\\
LESSJ033141.2-274441 & LESS\,50a & 03$^{\rm h}$31$^{\rm m}$41$\fs$11&$-27\degr$44$\arcmin$42$\farcs$4 & $0.85^{+0.16}_{-0.11}$ & $2.3$  & 17 [0] & $-21.91$ & $<0.3$ & $<15.9$ & M\\
LESSJ033141.2-274441 & LESS\,50b & 03$^{\rm h}$31$^{\rm m}$40$\fs$97&$-27\degr$44$\arcmin$34$\farcs$8 & $2.69^{+0.49}_{-0.25}$ & $7.6$  & 11 [5] & $-24.67$ & $15.1^{+9.0}_{-4.1}$ & $45.6^{+6.7}_{-4.3}$ & RM\\
LESSJ033243.6-273353 & LESS\,54 & 03$^{\rm h}$32$^{\rm m}$43$\fs$62&$-27\degr$33$\arcmin$56$\farcs$6 & $1.84^{+0.62}_{-0.25}$  & $3.7$  & 7 [6]  & $-23.35$ & $<1.6$ & $<24.1$ & M\\
LESSJ033153.2-273936 & LESS\,56 & 03$^{\rm h}$31$^{\rm m}$53$\fs$11&$-27\degr$39$\arcmin$37$\farcs$3 & $2.46^{+0.41}_{-0.24}$  & $0.6$  & 9 [8]  & $-24.38$ & $5.1^{+3.8}_{-2.0}$ & $34.3^{+5.1}_{-3.9}$ & RM\\
LESSJ033152.0-275329 & LESS\,57 & 03$^{\rm h}$31$^{\rm m}$51$\fs$93&$-27\degr$53$\arcmin$26$\farcs$8 & $2.94^{+0.14}_{-0.11}$  & $10.8$ & 11 [6] & $-24.26$ & $11.8^{+3.2}_{-2.6}$ & $43.6^{+3.6}_{-3.5}$ & RM\\
LESSJ033303.9-274412 & LESS\,59 & 03$^{\rm h}$33$^{\rm m}$03$\fs$62&$-27\degr$44$\arcmin$12$\farcs$6 & $1.40^{+0.29}_{-0.13}$  & $1.9$  & 13 [4] & $-23.52$ & $1.3^{+1.2}_{-0.5}$ & $23.5^{+3.6}_{-2.5}$ & RM\\
LESSJ033317.5-275121 & LESS\,60 & 03$^{\rm h}$33$^{\rm m}$17$\fs$53&$-27\degr$51$\arcmin$27$\farcs$5 & $1.64^{+0.10}_{-0.24}$  & $5.1$  & 17 [0] & $-24.01$ & $4.0^{+1.1}_{-1.6}$ & $31.8^{+2.7}_{-3.8}$ & RM\\
LESSJ033236.4-273452 & LESS\,62 & 03$^{\rm h}$32$^{\rm m}$36$\fs$52&$-27\degr$34$\arcmin$53$\farcs$0 & $1.52^{+0.10}_{-0.21}$  & $0.8$  & 16 [1] & $-24.45$ & $7.9^{+1.7}_{-2.6}$ & $37.5^{+3.1}_{-4.2}$ & RM\\
LESSJ033308.5-280044 & LESS\,63 & 03$^{\rm h}$33$^{\rm m}$08$\fs$49&$-28\degr$00$\arcmin$42$\farcs$8 & $1.39^{+0.07}_{-0.05}$  & $2.5$  & 15 [1] & $-23.50$ & $1.3^{+0.6}_{-0.4}$ & $23.8^{+2.4}_{-2.4}$ & RM\\
LESSJ033201.0-280025 & LESS\,64 & 03$^{\rm h}$32$^{\rm m}$00$\fs$98&$-28\degr$00$\arcmin$25$\farcs$3 & $4.19^{+0.04}_{-0.32}$  & $1.9$  & 11 [4] & $-24.31$ & $12.4^{+4.4}_{-5.4}$ & $48.3^{+5.4}_{-6.1}$ & RM\\
LESSJ033331.7-275406 & LESS\,66 & 03$^{\rm h}$33$^{\rm m}$31$\fs$92&$-27\degr$54$\arcmin$10$\farcs$3 & $2.39^{+0.04}_{-0.05}$  & $37.2$ & 14 [0] & $-25.78$ & $10.0^{+1.8}_{-1.7}$ & $41.0^{+3.4}_{-3.4}$ & RM\\
LESSJ033243.3-275517 & LESS\,67 & 03$^{\rm h}$32$^{\rm m}$43$\fs$18&$-27\degr$55$\arcmin$14$\farcs$2 & $2.27^{+0.05}_{-0.11}$  & $3.2$  & 16 [1] & $-24.82$ & $11.9^{+1.6}_{-2.1}$ & $42.8^{+3.2}_{-3.4}$ & RM\\
LESSJ033144.0-273832 & LESS\,70 & 03$^{\rm h}$31$^{\rm m}$43$\fs$92&$-27\degr$38$\arcmin$35$\farcs$2 & $2.31^{+0.15}_{-0.06}$  & $3.8$  & 17 [0] & $-24.48$ & $44.1^{+7.9}_{-3.7}$ & $61.0^{+5.2}_{-4.6}$ & RM\\
LESSJ033229.3-275619 & LESS\,73 & 03$^{\rm h}$32$^{\rm m}$29$\fs$28&$-27\degr$56$\arcmin$18$\farcs$9 & $4.61^{+0.94}_{-0.59}$  & $1.1$  & 8 [9]  & $-24.42$ & $12.3^{+12.2}_{-6.2}$ & $49.3^{+9.9}_{-7.6}$ & R\\
LESSJ033309.3-274809 & LESS\,74a & 03$^{\rm h}$33$^{\rm m}$09$\fs$34&$-27\degr$48$\arcmin$15$\farcs$9 & $1.84^{+0.32}_{-0.49}$ & $0.9$  & 10 [6] & $-23.49$ & $2.8^{+2.2}_{-1.8}$ & $29.2^{+4.3}_{-5.8}$ & RI\\
LESSJ033309.3-274809 & LESS\,74b & 03$^{\rm h}$33$^{\rm m}$09$\fs$14&$-27\degr$48$\arcmin$16$\farcs$6 & $1.71^{+0.20}_{-0.17}$ & $2.5$  & 10 [6] & $-23.29$ & $3.00^{+1.6}_{-1.1}$ & $29.6^{+3.4}_{-3.1}$ & RI\\
LESSJ033126.8-275554 & LESS\,75 & 03$^{\rm h}$31$^{\rm m}$27$\fs$17&$-27\degr$55$\arcmin$50$\farcs$9 & $2.46^{+0.06}_{-0.09}$  & $33.2$ & 15 [0] & $-25.39$ & $11.5^{+2.1}_{-2.2}$ & $43.1^{+3.4}_{-3.5}$ & RM\\
LESSJ033221.3-275623 & LESS\,79 & 03$^{\rm h}$32$^{\rm m}$21$\fs$61&$-27\degr$56$\arcmin$23$\farcs$1 & $1.41^{+0.23}_{-0.17}$  & $2.2$  & 16 [1] & $-23.89$ & $1.5^{+1.1}_{-0.6}$ & $25.4^{+3.4}_{-3.0}$ & RM\\
LESSJ033127.5-274440 & LESS\,81 & 03$^{\rm h}$31$^{\rm m}$27$\fs$54&$-27\degr$44$\arcmin$39$\farcs$5 & $2.23^{+0.13}_{-0.15}$  & $27.9$ & 14 [1] & $-24.89$ & $27.7^{+5.0}_{-4.9}$ & $54.4^{+5.0}_{-5.1}$ & RM\\
\hline
\vspace{-6mm}
\end{tabular}

\label{tab:info}
\end{minipage} 
\end{table*}

\begin{table*}
\begin{minipage}{175mm} 
\contcaption{}  
\setlength{\tabcolsep}{1.4 mm}
\begin{tabular}{llccccccccl}
\hline
SMG$^{a}$ & Short name & RA$^{b}$ & Dec$^{b}$ & $z_{\rm phot}$$^{c}$ & $\chi^{2}_{\rm red}$$^{d}$ & Filters$^{e}$ & $M_H$$^{f}$ & $L_{FIR}$$^{g}$ & $T_D$$^{h}$ & ID type$^{i}$ \\
 &  &  &  &  &  &  & (mag) & ($10^{12}L_{\sun}$) & (K) & \\
\hline
LESSJ033154.2-275109 & LESS\,84 & 03$^{\rm h}$31$^{\rm m}$54$\fs$49&$-27\degr$51$\arcmin$05$\farcs$3 & $2.29^{+0.15}_{-0.07}$  & $3.6$  & 14 [3] & $-24.14$ & $4.5^{+1.8}_{-1.2}$ & $34.5^{+3.6}_{-3.4}$ & I\\
LESSJ033251.1-273143 & LESS\,87 & 03$^{\rm h}$32$^{\rm m}$50$\fs$83&$-27\degr$31$\arcmin$41$\farcs$2 & $3.20^{+0.10}_{-0.81}$  & $0.1$  & 5 [0]  & $-24.84$ & $37.0^{+5.6}_{-19.3}$ & $60.1^{+5.9}_{-12.9}$ & RM\\
LESSJ033155.2-275345 & LESS\,88 & 03$^{\rm h}$31$^{\rm m}$54$\fs$81&$-27\degr$53$\arcmin$40$\farcs$9 & $2.35^{+0.11}_{-0.10}$  & $1.1$  & 16 [1] & $-24.37$ & $11.7^{+2.3}_{-1.9}$ & $44.0^{+4.0}_{-3.9}$ & R\\
LESSJ033313.0-275556 & LESS\,96 & 03$^{\rm h}$33$^{\rm m}$12$\fs$62&$-27\degr$55$\arcmin$51$\farcs$6 & $2.71^{+0.03}_{-0.09}$  & $22.0$ & 17 [0] & $-26.30$ & $16.0^{+1.9}_{-2.5}$ & $49.7^{+4.4}_{-4.5}$ & RM\\
LESSJ033130.2-275726 & LESS\,98 & 03$^{\rm h}$31$^{\rm m}$29$\fs$89&$-27\degr$57$\arcmin$22$\farcs$4 & $1.55^{+0.17}_{-0.16}$  & $1.0$  & 10 [4] & $-24.40$ & $7.8^{+2.8}_{-2.1}$ & $39.9^{+4.4}_{-4.3}$ & RM\\
LESSJ033151.5-274552 & LESS\,101 & 03$^{\rm h}$31$^{\rm m}$51$\fs$53&$-27\degr$45$\arcmin$53$\farcs$1 & $2.39^{+0.36}_{-0.52}$ & $2.5$  & 10 [7] & $-23.51$ & $3.8^{+2.9}_{-2.3}$ & $33.8^{+5.3}_{-6.5}$ & R\\
LESSJ033335.6-274020 & LESS\,102 & 03$^{\rm h}$33$^{\rm m}$35$\fs$56&$-27\degr$40$\arcmin$23$\farcs$2 & $1.68^{+0.13}_{-0.25}$ & $1.1$  & 11 [2] & $-24.34$ & $<1.3$ & $<24.9$ & M\\
LESSJ033325.4-273400 & LESS\,103 & 03$^{\rm h}$33$^{\rm m}$25$\fs$37&$-27\degr$33$\arcmin$58$\farcs$5 & $1.84^{+0.59}_{-0.87}$ & $0.3$  & 5 [7]  & $-23.44$ & $<1.6$ & $<26.3$ & M\\
LESSJ033140.1-275631 & LESS\,106 & 03$^{\rm h}$31$^{\rm m}$40$\fs$17&$-27\degr$56$\arcmin$22$\farcs$4 & $1.96^{+0.31}_{-0.48}$ & $2.1$  & 11 [5] & $-25.00$ & $6.3^{+3.4}_{-3.4}$ & $38.5^{+5.4}_{-7.2}$ & RI\\
LESSJ033316.4-275033 & LESS\,108 & 03$^{\rm h}$33$^{\rm m}$16$\fs$51&$-27\degr$50$\arcmin$39$\farcs$3 & $0.20^{+0.03}_{-0.05}$ & $6.3$  & 15 [0] & $-22.75$ & $0.2^{+0.08}_{-0.08}$ & $24.5^{+2.3}_{-2.4}$ & RM\\
LESSJ033122.6-275417 & LESS\,110 & 03$^{\rm h}$31$^{\rm m}$22$\fs$63&$-27\degr$54$\arcmin$17$\farcs$0 & $2.35^{+4.65}_{-0.44}$ & $0.0$  & 4 [0]  & $-23.22$ & $<2.8$ & $<31.4$ & MI\\
LESSJ033325.6-273423 & LESS\,111 & 03$^{\rm h}$33$^{\rm m}$25$\fs$21&$-27\degr$34$\arcmin$25$\farcs$9 & $2.61^{+0.14}_{-0.06}$ & $14.4$ & 13 [0] & $-24.49$ & $9.8^{+3.4}_{-2.3}$ & $44.1^{+5.0}_{-4.8}$ & RM\\
LESSJ033249.3-273112 & LESS\,112 & 03$^{\rm h}$32$^{\rm m}$48$\fs$85&$-27\degr$31$\arcmin$12$\farcs$8 & $1.81^{+0.42}_{-0.30}$ & $0.7$  & 5 [0]  & $-24.02$ & $2.3^{+2.5}_{-1.2}$ & $28.5^{+5.8}_{-4.9}$ & RI\\
LESSJ033150.8-274438 & LESS\,114 & 03$^{\rm h}$31$^{\rm m}$51$\fs$08&$-27\degr$44$\arcmin$37$\farcs$0 & $1.57^{+0.08}_{-0.07}$ & $1.6$  & 17 [0] & $-24.61$ & $5.3^{+1.1}_{-0.9}$ & $36.8^{+3.7}_{-3.7}$ & RM\\
LESSJ033128.0-273925 & LESS\,117 & 03$^{\rm h}$31$^{\rm m}$27$\fs$62&$-27\degr$39$\arcmin$27$\farcs$3 & $1.73^{+0.29}_{-0.34}$ & $3.3$  & 9 [4]  & $-24.23$ & $5.7^{+3.4}_{-2.6}$ & $37.7^{+5.5}_{-6.0}$ & R\\
LESSJ033121.8-274936 & LESS\,118 & 03$^{\rm h}$31$^{\rm m}$21$\fs$91&$-27\degr$49$\arcmin$34$\farcs$0 & $2.17^{+4.83}_{-1.49}$ & $1.7$  & 5 [1]  & $-22.21$ & $2.8^{+47.5}_{-2.7}$ & $31.8^{+48.7}_{-15.6}$ & R\\
LESSJ033328.5-275655 & LESS\,120 & 03$^{\rm h}$33$^{\rm m}$28$\fs$55&$-27\degr$56$\arcmin$54$\farcs$1 & $1.43^{+0.30}_{-0.21}$ & $2.2$  & 13 [3] & $-23.41$ & $2.1^{+1.8}_{-0.9}$ & $29.2^{+5.0}_{-4.3}$ & RM\\
LESSJ033139.6-274120 & LESS\,122 & 03$^{\rm h}$31$^{\rm m}$39$\fs$52&$-27\degr$41$\arcmin$19$\farcs$4 & $2.08^{+0.08}_{-0.08}$ & $5.2$  & 17 [0] & $-25.14$ & $22.4^{+2.8}_{-2.7}$ & $55.2^{+6.2}_{-6.2}$ & RM\\
LESSJ033209.8-274102 & LESS\,126 & 03$^{\rm h}$32$^{\rm m}$09$\fs$60&$-27\degr$41$\arcmin$06$\farcs$9 & $2.02^{+0.17}_{-0.13}$ & $2.4$  & 12 [4] & $-23.84$ & $2.3^{+1.3}_{-0.9}$ & $30.6^{+4.2}_{-4.1}$ & MI\\
\hline
\end{tabular}
%\vspace{3mm}
$^{a}$The SMG names correspond to those in \citet{Weiss09} and \citet{Biggs10}.\\
$^{b}$Coordinates are the J2000 position of the optical/near-infrared counterpart.\\
$^{c}$Since \hyperz\ was restricted to $0<z<7$ the six galaxies whose upper redshift limits yield a formal maximum redshift of $z_{\rm max}=7$ are actually only constrained in the lower redshift limit. Therefore, throughout this paper the redshifts of these galaxies are plotted as lower limits. \\
$^{d}$The reduced $\chi^2$ of the best-fit SED at the derived photometric redshift.\\
$^{e}$The number of photometric filters in which each SMG counterpart was detected [and the number of filters in which the SMG was observed but not detected, providing a limiting flux].\\
$^{f}$$M_H$ is the absolute magnitude in the rest-frame {\it H}-band.\\
$^{g}$As discussed in \S\ref{sec:tdlirsfrd} the far-infrared luminosity ($L_{FIR}$) is derived from the infrared-radio correlation using the radio flux and the photometric redshift of each SMG.\\
$^{h}$The characteristic dust temperature ($T_D$) is derived as discussed in \S\ref{sec:tdlirsfrd} from radio and submillimetre fluxes and the photometric redshift of each SMG.\\
$^{i}$ID types R, M and I indicate radio, 24\,\um\ and IRAC identified counterparts respectively (see \citealt{Biggs10} for details). \\
$^{j}$As shown in \S\ref{sec:photoz} LESS\,20 appears to contain a radio-loud AGN. Therefore, the $L_{FIR}$ and $T_D$ presented here are likely significantly overestimated due to the AGN contribution to the radio flux, as such LESS\,20 is excluded from our studies of the luminosity function, star-formation rates and star-formation history of SMGs (\S\ref{sec:tdlirsfrd}).\\

\end{minipage} 
\end{table*}

\subsection{Reliability of photometric redshifts}
\label{sec:testphotz}

\begin{figure}
\includegraphics[width=8.5cm]{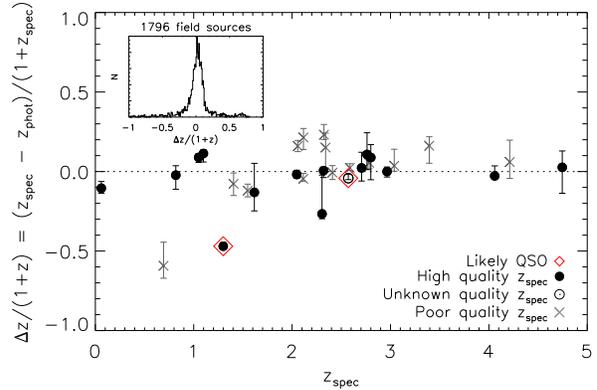}
\caption{Spectroscopic redshift against $\Delta z/(1+z)$ for robust
  counterparts to LESS SMGs. We distinguish between high and low
  quality spectroscopic redshifts as determined by the flags provided
  in most archival catalogues, and highlight the two likely quasars
  (LESS\,66 and LESS\,96; see Appendix \ref{sec:sources}). The median
  (mean) $\Delta z/(1+z)$ for the all the SMGs is $0.023\pm0.021$
  ($-0.013\pm0.178$).  The inset plot shows the histogram of $\Delta
  z/(1+z)$ for 1796 galaxies and AGN in the ECDFS with spectroscopic
  redshifts.  The distribution is centered on $0.016\pm 0.002$ and has
  a $1\sigma$ dispersion of $0.05$.  We conclude that our photometric
  redshifts are a good proxy for spectroscopic redshifts for both
  samples. }
\label{fig:zphotzspec}
\end{figure}

To test the reliability of our photometric redshifts ($z_{\rm phot}$)
we first compare them to the spectroscopic redshifts ($z_{\rm spec}$)
for 1796 galaxies and AGN in the ECDFS and calculate $\Delta z= z_{\rm
spec}-z_{\rm phot}$ for each source. The histogram of $\Delta z/(1+z)$
for these 1796 sources is shown as an inset in
Fig.\ref{fig:zphotzspec}; the sample is centered on $\Delta z/(1+z)
=0.016\pm0.002$ and has a $1\sigma$ dispersion of $0.05$.  We define
outliers as sources with $|\Delta z|/(1+z)>0.3$; the outlier fraction
for these 1796 field galaxies and AGN is 0.15.  We also calculate the
outlier resistant normalised median absolute deviation (NMAD) of
$\Delta z$, $\sigma_{\rm NMAD} = 1.48\times{\rm median}(|\Delta z -
{\rm median}(\Delta z)|/(1+z))=0.097$. These statistics show that our
photometric redshifts are a good proxy for spectroscopic redshift for
these sources. However, the median redshift, $z=0.84$, is lower than
that expected for SMGs and the targets are typically brighter at
optical wavelengths, limiting the usefulness of these comparisons for
the SMGs. Therefore, we also test our photometric redshift calculation
on the 30 robust SMG counterparts with available spectroscopic
redshifts.

Fig.~\ref{fig:zphotzspec} shows spectroscopic redshift against $\Delta
z/(1+z)$ for the 30 robust LESS SMG counterparts with spectroscopic
redshifts from archival surveys of the ECDFS (10 counterparts) and our
spectroscopic survey of SMGs in the ECDFS (20 counterparts).  Quality
flags are published in many catalogues and where possible we
distinguish between high- and low-quality spectroscopic redshifts. 
The median $\Delta z/(1+z)$ for SMGs is $0.023\pm 0.021$ (here are
throughout this paper errors on median measurements are from
bootstrapping) the mean $\Delta z/(1+z) = -0.013\pm0.178$, and
$\sigma_{\rm NMAD}=0.037$, suggesting that for SMGs our photometric
redshifts are reliable. We caution that the SMGs without reliable
spectroscopic redshifts are fainter on average than the SMGs with
spectroscopic redshifts, which could affect the quality of their
photometric redshifts. Although the median $R$-band magnitude of the SMGs
detected in the MUSYC survey is the same for the counterparts with and
without spectroscopic redshifts, all of the 30 SMGs with
spectroscopic redshifts are detected in the MUSYC $R$ band,
while only  18 of the 45 SMGs without spectroscopic redshifts are detected.

\citet{Dunlop10} have independently calculated photometric redshifts
for the six LESS SMGs (five with robust counterparts) in GOODS-South
that were also detected by BLAST at 250\,\um\ \citep{Devlin09}. Their
photometry uses imaging from \HST\ ($B_{435}$, $V_{606}$, $i_{775}$,
$z_{850}$), the VLT ($J$, $H$, $K$) and \Spitzer\ (3.6, 4.5, 5.8 and
8\,\um) and they use \hyperz\ with the stellar population models of
Charlot \& Bruzual \citep[e.g.][]{Bruzual07} which have a Salpeter
IMF. In all cases the \citep{Dunlop10} photometric redshifts agree with
those presented in Table \ref{tab:info}, providing further confidence
that our photometric redshifts are reasonable for SMGs.

To assess the level of systematic uncertainties for the derived
photometric redshifts due to the adopted methodology, SED templates,
and/or photometric data, we also use the Zurich Extragalactic Bayesian
Redshift Analyzer (ZEBRA; \citealt{Feldmann06}) to calculate
photometric redshifts. Our adopted procedure is similar to that
discussed in \S3.2 of \citet{Luo10}. Briefly, we use ZEBRA to obtain a
maximum-likelihood estimate for the photometric redshifts of
individual galaxies or AGNs using an initial set of 265 galaxy, AGN,
and galaxy/AGN hybrid SED templates.  These SED templates were then
expanded to 463 templates during the template-training mode of ZEBRA
to best represent the SEDs of the $\approx2$~Ms CDFS X-ray sources
\citep{Luo10}, including AGN.  Besides the different SED templates
used, this method differs from the \hyperz\ approach described in
\S\ref{sec:calcphotz} in some additional details such as how the
redshift intervals and minimum photometric errors are determined; see \S3.2
of \citet{Luo10} for details.

The ZEBRA-derived photometric redshifts ($z_{\rm phot,check}$) were
compared to those listed in Table 1 ($z_{\rm phot}$); the difference
was measured by $\delta z_{\rm phot}=(z_{\rm phot,check}-z_{\rm
phot})/(1+z_{\rm phot})$.  For sources with secure spectroscopic
redshifts, individual $|\delta z_{\rm phot}|$ values range from
$\approx 0.01$ to 0.10, indicating that both methods are able to
deliver photometric redshifts to a similar accuracy. For the full
sample, the mean (median) value of $\delta z_{\rm phot}$ is $-0.006$
($0.011$), with an rms scatter of 0.028, suggesting that the
photometric redshifts in Table 1 are fairly robust.
After accounting for the effective
$1\sigma$ errors of the photometric redshifts, only three (sources
LESS\,7, LESS\,37 and LESS\,111) of the 74 sources have
inconsistent $z_{\rm phot}$ and $z_{\rm phot,check}$.
As some sources have photometry data in addition to those presented in
Table 2, we also tested the effect of including more data points. The
photometric redshifts differ by a mean value of $|\delta z_{\rm
phot}|$ of $0.024$ with an rms error of 0.030, after including the
WFI $R$-band data \citep{Giacconi02, Giavalisco04} for 25 sources and
the {\it GALEX} near-UV and far-UV data \citep{Morrissey07} for 3
sources. Given the small difference caused by the additional data, we
consider the consistent aperture photometry in Table 2 suitable for
the purpose of deriving reliable photometric redshifts.

\subsection{Sample subsets}
\label{sec:nz_cf}

\begin{figure}
\includegraphics[width=8.5cm]{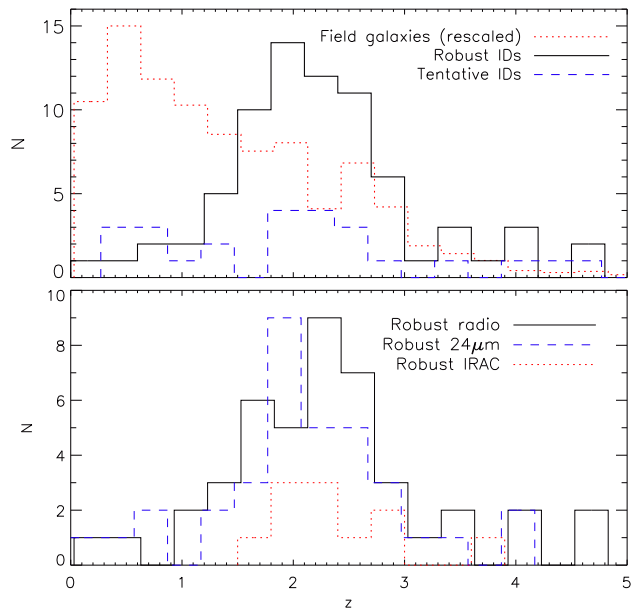}
\caption{ In the top panel we present a comparison between robustly
and tentatively identified SMG counterparts and the field population
of the ECDFS; tentative counterparts and field galaxies are offset
slightly in redshift for clarity. The robust
counterparts have a median redshift of $z=2.2\pm0.1$, compared to $z=2.0\pm0.2$
for the tentative counterparts. Tentative counterparts have a larger
interquartile range -- 1.2--2.8 compared to 1.8--2.7 for the robust
SMG counterparts. We interpret these distributions as evidence that
tentative counterparts are mainly drawn from the same parent
population as the robust counterparts, but with the addition of some
contamination, particularly at low redshifts.
In the lower panel we compare the redshift distributions of robust
counterparts (with $p\le0.05$) in radio, 24\,\um\ and IRAC data; for
clarity radio and 24\,\um\ counterparts are plotted offset slightly in
redshift.  The radio, 24\,\um\ and IRAC samples have median redshifts
of $2.3\pm0.1$, $2.1\pm0.2$ and $2.3\pm0.2$ and interquartile ranges of
1.8--2.8, 1.6--2.5 and 1.9--2.7 respectively. Therefore, we find no
significant differences in the redshift distributions of the three
identification methods.  }
\label{fig:nz_cf}
\end{figure}

As discussed in \S\ref{sec:intro}, \citet{Biggs10} identified robust
 counterparts to 71 SMGs (of which 68 have detectable optical
counterparts) and  tentative counterparts to 26 LESS SMGs, respectively. In Fig.~\ref{fig:nz_cf}
we compare the redshift distributions of the robust and tentative
counterparts to determine whether our results may be biased by the
exclusion of tentative counterparts in our main analysis.  The median
redshifts of robust and tentative counterparts are $2.2\pm0.1$ and
$2.0\pm0.2$ and the interquartile ranges are $z=1.8$--2.7
and $z=1.2$--2.8, respectively. We also
use a Kolmogorov-Smirnov test (KS-test), which calculates the
probability that two samples are drawn from the same parent population
($P_{KS}$), to compare the redshift distributions of robust and
tentative counterparts statistically. We find $P_{KS}=0.09$ and
conclude that it is likely that there is some contamination from
physically unassociated foreground ($z\la1$) galaxies in the
tentative identifications. 
Therefore, throughout the remainder of this
paper we restrict our analysis to robust counterparts. 

We note that our identified sample may contain a small number of potential gravitational lenses. 
These are typically low-redshift counterparts where the radio or mid-infrared emission is offset from the optical source. These are discussed
individually in Appendix \ref{sec:sources} and we have confirmed that
their inclusion does not affect
our results.

In Fig.~\ref{fig:nz_cf} we also compare the redshift distributions of
counterparts with $p\le 0.05$ in the radio, 24\,\um\ and IRAC
data. The median redshifts are $z=2.3\pm0.1$, $2.1\pm0.2$ and
$2.3\pm0.2$, and the interquartile ranges are $z=1.8$--2.8, 1.6--2.5 and
1.9--2.7 for the radio, 24\um\ and IRAC samples respectively.  
A comparison of the three redshift samples shows that they are
statistically indistinguishable. We conclude that the three
counterpart identification methods select galaxies with similar
redshift distributions, and are not significantly biased with respect
to each other.

\subsection{The effect of AGN on photometric redshifts}
\label{sec:agncontamination}

As discussed in \S\ref{sec:calcphotz} our photometric redshift
calculations are based on fitting stellar templates to the SMG
photometry.  However, studies have shown that the 8\,\um\ flux
in SMGs with a luminous AGN component can be dominated by the AGN and
therefore fitting stellar templates may yield misleading results
\citep[][]{Hainline09, Hainline10, Coppin10b}.

We employ two methods to identify potential AGN in the LESS SMGs.
Firstly, we cross-correlate the LESS SMG counterparts with the {\it
Chandra} X-ray catalogues of the CDFS \citep{Luo08} and ECDFS
\citep{Lehmer05} with a matching radius of $1^{\prime\prime}$. This
yields 12 X-ray luminous SMG counterparts (16\% of the robust LESS
counterparts).  Secondly, we identify nine SMG counterparts (12\% of
the robust LESS counterparts) with a large excess of 8\,\um\ flux
compared to the best-fit SED template, which potentially indicates
obscured power-law emission from an AGN. In total this yields 16 SMG
counterparts (22\% of the robust LESS counterparts) which may contain
AGN (five are both X-ray detected and have an 8\,\um\ excess).

To determine whether AGN contamination in the 8-\um\ filter reduces the
accuracy of our photometric redshift estimates we re-fit the photometry
of the counterparts which display an 8-\um\ excess whilst excluding the
8\,\um\ photometry. Spectroscopic redshifts are available for eight of
the affected galaxies and we find that the average $|\Delta z|$ of
these eight galaxies is not significantly decreased, while the median
reduced $\chi^2$ drops by 1.2, when the 8\,\um\ photometry is excluded
from the fitting. This result does not change if we also exclude the
8\,\um\ photometry of the SMGs which are X-ray detected.

It is also possible that our sample of SMG counterparts contains AGN
which enhance the 8\,\um\ flux but do not cause a detectable excess.
Therefore, we also exclude the 8\,\um\ photometry of {\it all} the SMG
counterparts during the fitting procedure. Once again the average
$|\Delta z|$ for the spectroscopic sample does not change significantly
(median $\Delta z/(1+z)=0.032\pm0.021$, compared to median $\Delta
z/(1+z)=0.023\pm0.021$ originally). These results indicate that when calculating
photometric redshifts the benefit of including the longer-wavelength
data is greater than the bias which is removed by ignoring the 8\,\um\
photometry. Therefore, we include the 8\,\um\ photometry in the SED
fitting.

\subsection{Reliability of SED parameters}
\label{sec:params}
In addition to calculating photometric redshifts, \hyperz\ also returns
the spectral type, age and reddening of the best-fit SED template. To
test the sensitivity of the choice of template we refit the photometry
of the SMGs allowing only the Burst template, and then only a constant
star-formation rate history (Im). These two templates represent the
extremes of the star-formation histories and so they will let us
gauge the sensitivity of the derived parameters to the choice of the
best-fit template.

We compare the quality of the Burst and Im fits of each galaxy with
$\Delta\chi^2_{\rm red}$ -- the difference in the reduced $\chi^2$ of
the Burst and Im fits. We find that 63\% of the SMGs have
$|\Delta\chi^{2}_{\rm red}|\le1$ and as such the Burst and Im templates
are indistinguishable at the 99\% level for these SMGs.  61\% (17) of
the SMGs with $|\Delta\chi^{2}_{\rm red}|>1$ between the two template
fits are best-fit by Bursts and 39\% (11) by Im templates. If three
templates -- Burst, Sb and Im -- are considered the star-formation
histories of only 23\% (17) of all the SMG counterparts can be
distinguished. Therefore, although it may be possible to
crudely distinguish the star-formation histories of a fraction ($\sim
20$--40\%) of SMG counterparts with \hyperz, the star-formation
histories of most SMGs counterparts cannot be reliably established.

We find that the SMGs that have SED fits with $|\Delta\chi^{2}_{\rm
red}|\le1$ and can be equally well fit by either Burst or Im templates
have different age estimates depending on the template.  The age and
star-formation history of a stellar population affects the
light-to-mass ratio; thus our inability to distinguish
between star-formation histories  leads to
uncertainties in the light-to-mass ratios and stellar mass estimates.
Using the $H$-band light-to-mass ratios for Burst and Im models
from the {\sc Starburst99} stellar population model \citep{Leitherer99} we
calculate that the uncertainties in SED fitting parameters result in a
$1\sigma$ dispersion of a factor of 4.6 range in the 
light-to-mass ratios and consequently  in
the stellar mass estimates (see also Fig.~\ref{fig:stackopt}).  

We also compare reddening measurements for those galaxies with SED fits
with $|\Delta\chi^{2}|\le1$ and find that on average the difference
between $A_V$ for the best-fit Im and Burst templates, $\Delta A_V$, is
equal to $0.32\pm0.16$. Since average estimates based on either
template return the same value of $A_V$ (to $\sim2\sigma$) we conclude
that average reddening measurements for the SMG population are not
strongly sensitive to the adopted star-formation history and are likely
to be statistically meaningful (although we caution against trusting
values for individual SMGs).
When fitting E, Sb, Burst, or Im templates and allowing $A_V=0$--5, as
in our photometric redshift calculations we determine a median
$A_V=1.5\pm0.1$ and 96\% of SMGs have $A_V\le3$. We conclude that in
most instances restricting to $A_V\le3$ is sufficient for calculating
photometric redshifts of SMGs. We note that these values of $A_V$ are
integrated across the whole galaxy and that obscuration in the regions
responsible for the majority of the far-infrared/submillimetre emission
is considerably larger \citep[e.g.][]{Chapman04a, Takata06, Ivison10}.

\section{Results and Discussion}
\label{sec:results}

In \S\ref{sec:analysis} we derived reliable photometric redshifts for
74 counterparts to 68 robustly identified SMGs. We now use these
photometric redshifts and the SED fits to further investigate the
properties of SMGs.

\subsection{Photometric redshifts}
\label{sec:photoz}

% N(z)
\begin{figure*}
\begin{minipage}{17.5cm}
  \subfigure{
\includegraphics[width=8.5cm]{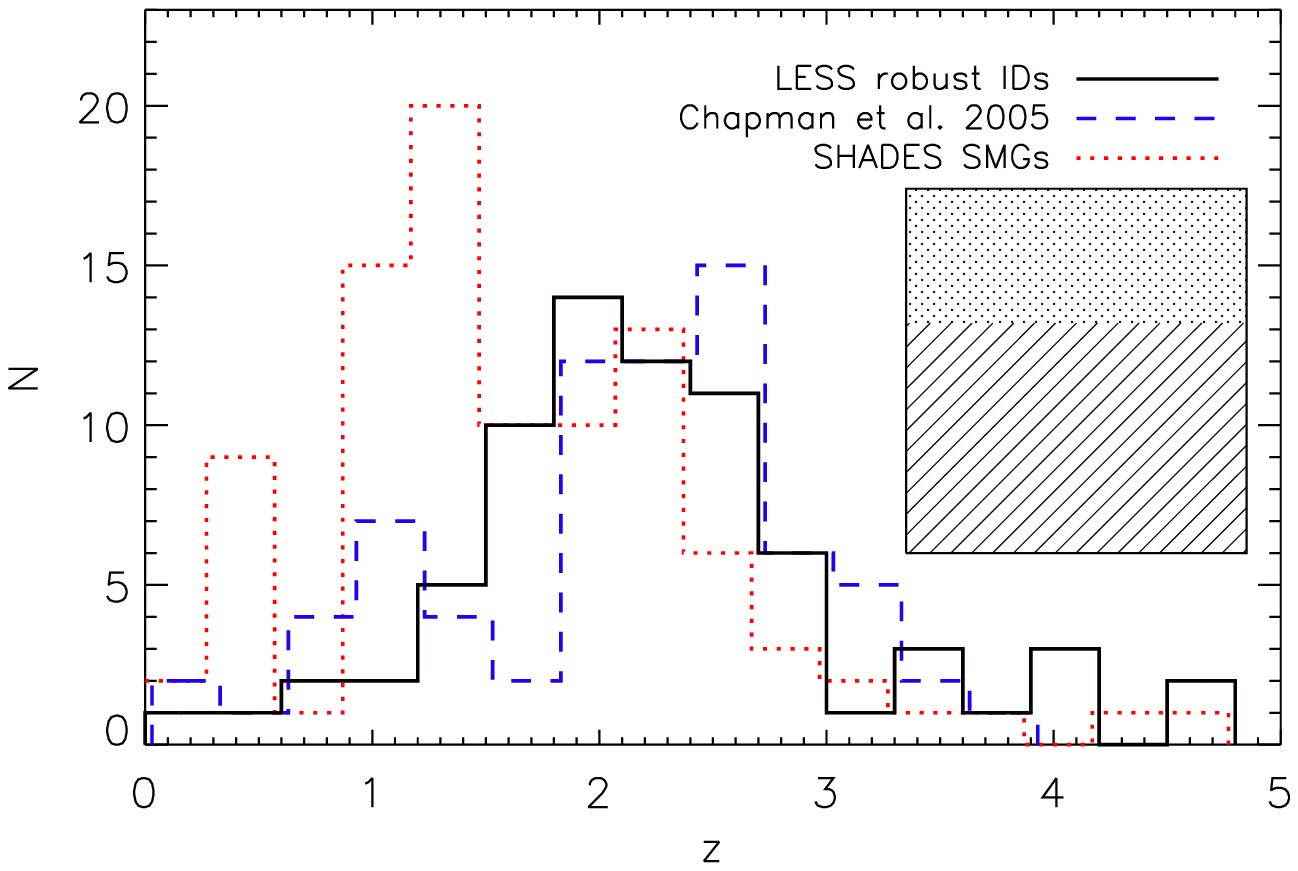}}
  \subfigure{
\includegraphics[width=8.5cm]{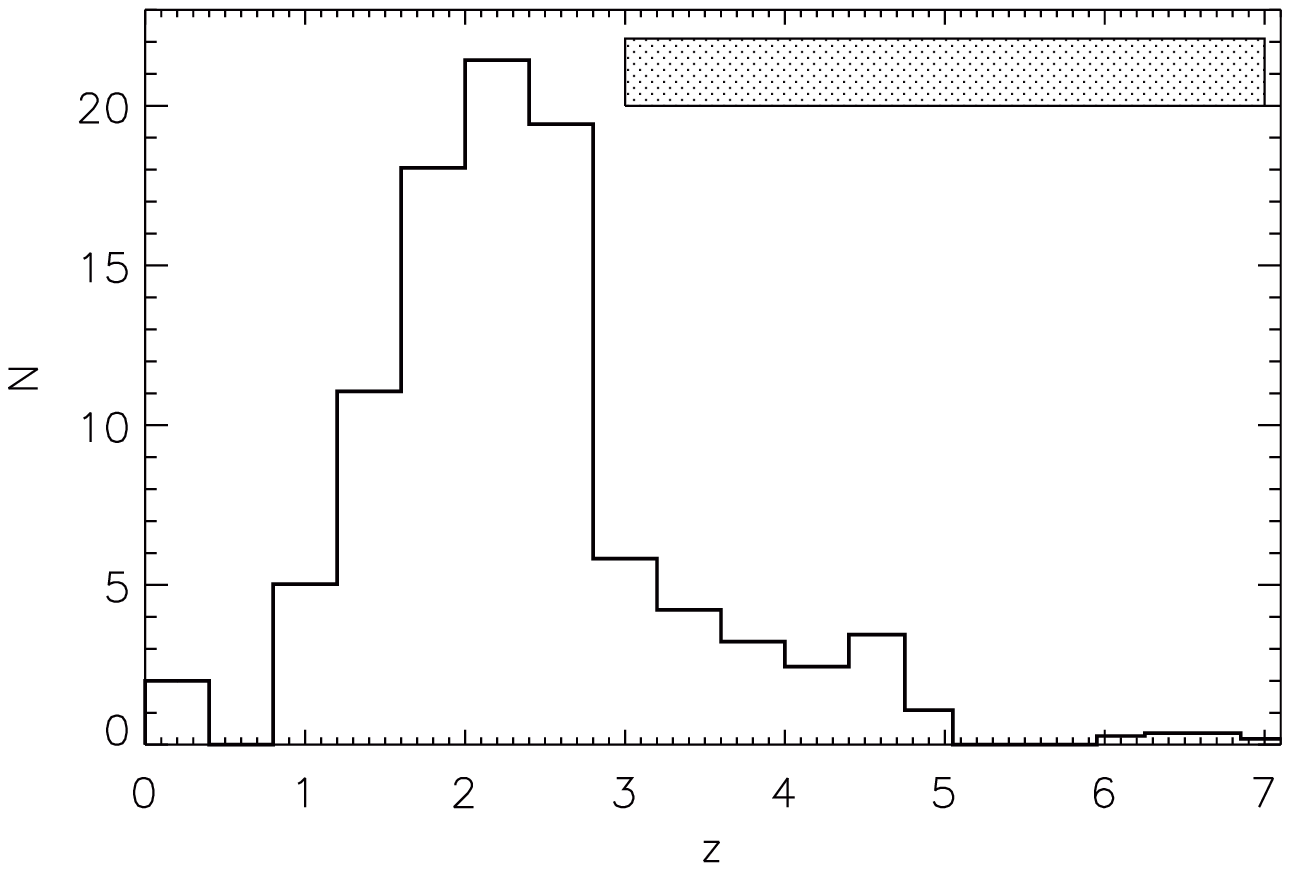}}
\caption{[left] The photometric redshift distribution of the LESS SMGs.
  We compare this to the photometric redshift distribution of SHADES
  SMG counterparts \citep{Clements08, Dye08}, and the spectroscopic
  redshift distribution of SMGs from \citet{Chapman05}; for clarity the
  SHADES and \citet{Chapman05} samples are offset slightly in redshift.
  The median redshift of identified SMGs in LESS is $z=2.2\pm0.1$,
  which is the same as that from the \citet{Chapman05} spectroscopic
  survey; SHADES has a lower median redshift of $z=1.5\pm0.1$.  There
  is a slightly larger high-redshift tail in the LESS SMG population
  than the \citet{Chapman05} SMG population. Additionally, the
  so-called `redshift desert' at $z\sim1.5$, which is evident in the
  \citet{Chapman05} study does not affect our photometric redshifts and
  as such, in contrast with \citet{Chapman05}, the increase in the
  number of galaxies from $z\sim1$ to $z\sim2$ is smooth. Statistical
  comparisons show that the \citet{Chapman05} and LESS SMGs are most
  likely drawn from populations with similar redshift distributions,
  but that the SHADES SMGs are biased to low redshifts either from
  systematic errors in the photometric redshift calculations, sample
  selection, or cosmic variance. The shaded region represents the area
  that would be added to the histogram were the redshifts of the 57
  statistically identified or completely unidentified SMGs known and is
  designed to give an impression of the potential contribution of the
  unidentified SMGs to this figure.  The lower region corresponds to
  the unidentified SMGs that we statistically identify in
  \S\ref{sec:noidzs}, and which have redshifts similar to the
  identified SMGs.  The upper shaded region represents the SMGs which
  remain unaccounted for after the statistical analysis and likely have
  $z\ga3$.
[right] The same distribution but now including the statistically
identified SMG population from \S\ref{sec:noidzs} in addition to the
identified sample.  These are distributed uniformly within the relevant
$\Delta z=1$ ranges.  The shaded area now represents the remaining
unidentified SMGs, which are likely to lie at $z\ga 3$.  We conclude
that the median redshift of the $S_{870\mu m}\ga 4$\,mJy SMG population
is likely to be $z=2.5\pm 0.6$.  }
\label{fig:nz}
\end{minipage}
\end{figure*}

In Fig.~\ref{fig:nz} we show our photometric redshift distribution for
the 74 robust SMG counterparts; it peaks at $z=2.2\pm0.1$ and has an
interquartile range of 1.8--2.7. We compare to the photometric redshift
distribution of SMG counterparts in the SCUBA Half-Degree Extragalactic
Survey (SHADES) \citep{Clements08, Dye08}, median $z=1.5\pm0.1$, and
the spectroscopic sample from \citet{Chapman05}, median $z=2.2\pm0.1$,
both of which have similar submillimetre flux limits as our survey.
The LESS SMGs have a similar redshift distribution to
\citet{Chapman05}, although in LESS the spectroscopic `redshift desert'
at $z\sim 1.2$--1.8 is filled and there is a larger high-redshift
tail. The redshift distributions of both LESS and \citet{Chapman05}
SMGs are peaked at higher redshifts than the SHADES SMGs.  A KS-test
between \citet{Chapman05} and LESS SMGs yields $P_{KS}=0.44$,
suggesting the two samples appear to be drawn from the same parent
population. However, a KS-test between the LESS and SHADES SMGs gives
$P_{KS}=1.3\times10^{-5}$ indicating that these samples are likely
drawn from intrinsically different populations.

We conclude that the global properties of our photometric redshifts
are consistent with the largest previous spectroscopic survey, albeit
with a higher-redshift tail -- we find 10 (14\%) SMGs with $z\ge 3$
and eight (11\%) with $z\ge3.5$, of which LESS\,73 is spectroscopically
confirmed at $z=4.76$ (see \citealt{Coppin09, Coppin10a}). It is
likely that the larger number of high-redshift sources in the LESS survey
compared to \citet{Chapman05}, where there
are 10\% at $z\ge 3$ and just 1\% at $z\ge 3.5$, is due to deeper radio data (on
average) and the inclusion of 24-\um\ 
counterparts in LESS \citep{Biggs10} and most critically the use
of photometric redshifts covering the UV to mid-infrared
which are less reliant on the detection of  spectral features
in the optical.  Conversely, the SHADES SMGs
appear to typically lie at lower redshifts. We stress that compared to
the SHADES analyses we have used about twice as many photometric
bands, tested against a larger spectroscopic sample of SMGs, and
obtained qualitatively better fits to the SEDs. We suggest that either
there is a systematic error in the original SHADES photometric
redshifts or their counterpart identifications, or that cosmic
variance is the cause of the different redshift distributions.
However, we note that a re-analysis of the optical-to-infrared
photometry of the SHADES SMGs yields a median photometric redshift of
$z=2.05$ (Schael et al.\ in prep.) -- more similar to
LESS and \citet{Chapman05} than the original SHADES analyses.

%{Submilimeter flux vs z}
\begin{figure}
\includegraphics[width=8.5cm]{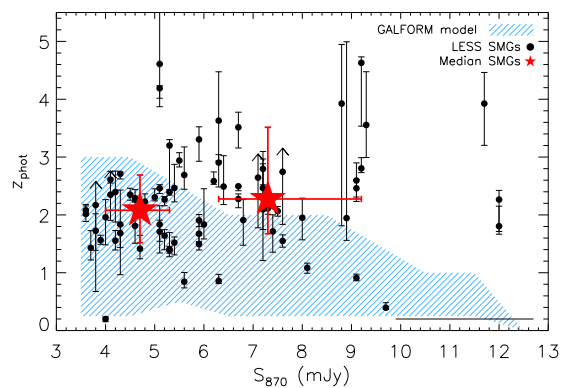}
\caption{ Photometric redshift versus submillimetre flux for LESS
  SMGs; the median $S_{870\mu m}$ error bar is shown in the bottom right.
  The median $S_{870\mu m}$ and redshift, with $1\sigma$ error bars, are
  presented for SMGs with $S_{870\mu m}\le5.6$ mJy and $S_{870\mu m}>5.6$
  mJy.  Previous studies have
  suggested that the brightest SMGs may lie at the highest redshifts.
  This work contains optical-infrared photometric redshifts for a
  larger sample of SMGs than previous studies of the phenomenon and
  finds no evidence for a trend. For comparison we also highlight the 1-$\sigma$ distribution of
  SMGs in flux bins of 1 mJy in the $\Lambda$CDM {\sc galform} model
  \citep{Baugh05, Lacey08, Swinbank08}.  }
\label{fig:zvss870}
\end{figure}

% nz non-id'd plot
\begin{figure*}
\begin{minipage}{17.5cm}
\includegraphics[width=17.5cm]{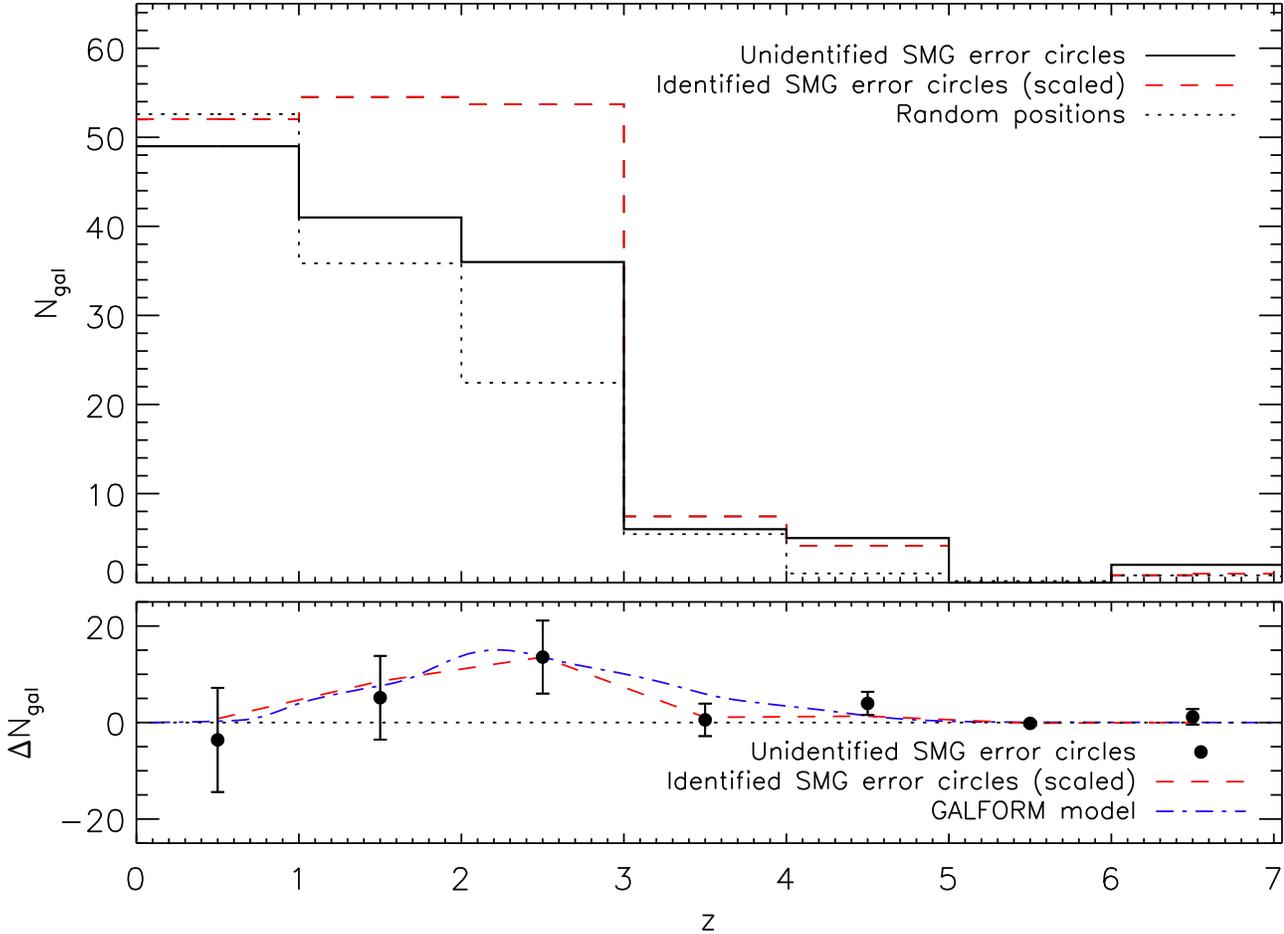}
\caption{ The top panel shows the redshift histogram of sources within
the submillimetre positional error circles of SMGs without robust
radio, 24\,\um\ or IRAC counterparts compared to the same number of
random positions in the field. For comparison we also plot the redshift
histogram for sources in the submillimetre positional error circles of
SMGs {\it with} robust counterparts, also scaled to the same number of
error circles.  In the bottom panel we show the difference between the
redshift histogram of galaxies around unidentified SMGs and the field
population (from random positions). We also plot both the difference in
redshift of galaxies around identified SMGs and the field population,
and the redshift distribution of radio-undetected SMGs in the
$\Lambda$CDM {\sc galform} model \citep{Swinbank08}. 
In order to highlight potential differences in the
redshift distributions of the populations the latter two datasets are
scaled to match the value of $\Delta N_{\rm gal}$ of unidentified SMGs
in the $z=2$--3 bin. By using $\Delta N_{\rm gal}$ and assuming a
uniform distribution of galaxies within the bins we calculate that the
average redshift of unidentified SMGs is $z=2.5\pm0.3$.  }
\label{fig:noidzs}
\end{minipage}
\end{figure*}

Studies have suggested that the brightest SMGs may have higher
redshifts than those with lower submillimetre fluxes
\citep[e.g.][]{Ivison02, Pope05, Biggs10}.  In Fig.~\ref{fig:zvss870}
we plot the photometric redshift against 870\,\um\ flux ($S_{870\mu m}$) for
robust LESS SMG counterparts.  We split the galaxies into those
brighter and fainter than the median deboosted submillimetre flux of the sample,
$S_{870\mu m}=5.6$\,mJy (we use deboosted 870-$\mu$m fluxes
through out). SMGs with $S_{870\mu m}\le5.6$ mJy have a median redshift of
$z=2.1\pm 0.2$ and SMGs with $S_{870\mu m}>5.6$ mJy have a median redshift of
$z=2.3\pm 0.2$, where the errors are bootstrap uncertainties
on the medians. Spearman's rank correlation coefficient between $S_{870\mu m}$
and $z_{\rm phot}$ is 0.20, which corresponds to a probability of zero
correlation of 0.08 and indicates that there is no significant
 correlation between submillimetre flux and redshift for SMGs in our sample.
We have verified that the result is not dependent on the choice of the
flux limit between the two bins. Additionally, if all the unidentified
SMGs lie at $z=5$ or $z=1$ (in \S\ref{sec:noidzs} both of these
scenarios are shown to be unlikely) we still find no statistically
significant difference between the redshifts of SMGs in the two flux
bins.  The sample of SMGs with optical-infrared photometric redshifts
in this work is larger than previous studies of this phenomenon and our
analysis finds no significant correlation between $S_{870\mu m}$ and
redshift for robustly identified sources, also implying that $S_{870\mu m}$
is not a good proxy for redshift. This result agrees with
\citet{Knudsen10}, who find no difference in the redshift distributions
of faint lensed SMGs ($S_{850\mu m}<2$ mJy) and the brighter ($S_{850\mu m}\ga3$
mJy) SMGs from \citet{Chapman05}.  We note that SMGs in the
semi-analytic $\Lambda$CDM {\sc galform} model \citep{Baugh05, Lacey08, Swinbank08}
also show no correlation between $S_{870\mu m}$ and redshift (although the
error range decreases at high fluxes where there are few
galaxies in the model).

\subsection{Redshift distribution of unidentified SMGs}
\label{sec:noidzs}

To date, redshift surveys of SMGs have focused on the $\sim 60$--80\%
of the population with counterparts identified from radio and 24\,\um\
imaging, and a few located using high-resolution (sub-)millimetre
interferometry.  The requirement for radio or infrared counterparts to
SMGs can bias the redshift or the dust temperature distributions of
identified SMGs \citep{Chapman05} and it is currently unknown if the
identified population is representative of the $\sim20$--40\% of SMGs
without identified counterparts. In particular, it is unclear whether
they have the same redshift distribution. In order to investigate the
redshift distribution of the unidentified SMGs we utilise our
extensive 17-band photometric redshifts in the ECDFS to investigate 
the photometric redshifts of sources around SMGs without
robustly identified counterparts to the field population.

In Fig.~\ref{fig:noidzs} we show the redshift histogram of galaxies
within the error circles of the 55 unidentified SMGs in the ECDFS,
where the region considered is that used by \citet{Biggs10} to
identify SMG counterparts (both the completely unidentified and those with only
tentative identifications).  For comparison, we show the redshift
histogram of all the galaxies in the submillimetre error circles of
SMGs with robustly identified counterparts, scaled such that the
number of error circles examined is the same as the unidentified SMG
sample.  We also consider the photometric redshifts of galaxies in the
same area around random positions in the field, which are required to
be $>15^{\prime\prime}$ from any LESS SMGs.  We consider 50 Monte Carlo simulations
of 55 random field positions (equal to the number of unidentified
SMGs), and employ the mean and standard deviation of the 50 simulations
in each redshift bin for our statistical analyses.  As discussed in
\S\ref{sec:dr_opt} our photometric source extraction procedure
included manually examining the regions around the SMGs and adding to
the catalogue potential sources which may have been missed by the
automated procedure. To remove any bias and ensure a fair comparison
between the SMGs and random positions, we exclude these additional
sources from this analysis.

In the lower panel of Fig.~\ref{fig:noidzs} we show the difference
between the redshift distributions of the field and SMGs without
robustly identified counterparts. Compared to the field there is an
excess of $26\pm12$ ($2.2\sigma$) at $z>1$ around unidentified SMGs.  
There are positive excesses of galaxies in the $z=2$--3
($14\pm8$; $1.8\sigma$) and $z=4$--5 ($4\pm2$; $1.6\sigma$) bins around
unidentified SMGs. This suggests that the peak of the redshift
distribution of the population of radio, 24-\um\ and IRAC unidentified
SMGs is at $z=2$--3.
 
To crudely compare the redshift distributions of the identified and
unidentified SMGs we also plot the difference between redshifts of
sources in the submillimetre error circles of the identified SMGs to
the field (scaled to the value in the $z=2$--3 bin of the unidentified
SMGs). We conclude that the redshift distribution of unidentified SMGs
is broadly similar to that of robustly identified SMGs.

To provide a more reliable estimate of the average redshift of the unidentified
SMGs we evenly distribute the excess galaxies in the SMG error circles 
in each redshift bin. 
We verify that this method is valid by using it to calculate the
average redshift for identified SMGs, which yields $z=2.2\pm0.1$, in
agreement with that derived using the robust counterparts alone
(\S\ref{sec:photoz}). For unidentified SMGs we derive an average
redshift of $z=2.5\pm0.3$.  This suggests that unidentified
SMGs may lie at marginally higher redshifts than the identified sample,
although we stress that the difference is not statistically
significant.  This conclusion is consistent with the
predicted redshift distribution of radio undetected
SMGs from the semi-analytic $\Lambda$CDM {\sc galform} model,
which predicts they should lie at $z\sim2.2$,
similar to the observed SMGs \citep{Swinbank08}.

There are $14\pm8$ more galaxies at $z=2$--3 in the error
circles of unidentified SMGs
than expected from comparing to the field population. We showed in
\S\ref{sec:nz_cf} that there are nine tentative SMG counterparts with
$z=2$--3 and two with $z=4$--5.  Thus potentially half of the
excess seen around the unidentified SMGs could be attributed
to tentative counterparts.    Indeed, if
tentative SMGs are removed from this analysis a $0.6\sigma$
($3\pm5$) excess of $z=2$--3 galaxies and a $1.3\sigma$
($2.4\pm1.9$) excess of $z=4$--5 galaxies around SMGs remains.

Finally, we use statistical arguments to estimate how many SMGs are
still unaccounted for. There are 55 out of the sample of 126 LESS SMGs
without robust radio, 24\,\um\ or IRAC counterparts
\citep{Biggs10}. Due to the signal-to-noise ratio limit on the
submillimetre catalogue ($S/N\ge3.7\sigma$) five of the 126 SMGs are
expected to be false detections \citep{Weiss09}; an additional 1--2 are
expected to have counterparts outside of the search radii used
\citep{Biggs10}. This leaves 48--49 SMGs with currently unidentified
counterparts that are expected to lie within our search area.  We then
calculate the total excess of galaxies in unidentified error circles
over the field. Due to clustering, an error circle can contain more
than one galaxy associated with the SMG.  We then compare the sources
around identified SMGs with the number of identified counterparts to
determine this ``overcounting factor'': $\sim 1.2\times$. We scale the
difference between the field and unidentified SMG regions by this
factor to estimate that there are $21\pm19$ LESS SMGs ($17\pm15\%$ of
the total) which are still unaccounted for. These have no robust radio,
24\,\um\ and IRAC counterparts and have mid-infrared fluxes below the
limits of our imaging.

These counterparts could lie at $z\sim 1$--3 and be fainter than
$M_H\la -23$ (Fig.~\ref{fig:zhmag}).  However, the specific
star-formation rates of such sources would be $\ga10^{-7}\,{\rm
yr}^{-1}$, corresponding to life times of $\la10$\,Myr. The
corresponding duty cycle for such short-lived sources means that to
detect $\sim20$ sources in our 0.5\,deg$^2$ survey, we require a parent
population with a space density of $\ga 0.02\,{\rm Mpc}^{-3}$, which we
consider unlikely.  Alternatively, if they have restframe near-infrared
luminosities similar to the identified SMG population, then
Fig.~\ref{fig:zhmag} suggests that they must lie at $z\ga3$.  If
correct we should add these sources to the SMGs identified at $z\ga 3$.
We have identified 10 SMGs at $z>3$, as well as $4\pm 2$ which have
been statistically identified in our IRAC sample. To these we add the
$21\pm19$ SMGs which are unaccounted for in our statistical analysis,
to derive a total of $35\pm19$ SMGs ($28\pm15\%$ of the whole
population) at $z\ge3$ in our survey.  We conclude that $\sim30\%$, and
at most $\sim45\%$ of the SMG population could reside at $z\ga3$.  This
corresponds to a volume density of $2.8\times 10^{-6}$\,Mpc$^{-3}$
(assuming they span the range $z=3$--7, or 80\% higher if they only
span $z=3$--5).  For comparison, the equivalent volume density of
$z=2$--3 SMGs, including identified counterparts and the $14$ that are
statistically identified in this redshift range, is
$1.2\times10^{-5}$\,Mpc$^{-3}$, signifying strong evolution in the
abundance of SMGs from $z>3$ to $z\sim2.5$.

We have statistically identified the redshifts of $\sim60\%$ of the
unidentified SMGs, and shown that the remainder likely lie at $z\ga3$.
In the right-hand panel of Fig.~\ref{fig:nz} we combine the redshift
distributions of the identified and unidentified SMGs to provide the
probable redshift distribution of the {\it entire} $S_{870\mu
m}\ga4$~mJy SMG population.  We conclude that the most likely median
redshift for the $S_{870\mu m}\ga4$~mJy SMG population is
$z=2.5\pm0.6$.

\subsection{Simple redshift estimators for SMGs}
\label{sec:simplez}

Previous studies have investigated and used optical ($BzK$),
ultraviolet (BX/BM; \citealt{Steidel04, Chapman05}) and IRAC colours
\citep{Yun08, Hainline09} and radio-to-submillimetre flux ratios
\citep[e.g.][]{Carilli99, Ivison07, Biggs10}  as simple estimators of
the redshifts of SMGs. Here we use our 17-band photometric redshifts to
investigate the reliability of such the $BzK$ colours and
radio-to-submillimetre fluxes as redshift estimators and derive a
simple IRAC colour indicator of redshift.

% BzK
\begin{figure}
\includegraphics[width=8.5cm]{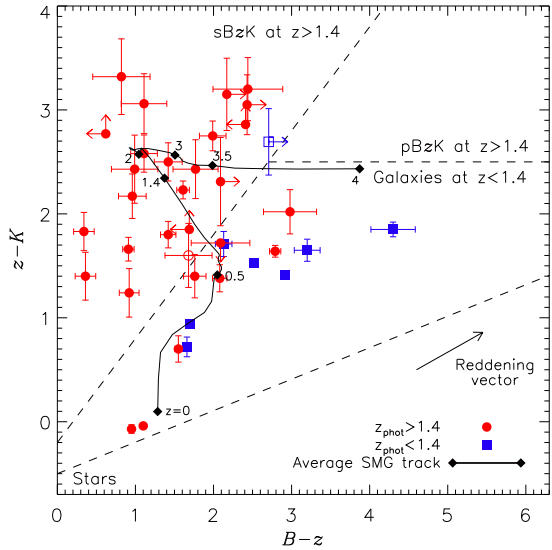}
\caption{$(B-z)$ versus $(z-K)$ colour-colour plot of LESS SMG
counterparts. The selection regions for star-forming and passive
$z>1.4$ $BzK$ galaxies (sBzK and pBzK respectively), $z<1.4$
galaxies, and stars \citep{Daddi04} are shown, and we distinguish
between SMG counterparts with $\zphot\ge1.4$ and $\zphot<1.4$. The
photometric redshifts typically agree with the $BzK$ colours and most
of the SMG counterparts have $BzK$ colours of $z>1.4$ star-forming
galaxies and none have colours of stars or $z>1.4$ passive galaxies
(similar to the result of \citealt{Bertoldi07}).  All of the
counterparts with sBzK colours are found to have $\zphot\ge1.4$, but
galaxies with BzK colours suggesting $z<1.4$ have a $\sim45\%$
contamination from counterparts with $\zphot\ge1.4$. We also show the
redshift track of the average SMG SED (\S\ref{sec:typicalsed}) from
$z=0$--4 and the reddening vector for $A_V=1$\,mag.  The scatter in the
photometry of individual SMGs compared to the redshift track of the
average SMG SED suggests that the SMGs have a  range in optical
SEDs. Open symbols show galaxies which lie in halos of bright stars in
the $z$-band, in these cases, for the purpose of this plot only, the 
$z$-band magnitude is extrapolated from the SED fit and the measured
$I$-band magnitude.  }
\label{fig:bzk}
\end{figure}

In Fig.~\ref{fig:bzk} we show the $BzK$ colour-colour plot
\citep{Daddi04}, which is designed to identify galaxies at
$1.4<z\la2.5$, for LESS SMG counterparts.  We have distinguished
between counterparts with photometric redshifts above and below
$z=1.4$ and find that all the SMGs with $\zphot<1.4$ lie in the
expected region of colour-colour space. However, whilst SMG
counterparts with $\zphot\ge1.4$ typically have the colours of
high-redshift star-forming galaxies, this population does scatter into
the low-redshift region.  Two galaxies with $\zphot>1.4$, $B-z\sim1$
and $z-K\sim-0.1$ lie near the separation between $z<1.4$
galaxies and stars and are both X-ray luminous, bright unresolved
sources which may be submillimetre-bright quasars (see
Appendix \ref{sec:sources}). 
We conclude that the $BzK$ analysis of SMG counterparts can
select clean but incomplete samples of $z>1.4$ SMGs, and that samples
selected to have $z<1.4$ will contain $\sim45\%$ contamination from
galaxies at higher redshift.  We compare the observed $BzK$ colours
with a redshift track of the average SMG SED (\S\ref{sec:typicalsed})
and note that the median
SED of SMGs has a redder restframe $(U-z)$ colour (corresponding to
observed $(z-K)$ at $z\sim 1.4$) than used to define the
selection areas for $z>1.4$ galaxies and that the SMGs at the highest
redshifts may fall in the passive $ BzK$ or $z<1.4$ region.

\begin{figure*}
\begin{minipage}{17.5cm}
\includegraphics[width=17.5cm]{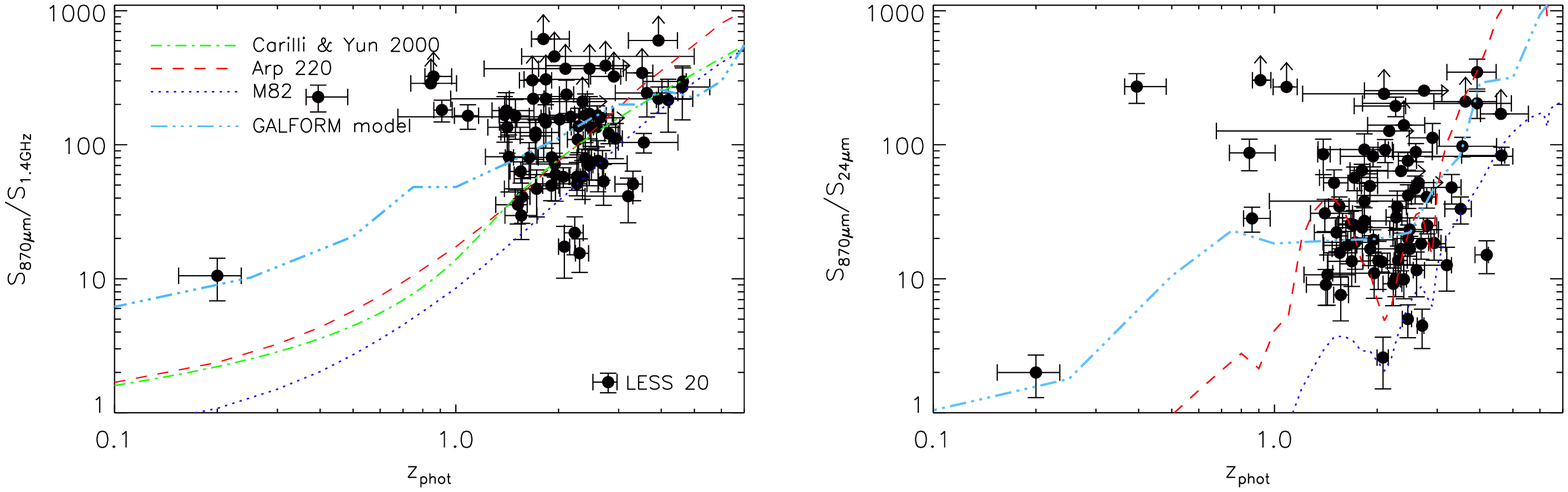}
\caption{The variation of submillimetre to radio (left) and
  submillimetre to mid-infrared (right) flux ratios with redshift,
  compared with Arp 220 and M82 (based on the SEDs of
  \citealt{Silva98}) and SMGs in the $\Lambda$CDM {\sc galform} model
  \citep{Baugh05} with $S_{850\mu m}\ge3$ mJy; we also
  show the relationship between redshift and radio-to-submillimetre
  spectral index derived by \citet{Carilli00} in the left-hand panel.
  The SMGs show two orders of magnitude dispersion in both $S_{870\umu
  m}/S_{\rm 1.4 GHz}$ and $S_{870\mu m}/S_{24\mu m}$.  The model
  track of $S_{870\mu m}/S_{\rm 1.4 GHz}$ for M82 lies below the
  majority of the SMGs, while that of Arp 220 more closely follows the
  SMGs suggesting that they typically have a characteristic dust
  temperature which is cooler than that in M82 and more like that in
  Arp 220.  Similarly, although studies have found that mid-infrared
  spectral properties of SMGs are similar to M82
  \citep{MenendezDelmestre09} we find that M82 does not describe the
  submillimetre to mid-infrared continuum flux ratios well, and that Arp 220 fits
  better to this data. LESS\,20 (labelled) is significantly brighter at 1.4\,GHz than
  expected from its submillimetre flux and redshift and is most likely
  a radio-bright AGN. }
\label{fig:fluxratios}
\end{minipage}
\end{figure*}

In Fig.~\ref{fig:fluxratios} we plot the photometric redshift against
$S_{870\mu m}/S_{\rm 1.4 GHz}$ for the LESS SMGs \citep{Weiss09,
Biggs10}, the tracks of Arp 220 and M82 (based on the SEDs of
\citealt{Silva98}). We also show the $\Lambda$CDM {\sc galform}
predictions \citep{Baugh05} and the \citet{Carilli00} relationship. The
wide range in $S_{870\mu m}/S_{\rm 1.4 GHz}$ at a fixed redshift limits
the usefulness of $S_{870\mu m}/S_{\rm 1.4 GHz}$ as a redshift
indicator for SMGs and indicates that SMGs have a variety of
submillimetre-to-radio flux ratios, suggesting a range in dust
temperatures \citep{Chapman05, Clements08}.  We also note that the
majority of SMGs lie above the redshift track of M82, suggesting higher
submillimetre-to-radio flux ratios (potentially due to the presence of
more cold dust).  LESS\,20 has $z_{\rm phot}\sim2.8$ and $S_{\rm 870\mu
m}/S_{\rm 1.4GHz}\sim1.7$, which is significantly lower than expected from
its redshift, indicating that it is most likely a radio-bright AGN, so
we remove it from our subsequent analyses of far-infrared luminosities,
star-formation rates, and characteristic dust temperature
(\S\ref{sec:tdlirsfrd}).

Studies of mid-infrared spectra of SMGs have shown that they are
similar to M82 with an additional power-law contribution from AGN
emission \citep{MenendezDelmestre07, MenendezDelmestre09, Pope08,
Coppin10b}.  In Fig.~\ref{fig:fluxratios} we plot photometric redshift
against $S_{870\mu m}/S_{24\mu m}$, which shows that the mid-infrared
to submillimetre flux ratios of SMGs are similar to Arp 220 and those
derived in the $\Lambda$CDM {\sc galform} model, but are poorly
represented by M82.  This suggests that although SMGs have mid-infrared
spectra similar to M82, the mid-infrared continuum emission is fainter
compared to the far-infrared emission and is more comparable to that of
Arp 220. We note that although $S_{870\mu m}/S_{24\mu m}$ for SMGs
varies with redshift in a manner comparable to Arp 220, the scatter and
the effect of PAH and silicate features passing through the $24$\,\um\
filter makes this measurement unsuitable for redshift derivation
\citep{Pope06}.

%{Infrared and radio flux ratios vs z}

\begin{figure}
\includegraphics[width=8.5cm]{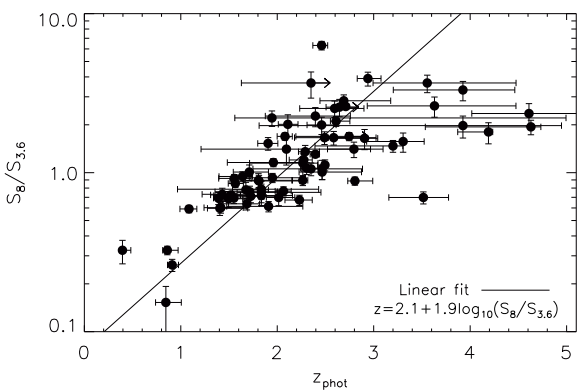}
\caption{ The correlation between redshift and the ratio of 8\,\um\ to
  3.6\,\um\ flux for the LESS SMGs. This shows a trend
  and so we plot a linear fit to the SMGs with $z<4$, which yields
  $z=2.1+1.9{\rm log}_{10}(S_{8}/S_{3.6})$, with a $1\sigma$ dispersion
  in redshift of $\sigma_{z}=0.44$. This relation may be useful as a
  crude redshift indicator for SMGs as we note that
$\sim 90$\% of all SMGs with $z>2$ have $S_8/S_{3.6}>1$, while
similarly $\sim 90$\% of all SMGs with $z<2$ have $S_8/S_{3.6}<1$.}
\label{fig:iracz}
\end{figure}

We expand on the work of \citet{Yun08} and \citet{Hainline09} and
propose a new redshift estimator for SMGs, which is based on the IRAC
8 and 3.6\,\um\ fluxes and exhibits less scatter than the
commonly-employed radio-to-submillimetre flux ratio. In
Fig.~\ref{fig:iracz} we plot this ratio against redshift for the LESS
SMGs and using the {\sc robust\_linefit} procedure from the {\sc
idl} Astronomy Library \citep{Landsman93} we fit an outlier-resistant
linear relationship to SMGs with $z_{\rm phot}<4$, which yields:
\begin{equation}\label{eqn:iracz}
z=(2.1\pm0.1) + (1.9\pm0.2){\rm log}_{10}(S_{8}/S_{3.6})  
\end{equation}
We exclude SMGs with $z_{\rm phot}\ge4$ from the fit because at high
redshifts the 1.6\,\um\ stellar peak passes through the 8\,\um\ filter
making this redshift estimator unreliable. For SMGs with $z<4$ the
$1\sigma$ dispersion in redshift estimated using Eqn. \ref{eqn:iracz}
is  $\sigma_{z}=0.44$ and we
find that $\sim 90$\% of SMGs at $z>2$ have $S_8/S_{3.6}>1$.

\subsection{Typical SMG SEDs}
\label{sec:typicalsed}

\begin{figure}
\includegraphics[width=8.5cm]{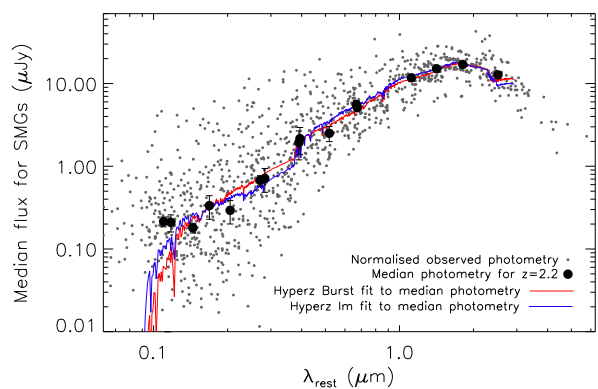}
\caption{ The photometry of SMG counterparts shifted to the rest frame
and normalised to the $H$ band (1.6\um). Redshifting this data to
$z=2.2$ we calculate the apparent fluxes in the 17 photometric filters
considered throughout this paper and use \hyperz\ to fit galaxy
templates at this redshift. The median photometric points are shown and
the resulting best \hyperz\ Burst and Im template fits are displayed. The
best fit \hyperz\ templates  have:  Burst,
$A_V=1.7$, an age of 33\,Myrs and a $L_H/M^*\sim 24$;  Im, $A_V=1.1$, an age of 3.5\,Gyrs
and an $L_H/M^*\sim 6$. It is clear that it is not possible to
distinguish between these two very different star-formation histories and
hence there is a factor of $\sim 5$ uncertainty in the resulting masses. The MUSYC U38
filter has an $\ge 50\%$ contribution from limiting magnitudes and is
excluded from the fit.}
\label{fig:stackopt}
\end{figure}

To investigate the SED of a typical SMG we show in
Fig.~\ref{fig:stackopt} the SMG photometry in the rest-frame and
normalised in the $H$-band. We also calculate the expected fluxes
expected in each of the 17 photometric filters used throughout this
paper as observed at $z=2.2$ and determine the median flux in each.

We begin by noting that the data show evidence for a break at $\sim
3500$--4000\AA\ suggestive of a Balmer or 4000\AA\ break. Closer
inspection hints at it being a Balmer break indicating that the blue
rest-frame light is dominated by stars older than 20\,Myr and younger
than $\sim2$\,Gyr.  Then, as in \S\ref{sec:calcphotz}, we use \hyperz\
to fit this photometry, with redshift fixed at $z=2.2$, and show the
best-fit templates for both the Burst and Im star-formation histories
on Fig.~\ref{fig:stackopt}.  Comparing the $\chi^2$ for these two
models in the same manner as \S\ref{sec:params}, we find that we cannot
accurately distinguish between different star-formation histories (and
hence ages or light-to-mass ratios).  The best-fit Burst model has an
age of just 33\,Myrs, an $A_V=1.7$ and a resulting light-to-mass ratio
of $L_H/M^*\sim 24$, in contrast the Im template yields an age of
3.5\,Gyrs, $A_V=1.1$ and an $L_H/M^*\sim 6$.  The reddening derived
from these two template fits are agreement with the median of the
individual SED fits ($A_V=1.5\pm0.1$; \S\ref{sec:params}), while the
$L_H/M^*$ span a range of $4\times$.

We also estimate the extinction in LESS SMGs by comparing the
star-formation rate (SFR) derived from the rest-frame far-ultraviolet
luminosity (median ${\rm SFR_{UV}}=2$\,\myr; \citealt{Kennicutt98})
with the SFR derived from the far-infrared luminosity (median ${\rm
SFR_{FIR}}=1400$\,\myr; \S\ref{sec:tdlirsfrd}). Comparing the two
values yields $A_V=2.6\pm0.2$, corresponding to reddening $\sim4$ times
higher than the SED fit and indicating that the 
majority of the star formation within SMGs occurs in
totally obscured regions.  As discussed in
\citet{Kennicutt98} the conversion from far-ultraviolet luminosity to
${\rm SFR_{UV}}$ assumes that the star-formation rate has been constant for
$>10^8$ years.  SMGs are likely to be shorter bursts of activity and
therefore for a fixed SFR they will brighter at ultraviolet wavelengths
and likely have higher $A_V$ than estimated above. 

% z vs irac colours

% sed plot

\subsection{Stellar masses}
\label{sec:mass}

\begin{figure}
\includegraphics[width=8.5cm]{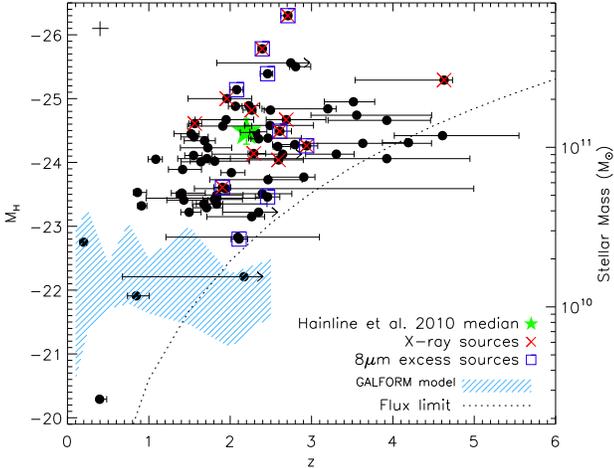}
\caption{A plot of photometric redshift against rest-frame $H$-band
  absolute magnitude for LESS SMGs, and the approximate correspondence
  with stellar mass (as described in the text). 
The median $M_H$ is $-24.1\pm0.1$ with
  an interquartile range of $-24.7$ to $-23.6$ which corresponds to a
  median stellar mass of $\sim 9.2\times10^{10}M_{\sun}$ and
  interquartile range of $(4.7$--$14)\times10^{10}M_{\sun}$
in good agreement with the median $M_H$ from
  \citet{Hainline10}.  We also highlight SMGs with evidence for AGN activity
  from X-ray detections or 8\,\um\ excesses, which appear
brighter than the average SMG. We also show the
  1-$\sigma$ distribution of absolute $H$-band magnitudes of SMGs with
  $S_{850\mu m}\ge 3$ mJy that are brighter than our flux limit at 4.5\,\um\
  (approximately the rest-frame $H$-band at $z=2$) from {\sc
  galform} \citep{Baugh05, Lacey08, Swinbank08} and note that $M_H$ is
  under-predicted in the model.  Errors in $M_H$ are dominated by the
  error in the photometric redshift; we calculate the error in $M_H$
  for SMGs with median redshift error by re-running the \hyperz\ with
  the redshift forced to the extremes of the error range; the
  corresponding errors are shown in the top left-hand corner of the
  plot and can be scaled with the error in redshift. The dotted line
  illustrates the trend in $M_H$ with redshift resulting from the flux
  limited nature of our survey. }
\label{fig:zhmag}
\end{figure}

We use \hyperz\ to estimate the rest-frame $H$-band absolute
magnitudes ($M_H$) from our SED fits and find the median
$M_H=-24.1\pm0.1$, with an interquartile range of $-24.7$ to
$-23.6$. In Fig.~\ref{fig:zhmag} we plot $M_H$ against photometric
redshift for the LESS SMG counterparts.  There is the suggestion of a
weak trend of $M_H$ with redshift. However, as the plotted detection
limit shows this is most likely a selection effect, with the higher
redshift galaxies needing to be more luminous to be detected. The average
$z=2$--3 SMG has $M_H=-24.4$ and would be detected in our survey out
to $z\sim3$--4, and therefore, any incompleteness in our SMG sample due
to the IRAC flux limits is
only significant at $z\ga3$.

$M_H$ is used to estimate the stellar mass of galaxies because it is
less influenced by young stars than optical bands and is relatively
unaffected by dust.   As discussed in
\S\ref{sec:params} the uncertainties in the derived spectral types and
ages result in an estimated factor of $\sim5$ uncertainty in assumed
mass-to-light ratios and thus stellar masses derived from
$M_H$. Therefore, we only consider the stellar masses of the LESS SMGs
statistically.

\citet{Hainline10} estimated $H$-band mass-to-light ratios for SMGs
with Burst and Im templates, based on a \citet{Chabrier03} IMF . We use
the average of their values converted to a Salpeter IMF (with a lower
mass limit of 0.1\,$M_{\sun}$ and an upper mass cutoff of
100\,$M_{\sun}$) for our stellar mass calculation:
$L_H/M^*=3.8\,L_{\sun}M_{\sun}^{-1}$. We estimate that the median
stellar mass of the SMGs in our sample is
$M^*=(9.2\pm0.9)\times10^{10}M_{\sun}$ and the interquartile range is
$(4.7$--$14)\times10^{10}M_{\sun}$. The quoted errors do not include
the systematic uncertainty from the star-formation histories and
mass-to-light ratios, which adds a factor of $\sim5$ uncertainty to the
values (\S\ref{sec:params}; Fig.~\ref{fig:stackopt}).  We also caution
that the choice of IMF coupled with the assumption that all the light is
from the current burst can affect the derived stellar masses
by an additional factor of $\sim2$. Finally, we note that on
average we observe the SMGs approximately halfway through the burst and
typical SMG gas masses \citep{Greve05} suggest an additional
$\sim3\times10^{10}M_{\sun}$ could be added by the end of the burst.

We find that galaxies with evidence for AGN activity from an 8\,\um\
excess or X-ray emission have median $M_H=-24.6\pm0.3$, compared to
$M_H=-24.1\pm0.1$ for the remainder of the SMGs. The two SMGs with the
brightest $M_H$ are the two submillimetre bright quasars (LESS\,66 and
LESS\,96; Appendix \ref{sec:sources}) in which the observed emission is
expected to be dominated by the AGN rather than starlight \citep{Hainline10}. If these are
excluded the median $M_H$ of SMGs containing AGN is
$M_H=-24.5\pm0.3$. 

The median stellar mass for SMGs in SHADES Lockman Hole was claimed to
be $M^*=(6.3^{+1.6}_{-1.3})\times 10^{11}M_{\sun}$ by \citet{Dye08}.
This is a factor of $\sim7$ higher than our estimate for LESS
SMGs. \citet{Dye08} use nine-band photometry for their photometric
redshift determination and claim to also be able to disentangle the
star-formation histories of the SMGs with sufficient accuracy to
identify a significant mass of old stars which underlies the current
burst. This leads to a higher effective mass-to-light ratio and
correspondingly higher stellar masses. 
In contrast, as discussed earlier (\S\ref{sec:params}), we do not believe 
that with existing data it is possible to untangle the influences 
of the potentially complex star-formation histories and dust 
distributions on the SEDs of SMGs. Hence, we do not believe that there is any
observational evidence for significant old stellar populations in these galaxies,
as required by the \citet{Dye08} results.
\citet{Hainline10} have used
optical and IRAC photometric data to calculated an average stellar mass
for the \citet{Chapman05} SMGs and they find $M^* = (1.4\pm0.3)\times
10^{11}M_{\sun}$ (converted to Salpeter IMF), comparable to our survey and a factor of $\sim 5$
lower than \citet{Dye08}.

In Fig.~\ref{fig:zhmag} we also show the absolute $H$-band magnitudes
of SMGs in the $\Lambda$CDM {\sc galform} model \citep{Baugh05}, which
assumes a top-heavy IMF with slope $x=0$. We consider only galaxies
with $S_{850\mu m}\ge3$ mJy and fluxes in the IRAC 4.5\,\um\ filter
brighter than our detection limit (4.5\,\um\ corresponds to the
rest-frame $H$-band at $z\sim2$). \citet{Swinbank08} showed that {\sc
galform} predicts rest-frame absolute $K$-band luminosities of SMGs
which are a factor of ten lower than observed. This arises primarily
due to an order of magnitude lower stellar masses than implied by
observations for SMGs (see also \citealt{Lacey10}). As
Fig.~\ref{fig:zhmag} shows the predicted rest-frame $H$-band magnitudes
of the model SMGs are also a factor of ten lower than our observations.
Indeed, if SMGs formed stars following the prescriptions used in the
\citet{Baugh05}, then few of the SMGs above a redshift of $z\sim$2
would have be detected.

\subsection{Dust temperatures, far-infrared luminosities and star-formation}
\label{sec:tdlirsfrd}

In order to further investigate the intrinsic properties of the 
LESS SMGs we next use our photometric redshifts and the observed radio
and submillimetre fluxes to derive the characteristic dust temperatures
($T_D$), far-infrared luminosities (8--1000\,\um; $L_{FIR}$) and star-formation
rates.

\citet{Blain02} showed that the submillimetre-to-radio flux ratio in
SMGs is mainly influenced by redshift and the characteristic dust
temperature. \citet{Chapman05} assumed a dust emissivity,
$\beta=1.5$, and the $z=0$ far-infrared--radio correlation, to determine
empirically that for their sample of SMGs:
\begin{equation}\label{eqn:td}
T_{D}= \frac{6.25(1+z)}{(S_{850\mu {\rm m}} / S_{1.4 {\rm GHz}})^{0.26}}
\end{equation}
We note that the most reliable method of calculating $T_D$ is to fit
template SEDs to multiple far-infrared and submillimetre photometric
points, but for simplicity and due to the absence of published deep
far-infrared photometry we use Eqn.~\ref{eqn:td} to calculate $T_D$ of
LESS SMGs (although we next use shallow far-infrared
observations to confirm the validity of this assumption).

We also use the infrared-radio correlation \citep{Helou85, Condon92},
\begin{equation} \label{eqn:lfir}
q_{FIR} = {\rm log}_{10}\left(\frac{L_{FIR}}{3.75\times10^{12} {\rm W}}\right) 
          - {\rm log}_{10}\left(\frac{L_{\rm 1.4GHz}}{\rm WHz^{-1}}\right)
\end{equation}
with radio spectral index $\alpha=0.8$ (where
$S_\nu\propto\nu^{-\alpha}$) and $q_{FIR}=2.64$ \citep[][for
star-forming galaxies]{Bell03}, to calculate far-infrared luminosities
of the LESS SMGs from their radio fluxes, as done by
\citet{Chapman05}. Although this approach was recently verified by
\citet{Magnelli10} who used {\it Herschel} data to show that the local
far-infrared radio correlation is consistent with SMGs, we caution that
there may be a factor of $\sim2$ uncertainty in the derived
luminosities due to possible evolution in
the far-infrared--radio correlation \citep{Ivison10a}
and hence the appropriate value of $q_{FIR}$.

%{Td vs Lir}
\begin{figure}
\includegraphics[width=8.5cm]{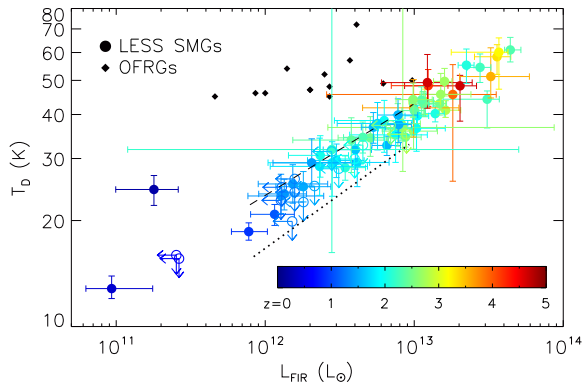}
\caption{ The characteristic dust temperature ($T_D$) versus
  far-infrared luminosity ($L_{FIR}$) for our SMGs. The SMGs are colour
  coded on the basis of their photometric redshifts and as expected the
  most luminous galaxies are the hottest, and are also tend to be those at the
  highest redshifts. This trend is driven in part by the radio luminosities of
the SMGs (which lack the positive K correction of the submillimetre waveband)
hence why there is a correlation between $L_{FIR}$ and $z$, but not
between $S_{870\mu m}$ and $z$ (Fig.~\ref{fig:zvss870}).
The regions above the dashed line and below the
  dotted line are illustrative of the regions excluded by our
  submillimetre and radio detection limits respectively.  The dashed
  line, which roughly demarcates the upper envelope of the data,
represents the derived temperature of galaxies at $z=2$ with
  various radio fluxes and $S_{870\mu m}=4.2$\,mJy. The dotted line,
which similarly demarcates the lower envelope, is
  derived for submillimetre-luminous ($S_{870\mu m}=16$\,mJy) sources with
  radio flux equal to our detection limit ($3\sigma=19.5\umu$Jy) at
  redshifts of $z=1.4$--4.  This reflects both the strong cut-off in
  the submillimetre luminosity function at high luminosities and the fact
  that our radio data is only just deep enough to detect counterparts
  to the majority of SMGs.  We conclude that the apparent correlation
  between $T_D$ and $L_{FIR}$ is in part caused by selection bias. We
  note that the OFRGs \citep{Chapman04b, Casey09, Magnelli10}, which
  are detected in the radio but not the submillimetre lie above the
  upper dashed line. }
\label{fig:tdlir}
\end{figure}

In Fig.~\ref{fig:tdlir} we plot the far-infrared luminosity against
$T_D$ for the LESS SMGs and optically faint radio galaxies (OFRGs;
\citealt{Chapman04b, Casey09, Magnelli10}. The LESS SMGs have a median $T_D = 35.9\pm
1.4$\,K, with an interquartile range of 28.5--43.3\,K, and
$L_{FIR}=(8.2\pm 1.2)\times 10^{12}$\,$L_{\sun}$, with an
interquartile range of $(3.0$--$13)\times 10^{12} L_{\sun}$, comparable
to previous surveys \citep[e.g.][]{Chapman05, Magnelli10}.

To check this result we also employ the 250, 350, and 500\,\um\ Balloon-borne Large Aperture
Submillimeter Telescope (BLAST) maps of the ECDFS \citep{Devlin09}.
We can stack the emission in these maps at
the positions of the LESS SMG counterparts and fit the stacked
fluxes with a modified black body with $\beta=1.5$ at $z=2.2$
and correct the luminosity of the fitted black body to {\it total}
infrared luminosity, 8--1000\,$\mu$m, based on \citet{Ivison10b}. From
this calculation the typical characteristic dust temperature of the
LESS SMGs is $T_D=33.6\pm1.1$, and the typical far-infrared luminosity
is $L_{FIR}=(7.6^{+1.7}_{-1.5})\times10^{12}L_{\sun}$.   These values are
in good agreement with those derived above from the local far-infrared--radio
correlation.

\begin{figure}
\includegraphics[width=8.5cm]{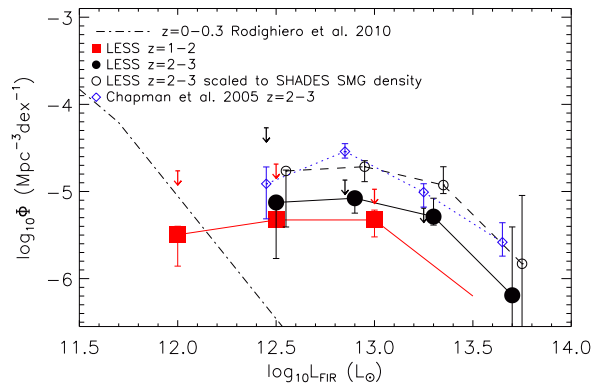}
\caption{ Far-infrared luminosity functions of the radio-detected LESS
  SMGs with $z=1$--2 and $z=2$--3.  Evolution is evident in the
  luminosity function of the LESS SMGs between the two redshift bins,
  and from the 24-\um\ selected low redshift comparison sample
  \citep{Rodighiero09}. The SMGs have higher luminosities than the
  $z<0.3$ 24-\um\ galaxies, and the $z=2$--3 SMGs having higher
  luminosities and $\Phi^*$.  We also show the $z=2$--3 luminosity
  function of \citet{Chapman05} SMGs (offset slightly in ${\rm
  log}_{10}L_{FIR}$ for clarity) for comparison.  The LESS $z=2$--3
  sample has a systematically lower luminosity density than
  \citet{Chapman05} SMGs in the same redshift range. \citet{Weiss09}
  showed that the ECDFS is underdense at submillimetre wavelengths and
  by scaling the LESS luminosity function such that the SMG number
  density matches that of SHADES survey \citep{Coppin06} we show that
  the disparity in $\Phi^*$ between LESS and \citet{Chapman05} is
  likely due to the relative density of SMGs in the two surveys (the
  scaled LESS luminosity function is offset slightly in ${\rm
  log}_{10}L_{FIR}$ SMGs for clarity). We calculate the maximum
  contribution from unidentified SMGs, by assigning them the redshift
  distribution that we measure in \S\ref{sec:noidzs} and radio fluxes equal
  to our detection limit. Including the contribution from unidentified
  SMGs the total maximum $\Phi$ in each luminosity bin is represented
  by an arrow (offset slightly in ${\rm log}_{10}L_{FIR}$ for $z=2$--3
  SMGs for clarity). }
\label{fig:lf}
\end{figure}

We find that the highest redshift galaxies also have the highest
luminosities due to a combination of the radio K-correction (preventing
the detection of low-luminosity galaxies at high redshifts) and
luminosity evolution (see Fig.~\ref{fig:lf}). There is an apparent
trend between the $T_D$ and $L_{FIR}$ but this is likely at least
partially a selection effect, although we note that locally {\it IRAS}
galaxies exhibit a tight correlation between $T_D$ and $L_{IR}$
\citep{Chapman03b, Chapin09}. To illustrate the selection effects we
also show in Fig.~\ref{fig:tdlir} OFRGs \citep{Chapman04b}, which are
detected at radio but not submillimetre wavelengths and have radio
luminosities similar to SMGs, but contain warmer dust \citep[$T_D\sim45$\,K;][]{Casey09,
Magnelli10}. Our $870$-\um\ detection limit misses warmer and lower
luminosity galaxies from the sample and the radio detection limit
excludes the colder luminous galaxies \citep[e.g.][]{Chapman05}.

%Luminosity function

\begin{table}
\caption{Far-infrared luminosity function for radio-detected LESS SMGs} 
\begin{tabular}{cccc}
\hline
\multicolumn{2}{c}{$z=1$--2 SMGs}&
\multicolumn{2}{c}{$z=2$--3 SMGs}\\
${\rm log_{10}}L_{FIR}$ & ${\rm log_{10}}\Phi$ &
${\rm log_{10}}L_{FIR}$ & ${\rm log_{10}}\Phi$  \\
$(L_{\sun})$ & ${\rm (Mpc^{-3}dex^{-1}})$ & $(L_{\sun})$ & ${\rm (Mpc^{-3}dex^{-1}})$ \\
\hline
12.0 & $-5.5_{-0.3}^{+0.2}$  & 12.5 & $-5.1_{-0.6}^{+0.1}$ \\
12.5 & $-5.3\pm0.1$          & 12.9 & $-5.1_{-0.2}^{+0.1}$ \\
13.0 & $-5.3_{-0.2}^{+0.1}$  & 13.3 & $-5.3_{-0.1}^{+0.2}$ \\
13.5 & $<-6.2$               & 13.7 & $-6.2_{-6.2}^{+0.7}$ \\
\hline
\end{tabular}
\label{tab:lf}
\end{table}

In Fig.~\ref{fig:lf} and Table \ref{tab:lf} we present the far-infrared
luminosity functions of the radio-detected LESS SMGs with $z=1$--2 and $z=2$--3, compared
to the $z=2$--3 result from \citet{Chapman05}.  We calculate the LESS SMG
luminosity function with an accessible volume technique where:
\begin{equation}\label{eqn:lf}
\Phi(L)\Delta L=\sum_{i}{\frac{1}{V_i}}
\end{equation}
which accounts for the flux limited nature of our
survey. $\Phi(L)\Delta L$ is the number density of sources with
luminosities between $L$ and $L+\Delta L$, and $V_i$ is the comoving
volume within which the $i$th source can be detected in the luminosity
bin under consideration. Since we derive the far-infrared luminosity
from the radio flux, $V_i$ is calculated using the radio
luminosity. Error bars are calculated by bootstrapping and account for
the redshift, luminosity and binning errors.  We use the same method to
calculate the luminosity function for \citet{Chapman05} SMGs based on
the redshifts and radio fluxes listed in that paper.  By assuming that
unidentified SMGs have radio fluxes equal to our detection limit and
the redshift distribution that we measure in \S\ref{sec:noidzs}, we also
calculate the maximum contribution of unidentified SMGs to the
far-infrared luminosity functions.

We observe strong evolution in the far-infrared luminosity function:
the $z=2$--3 SMGs are more luminous and have higher space densities
than the $z=1$--2 SMGs, which in turn are more luminous than the
$z<0.3$ 24\,\um-selected galaxies from \citet{Rodighiero09} (see
also the {\it Herschel} sample of \citealt{Vaccari10}). LESS SMGs at
$z=2$--3 have $\Phi^*\sim60\%$ and $L^*\sim2\times$ larger than
those at $z=1$--2.  

The $z=2$--3 LESS SMGs have systematically lower $\Phi^*$ than the
\citet{Chapman05} SMGs in the same redshift range. This may be due to
cosmic variance since \citet{Weiss09} showed that the ECDFS is a factor
of $\sim2$ underdense compared to other large submillimetre surveys at
flux densities $\ga 3$\,mJy. By rescaling
the LESS luminosity function so that the 870\,\um\ number counts agree
with those of the SHADES survey \citep{Coppin06}, which should be
similar to that of \citet{Chapman05} since both covered multiple
fields, Fig.~\ref{fig:lf} shows that the low surface density of SMGs in
the ECDFS is most likely the cause of the disparity in $\Phi^*$.

%Madau plot
\begin{figure}
\includegraphics[width=8.5cm]{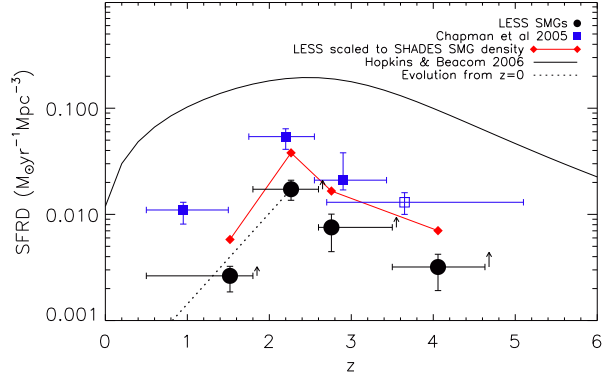}
\caption{ Evolution of the SFRD for the radio-detected LESS SMGs
  compared to \citet{Chapman05}. Arrows to the right of each LESS redshift
  bin indicate the maximum additional contribution from unidentified
  SMGs and the open symbol represents
  unidentified SMGs from \citet{Chapman05}.  We also show the modified
  Salpeter A IMF fit to the SFRD compilation from \citet{Hopkins06}
  and a line showing the evolution from {\it IRAS} ULIRGs at $z=0$
  \citep{Elbaz03} to LESS SMGs at $z=2.3$. The LESS SMG activity peaks at
  $z\sim2$ -- similar to that found by previous studies of
  star-forming galaxies and the peak activity of quasars
  \citep{Hopkins07}. The contribution from SMGs to the total SFRD also
  peaks at $z\sim2$ where they are responsible for $\sim10\%$ of the
  \citet{Hopkins06} SFRD.  The ECDFS is underdense at submillimetre
  wavelengths \citep{Weiss09} so similarly to Fig.~\ref{fig:lf} we
  also scale the SFRD of the LESS SMGs such that the number counts
  match the SHADES survey allowing a closer comparison to
  \citet{Chapman05}.}
\label{fig:madau}
\end{figure}

In Fig.~\ref{fig:madau} we show the evolution of the star-formation
rate density (SFRD) of the radio-detected LESS SMGs.  We use the same
accessible volume technique as in our luminosity function calculations
to account for the flux limited nature of the survey. Error bars are
calculated from bootstrapping and include the uncertainties in
binning, redshifts and SFRs. Since the SFRs are based upon radio
fluxes we exclude the suspected radio-bright AGN LESS\,20 from this
analysis.

We do not know the individual redshifts, infrared luminosities or SFRs
for 45\% of the LESS SMGs because they do not have robustly identified
optical counterparts. In Fig.~\ref{fig:madau} we account for this
population by assigning them the redshift distribution that we measure
in \S\ref{sec:noidzs}) and assuming radio fluxes equal to our detection
limit.  The calculated SFRD of the unidentified SMGs from this analysis
is an upper limit since the actual radio fluxes will typically be lower
than the detection limit. In Fig.~\ref{fig:madau} we indicate the
maximum contribution to the SFRD of unidentified SMGs in each redshift
bin.

The SFRD of the LESS SMGs appears to peak at $z\sim 2$, similar to
\citet{Chapman05}.   The LESS SMGs have a lower SFRD than the
SMGs from \citet{Chapman05} but we note that the lower number density
of SMGs in the ECDFS is sufficient to account for this effect. This
corresponds to the peak of quasar activity at $z=2.15\pm0.05$
\citep{Hopkins07}.  The fractional contribution of LESS SMGs to the
SFRD of the Universe also peaks at $z\sim2$ where they are responsible
for $\sim10\%$ of the SFRD as estimated by \citet{Hopkins06} from a
compilation of surveys that does not include any submillimetre surveys. We stress that
this only includes SMGs with $S_{870\mu m}\ga4$\,mJy. Assuming that fainter
sources have the same redshift distribution then the contribution of
SMGs with $S_{870\mu m}\ga1$\,mJy is $\sim100\%$ of the \citep{Hopkins07}
value. Thus, SMGs contribute $\sim50\%$ of the {\it total} SFRD of
the Universe at $z\sim2$.

%sSFR
\begin{figure}
\includegraphics[width=8.5cm]{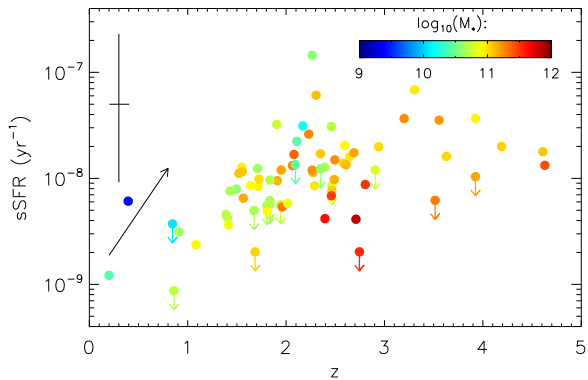}
\caption{ A plot of specific star-formation rate (sSFR) versus redshift
for the LESS SMGs.  Galaxies are colour-coded by mass and we show the
median error bar in the top-left hand corner and note that similarly
to Fig.~\ref{fig:zhmag} the error in sSFR is correlated with that in
redshift. The arrow represents the gradient of the trend in sSFR with
redshift for IRAC-selected galaxies with ${\rm
log}_{10}(M_*)=10.3$--$10.8\,M_{\sun}$, offset in sSFR by two orders of
magnitude for the purpose of display. We note that due to the
requirement for radio counterparts no SMGs are detected in the
high redshift and low sSFR region of this plot.  Similarly, the short lifetime of SMGs with sSFR $\ga
10^{-7}{\rm yr}^{-1}$ and the limited volume of our survey means that
few SMGs with very high sSFRs are detected.  However, the dearth of SMGs at
$z\la1.5$ with sSFR $\ga10^{-8}{\rm yr}^{-1}$ may indicate an upper-limit
to the sSFR of SMGs with a similar scaling to the trends
seen in lower activity galaxies at lower redshifts \citep{Damen09}. }
\label{fig:ssfr}
\end{figure}

We use \citet{Kennicutt98}, which assumes a Salpeter IMF with upper and
lower mass limits of 0.1 and 100\,$M_{\sun}$ respectively, to calculate
the SFRs of the LESS SMGs from their inferred far-infrared
luminosities. The median SFR is $1100\pm200$ \myr\ and the
interquartile range is 300--1900~\myr.  The median specific
star-formation rate ${\rm sSFR = SFR}/M^{*} =
(1.2\pm0.1)\times10^{-8}{\rm yr}^{-1}$ and the interquartile range is
(0.6--1.8)$\times10^{-8}{\rm yr}^{-1}$. Although again, we caution that
due to the uncertainties in the stellar mass estimates there is an
additional factor of $\sim5$ uncertainty in these values (see
\S\ref{sec:params} for a full discussion).  The median formation
timescale of the LESS SMGs is thus $\sim 100$\,Myr and it is feasible
that all of the stellar mass we see could be formed in a single burst.
In Fig.~\ref{fig:ssfr} we plot the trend of sSFR against the redshift
of the LESS SMGs. We note that the apparent lack of galaxies with low
sSFR at high redshifts is a selection effect due to the requirement for
a radio counterpart and that galaxies with sSFR $\ga10^{-7}{\rm
yr}^{-1}$ are rare due the brevity of the burst phase.  However, 
the dearth of SMGs at
$z\la1.5$ with sSFR $\sim10^{-8}$--$10^{-7}\,{\rm yr}^{-1}$ is not a
selection effect and this upper envelope may be following 
the same trend in sSFR with redshift seen in galaxies with
similar masses but lower SFRs \citep[e.g][]{Damen09}. 

Finally, we relate our new estimate of the redshift distribution of  
SMGs to constraints on the evolution of their likely descendants:  
massive early-type galaxies \citep{Swinbank06}.  As we have shown, the  
bulk of the SMG population with observed 870-$\mu$m fluxes above $\sim  
4$\,mJy lie at redshifts of $z\sim 1.5$--3 with a median redshift of $z 
\sim 2.5$. We estimate that the volume density of SMGs at $z=2$--3  
above our flux is $1.2\times 10^{-5}$\,Mpc$^{-3}$, where we include  
both the identified and statistically identified samples in this  
estimate. Using a characteristic lifetime of the SMG phase of $\sim  
100$\,Myr, we can correct this density for the burst duty cycle to  
derive a volume density for the remnants of $2\times 10^{-4}$\,Mpc 
$^{-3}$. As we have shown, the estimated baryonic masses of these  
galaxies are  $\sim 1.2\times 10^{11}$\,M$_\odot$ combining our best  
estimate of the stellar mass with the typical gas masses from  
\citet{Greve05}.  If the burst of star formation we are seeing in the  
SMG phase is the last major star formation event in these galaxies  
then we expect their descendants to appear as passive, red galaxies   
at $z\sim 1.5$ ($>1$\,Gyr after $z\sim 2$).

There have been various estimates of the volume density of massive,  
passive galaxies at $z\sim 1$--2 \citep{McCarthy04, Daddi05, Taylor09a}.   
For galaxies with masses of $\ga 10^{11}$\,M$_\odot$, the estimated  
space densities are  1--$2\times 10^{-4}$\,Mpc$^{-3}$ (at  
$z=1.5$--1.8; \citealt{Taylor09a}), $3\times 10^{-4}$\,Mpc$^{-3}$ (at  
$<\!z\!>= 1.7$; \citealt{Daddi05}) and $0.6\times 10^{-4}$\,Mpc 
$^{-3}$ (at $<\!z\!>= 1.5$; \citealt{McCarthy04}). These estimates,  
with their various uncertainties, are comparable to the predicted  
volume density of massive, passive galaxies if these all undergo an  
SMG-phase at an earlier epoch.  Hence, the starbursts in SMGs may be  
responsible for the formation of a large fraction of the passive,  
massive galaxies seen at $z\sim 1.5$.

We can attempt a similar calculation comparing the SMG population at  
$z>3$ with the constraints on massive galaxies at $z\ga 2$.  We  
estimate the volume density of $z>3$ SMGs from \S\ref{sec:noidzs} as  
$2.8\times 10^{-6}$\,Mpc$^{-3}$. This includes the 10 identified SMGs,  
$4\pm 2$ statistically identified SMGs and the remaining $21\pm 19$  
unidentified sources and assumes that they are contained within a  
redshift range of $z=3$--7.  Using a characteristic lifetime of the  
SMG phase of $\sim 100$\,Myr, we can correct this density for the  
burst duty cycle to derive a volume density for the remnants of  
$3.8\times 10^{-5}$\,Mpc$^{-3}$. Again the estimated baryonic masses  
of these galaxies are  $\sim 1.2\times 10^{11}$\,M$_\odot$.   
Unfortunately observable limits on the volume density of passive  
galaxies  are increasingly uncertain at $z>2$, but using the estimates  
from \citet{Coppin09} of the  volume density of massive galaxies of $ 
\sim 1$--$5\times 10^{-5}$\,Mpc$^{-3}$, we again conclude that it is  
possible that the SMG population we have identified is also  
responsible for the formation of a significant of the most massive  
galaxies at $z\sim 2.5$.

Thus we conclude that the presence of a sizable population of passive  
galaxies at high redshift  may be intimately linked to the strong  
evolution in dust obscured starbursts in the distant Universe.   
Theoretical attempts to match the properties of high-redshift  
galaxies  therefore need to focus on these observable constraints  as  
aspects of the same problem \citep{Swinbank08}.

\section{Summary and Conclusions}
\label{sec:conc}

We use deep multicolour imaging of the ECDFS in 17-bands to 
derive the photometric properties of the 
counterparts of SMGs in the LESS LABOCA survey of the
ECDFS \citep{Weiss09,Biggs10}. Our main results are as follows:

\begin{enumerate}

\item
LESS radio, 24\,\um\ and IRAC-identified SMGs have a median redshift of
$z=2.2\pm0.1$ and interquartile range of $z=1.8$--2.7.  Thus the peak
activity in SMGs corresponds to the epoch of maximal quasar and
star-formation activity in the Universe.  The redshift distribution of
LESS SMGs is consistent with the spectroscopic survey of
\citet{Chapman05}, but higher than the photometric studies of the
SHADES survey \citep{Clements08, Dye08}. We find a higher-redshift tail
to the distribution of LESS galaxies, with 10 (14\%) identified SMGs at
$z\ga 3$.  Counterparts identified through radio, $24$\,\um\ and IRAC
emission have statistically indistinguishable redshift distributions;
similarly robust and tentative counterparts have comparable redshift
distributions, albeit with some foreground contamination in the
tentative sample.  Previous studies provided tentative evidence that
SMGs with the highest submillimetre fluxes may be the highest redshift
sources \citep[e.g.][]{Ivison02, Pope05}, but with our extensive
photometric redshifts we find no such correlation.

\item
A statistical study of the source population in the error circles of
the 55 SMGs that lack robust radio, 24\,\um\ and IRAC counterparts
suggests that there is an excess of $26\pm12$ ($2.2\sigma$) $z>1$ galaxies
in these regions. Of these $14\pm8$ ($1.8\sigma$) are at $z=2$--3
and $4\pm2$ ($1.6\sigma$) are at $z=4$--5.  This excess population corresponds
to the counterparts or companions of the unidentified SMGs and
our analysis then suggests that
the redshift distribution of these unidentified SMGs peaks at
$z=2.5\pm0.3$.  This is similar to, but slightly higher than,
that of the identified population, 
suggesting that many of the unidentified SMGs are
at $z\sim 2$--3 and have radio or 24$\mu$m
fluxes just below our detection limits.

\item
We estimate that there are $21\pm19$ ($17\pm15\%$ of all the SMGs) LESS
SMGs that are not robustly identified, and which are not accounted
for in our statistical analysis of unidentified SMGs. These should be
galaxies without any detectable mid-infrared
emission and as a result are likely to lie at $z\ga3$. Including
the identified SMGs at $z>3$ we estimate that $28\pm15\%$
of all LESS SMGs lie at $z>3$.

\item
We combine the redshift distribution of identified SMGs with that
statistically determined for unidentified SMGs, including the
SMGs which are not detected in our survey and are likely to lie at $z\ga3$. 
We conclude that the likely median redshift of the {\it entire} population
of $S_{870\mu m}\ga4$~mJy SMGs is $z=2.5\pm0.6$.

\item
The separation of SMGs into those with redshifts above and below
$z=1.4$ broadly agrees with the $BzK$ colour-colour criteria
\citep{Daddi04}, making the $BzK$ colours a coarse but reliable
redshift indicator for SMGs.  Similarly the submillimetre-to-radio flux ratios
of LESS SMGs broadly agrees with the prediction of \citet{Carilli00},
although there is significant scatter, indicative of temperature
variations between SMGs. This means that redshifts derived from
submillimetre-to-radio flux ratios also exhibit significant scatter.
Instead, we show that the IRAC colours of the $\sim90\%$ of SMGs with $z<4$ follow
a trend with redshift of: $z=2.1+1.9{\rm log}_{10}(S_8/S_{3.6})$ 
and this may be a potentially
useful redshift indicator with a 1-$\sigma$ accuracy of
$\sigma_z=0.4$.

\item
The median rest-frame $H$-band absolute magnitude of the LESS SMGs
is $M_H=-24.1\pm0.1$ with an interquartile range of $-24.7$ to $-23.6$.
Using $M_H$ and the average mass-to-light ratio from \citet{Hainline10}
(converted to a Salpeter IMF) 
we calculate that the median stellar mass of the LESS SMGs is
$(9.2\pm0.9)\times 10^{10}M_{\sun}$, with an interquartile range of
$(4.7$--$14)\times10^{10}M_{\sun}$.  However, a $\chi^2$ analysis of
the best-fit star-formation histories  shows that,
even with 17-band photometry spanning the ultraviolet to mid-infrared, we cannot reliably
distinguish different  star-formation histories and ages for the SMGs. We
estimate that this results in an
additional factor of $\sim5$ uncertainty in the mass-to-light ratios and
hence  the derived stellar masses.

\item
Using our photometric redshifts, submillimetre and radio fluxes we
calculate that the median characteristic dust temperature of the SMGs
is $T_D=35.9\pm 1.4$\,K, with an interquartile range of
28.5--43.3\,K. The median far-infrared luminosity of the SMGs, derived
from the radio luminosity, is $L_{FIR}=(8.2\pm1.2)\times
10^{12}L_{\sun}$ and the interquartile range of
$L_{FIR}=(3$--$13)\times10^{12}L_{\sun}$.  For a Salpeter IMF this
corresponds to median SFR\,$=1100$\,\myr, with an interquartile range of
300--1900\,\myr.  We show that, for LESS SMGs, the apparent correlation
between the far-infrared luminosity and $T_D$ is in part a selection
effect.

\item
The far-infrared
luminosity function of the LESS SMGs exhibits a strong redshift
evolution, such that SMGs at $z=2$--3 are more numerous and have
higher luminosities than those at $z=1$--2. We find that the
normalisation of the luminosity function is lower for LESS than for the
\citet{Chapman05} SMG sample, and by scaling the ECDFS submillimetre
number counts we show that this is due to the underdensity of the
ECDFS at submillimetre wavelengths \citep{Weiss09}.  

\item
The SFRD and fractional contribution to the global SFRD of the LESS
SMGs with $S_{870\mu m}\ga4$\,mJy evolves with redshift and both peak at
$z\sim2$, where the LESS population contributes
a total SFR density of $0.02\,M_{\sun}{\rm yr^{-1}Mpc^{-1}}$.
If fainter submillimetre sources have the same redshift distribution
then SMGs with $S_{870\mu m}>1$\,mJy produce $\sim50\%$ of the SFRD of the
Universe at $z\sim2$.

\item
The masses and the volume density of LESS SMGs at $z=2$--3 are
comparable to those of massive, passive galaxies at $z\sim1$--2, and
similarly the volume density of $z>3$ SMGs is comparable to the limits
on the numbers of massive galaxies at $z\sim2$--3. This suggests that a
large fraction of the population of massive, passive galaxies at
high-redshifts form most of their stars during an earlier SMG phase.

\end {enumerate}

This analysis demonstrates the strengths and weaknesses of photometric
redshift analysis for SMGs in a field with excellent photometry. In the
impending era of SCUBA-2 and {\it Herschel} $\gg10^5$ SMGs will be
discovered; it will be impossible to obtain spectroscopic redshifts for
such large samples and hence the challenge is to obtain sufficient
photometric coverage of these survey fields to allow a photometric
analysis of the type described here.

\section*{Acknowledgements}

We thank Carlton Baugh, Mark Dickinson and Laura Hainline for help and
useful discussions.  J.L.W acknowledges the support of a Science and
Technology Facilities Council (STFC) studentship; K.E.K.C acknowledges
the support of an STFC fellowship; I.R.S acknowledges the support of
STFC. W.N.B, B.L and Y.Q.X acknowledge Chandra X-ray Observatory grant
SP8-9003A and NASA ADP grant NNX10AC99G.  J.S.D acknowledges the
support of the Royal Society through a Wolfson Research Merit award,
and the support of the European Research Council through the award of
an Advanced Grant.  This publication is based on data acquired with the
Atacama Pathfinder Experiment (APEX) under programme numbers
078.F-9028(A), 079.F-9500(A), 080.A-3023(A), and 081.F-9500(A). APEX is
a collaboration between the Max-Planck-Institut fur Radioastronomie,
the European Southern Observatory, and the Onsala Space Observatory.
Based on observations made with ESO Telescopes at the Paranal and La
Silla Observatories under programme numbers 171.A-3045, 168.A-0485,
082.A-0890 and 183.A-0666.

\begin{landscape}
\begin{table}
\begin{minipage}{240mm} 
\caption{Observed photometry for robust counterparts to LESS
SMGs. $3\sigma$ limiting magnitudes are presented where sources are
covered by imaging but not detected; SMGs which are not covered by
imaging in a given filter have no photometry listed in that filter.  }
\setlength{\tabcolsep}{0.7 mm} 
\begin{tiny}
\begin{tabular}{lccccccccccccccccc}
\hline
Source & MUSYC $U$ & $U38$ & VIMOS $U$ & $B$ & $V$ & $R$ & $I$ & $z$ & MUSYC $J$ & HAWKI $J$ & $H$ & MUSYC $K$ & HAWKI $K$ & $3.6$\um & $4.5$\um & $5.8$\um & $8$\um \\
\hline
LESS\,2a & $25.67\pm 0.13$ & $25.35\pm 0.32$ & $25.19\pm 0.10$ & $24.78\pm 0.05$ & $24.65\pm 0.05$ & $24.39\pm 0.09$ & $24.08\pm 0.15$ & $23.82\pm 0.17$ & $22.81\pm 0.16$ & $22.88\pm 0.04$ & $>23.02$ & $21.65\pm 0.13$ & $22.18\pm 0.03$ & $21.37\pm 0.07$ & $21.18\pm 0.07$ & $21.13\pm 0.14$ & $21.49\pm 0.09$\\
LESS\,2b & $>26.85$ & $>25.40$ & $>28.38$ & $>26.81$ & $26.56\pm 0.29$ & $>25.79$ & $>24.94$ & $>24.48$ & $>23.64$ & $24.44\pm 0.09$ & $22.94\pm 0.29$ & $>22.72$ & $23.26\pm 0.05$ & $22.38\pm 0.11$ & $22.08\pm 0.10$ & $21.86\pm 0.20$ & $22.23\pm 0.14$\\
LESS\,3 & $>26.85$ & $>25.40$ & ... & $>26.81$ & $>26.68$ & $>25.79$ & $>24.94$ & $>24.48$ & $>23.64$ & ... & $>23.02$ & $>22.72$ & $23.30\pm 0.05$ & $22.72\pm 0.12$ & $22.09\pm 0.10$ & $21.74\pm 0.19$ & $21.42\pm 0.09$\\
LESS\,6 & $26.57\pm 0.27$ & $>25.40$ & $28.11\pm 0.23$ & $25.95\pm 0.15$ & $24.94\pm 0.07$ & $24.00\pm 0.06$ & $22.91\pm 0.05$ & $22.75\pm 0.07$ & $21.83\pm 0.07$ & $22.09\pm 0.03$ & $21.73\pm 0.10$ & $21.10\pm 0.08$ & $21.15\pm 0.02$ & $21.14\pm 0.06$ & $21.52\pm 0.08$ & $21.92\pm 0.21$ & $22.36\pm 0.17$\\
LESS\,7 & $26.46\pm 0.25$ & $>25.40$ & $26.29\pm 0.12$ & $24.90\pm 0.06$ & $24.08\pm 0.03$ & $23.37\pm 0.04$ & $22.22\pm 0.03$ & $22.11\pm 0.04$ & $21.69\pm 0.06$ & $21.44\pm 0.02$ & $21.07\pm 0.06$ & $20.47\pm 0.04$ & $20.46\pm 0.01$ & $20.02\pm 0.04$ & $19.91\pm 0.04$ & $19.86\pm 0.08$ & $20.15\pm 0.05$\\
LESS\,9 & $>26.85$ & ... & $>28.38$ & $>26.81$ & $>26.68$ & $>25.79$ & $>24.94$ & ... & $>23.64$ & $24.19\pm 0.08$ & $>23.02$ & $>22.72$ & $22.47\pm 0.04$ & $21.85\pm 0.08$ & $21.43\pm 0.08$ & $21.29\pm 0.15$ & $21.13\pm 0.08$\\
LESS\,10a & $26.29\pm 0.21$ & $>25.40$ & $25.74\pm 0.11$ & $25.53\pm 0.11$ & $25.60\pm 0.13$ & $25.05\pm 0.16$ & $24.61\pm 0.23$ & $>24.48$ & $>23.64$ & $24.56\pm 0.10$ & $>23.02$ & $>22.72$ & $23.33\pm 0.05$ & $22.21\pm 0.10$ & $21.78\pm 0.09$ & $21.72\pm 0.19$ & $21.46\pm 0.09$\\
LESS\,10b & $>26.85$ & $>25.40$ & $27.45\pm 0.14$ & $26.66\pm 0.28$ & $25.79\pm 0.15$ & $24.28\pm 0.08$ & $22.63\pm 0.04$ & $22.36\pm 0.05$ & $21.41\pm 0.05$ & $21.18\pm 0.02$ & $20.81\pm 0.05$ & $20.51\pm 0.05$ & $20.29\pm 0.01$ & $19.87\pm 0.03$ & $20.14\pm 0.04$ & $20.75\pm 0.12$ & $21.32\pm 0.09$\\
LESS\,11 & $>26.85$ & ... & $27.86\pm 0.15$ & $>26.81$ & $>26.68$ & $>25.79$ & $>24.94$ & $>24.48$ & $>23.64$ & $25.04\pm 0.12$ & $>23.02$ & $>22.72$ & $23.58\pm 0.06$ & $22.16\pm 0.10$ & $21.63\pm 0.09$ & $21.29\pm 0.15$ & $21.15\pm 0.08$\\
LESS\,12 & $>26.85$ & $>25.40$ & $>28.38$ & $>26.81$ & $>26.68$ & $>25.79$ & $>24.94$ & $>24.48$ & $>23.64$ & $25.71\pm 0.18$ & $>23.02$ & $>22.72$ & $23.83\pm 0.06$ & $22.72\pm 0.12$ & $22.35\pm 0.12$ & $22.17\pm 0.23$ & $21.98\pm 0.12$\\
LESS\,14 & $>26.85$ & $>25.40$ & $>28.38$ & $>26.81$ & $26.71\pm 0.32$ & $>25.79$ & $>24.94$ & $>24.48$ & $>23.64$ & $25.56\pm 0.19$ & ... & $>22.72$ & $23.53\pm 0.06$ & $22.55\pm 0.11$ & $21.87\pm 0.09$ & $21.56\pm 0.17$ & $21.14\pm 0.08$\\
LESS\,15 & $>26.85$ & ... & ... & $>26.81$ & $>26.68$ & $>25.79$ & $>24.94$ & $>24.48$ & $>23.64$ & ... & ... & $>22.72$ & ... & $21.91\pm 0.09$ & $21.30\pm 0.07$ & $21.11\pm 0.14$ & $21.05\pm 0.08$\\
LESS\,16 & $25.04\pm 0.07$ & $24.86\pm 0.21$ & $24.83\pm 0.10$ & $24.49\pm 0.04$ & $24.01\pm 0.03$ & $23.29\pm 0.03$ & $21.79\pm 0.02$ & $21.57\pm 0.02$ & $20.95\pm 0.03$ & $20.93\pm 0.02$ & $20.44\pm 0.05$ & $20.16\pm 0.03$ & $20.15\pm 0.01$ & $19.51\pm 0.03$ & $19.71\pm 0.04$ & $20.03\pm 0.08$ & $20.08\pm 0.05$\\
LESS\,17 & $25.15\pm 0.08$ & $25.25\pm 0.29$ & $24.83\pm 0.10$ & $24.47\pm 0.04$ & $24.26\pm 0.04$ & $23.94\pm 0.06$ & $23.27\pm 0.07$ & $23.05\pm 0.09$ & $22.05\pm 0.08$ & $22.16\pm 0.03$ & $21.64\pm 0.10$ & $21.25\pm 0.09$ & $21.07\pm 0.02$ & $20.36\pm 0.04$ & $20.16\pm 0.04$ & $20.38\pm 0.10$ & $20.76\pm 0.07$\\
LESS\,18 & $26.11\pm 0.18$ & $>25.40$ & $25.71\pm 0.11$ & $25.52\pm 0.11$ & $25.34\pm 0.10$ & $25.30\pm 0.20$ & $24.64\pm 0.23$ & $24.41\pm 0.27$ & $23.21\pm 0.22$ & $22.81\pm 0.04$ & $22.05\pm 0.14$ & $21.35\pm 0.10$ & $21.45\pm 0.02$ & $20.38\pm 0.04$ & $20.06\pm 0.04$ & $20.01\pm 0.08$ & $20.67\pm 0.06$\\
LESS\,19 & $>26.85$ & $>25.40$ & $27.01\pm 0.12$ & $27.04\pm 0.35$ & $26.39\pm 0.25$ & $26.79\pm 0.62$ & $>24.94$ & $>24.48$ & $>23.64$ & $23.82\pm 0.07$ & ... & $>22.72$ & $24.22\pm 0.08$ & $22.74\pm 0.12$ & $22.23\pm 0.11$ & $21.83\pm 0.19$ & $21.98\pm 0.12$\\
LESS\,20 & $>26.85$ & $>25.40$ & ... & $26.25\pm 0.20$ & $25.66\pm 0.13$ & $25.89\pm 0.33$ & $>24.94$ & $>24.48$ & $>23.64$ & $24.56\pm 0.09$ & $>23.02$ & $>22.72$ & $22.57\pm 0.03$ & $21.83\pm 0.08$ & $21.48\pm 0.08$ & $21.16\pm 0.14$ & $21.46\pm 0.09$\\
LESS\,22 & $>26.85$ & ... & ... & $>26.81$ & $26.20\pm 0.21$ & $25.60\pm 0.26$ & $>24.94$ & $>24.48$ & ... & ... & ... & ... & ... & $20.47\pm 0.04$ & $20.16\pm 0.04$ & $20.04\pm 0.08$ & $20.55\pm 0.06$\\
LESS\,24 & $>26.85$ & $>25.40$ & ... & $25.53\pm 0.11$ & $25.71\pm 0.14$ & $25.15\pm 0.18$ & $24.31\pm 0.18$ & $24.71\pm 0.35$ & $22.71\pm 0.14$ & ... & ... & $21.39\pm 0.10$ & ... & $20.67\pm 0.05$ & $20.41\pm 0.05$ & $20.73\pm 0.12$ & $20.97\pm 0.08$\\
LESS\,25 & ... & $>25.40$ & $26.78\pm 0.13$ & $26.02\pm 0.16$ & $25.46\pm 0.11$ & $25.04\pm 0.16$ & $24.45\pm 0.20$ & $24.25\pm 0.24$ & $>23.64$ & $23.23\pm 0.05$ & ... & $21.82\pm 0.15$ & $21.67\pm 0.02$ & $21.03\pm 0.06$ & $20.73\pm 0.05$ & $20.58\pm 0.11$ & $20.83\pm 0.07$\\
LESS\,27a & $>26.85$ & $>25.40$ & ... & $>26.81$ & $>26.68$ & $>25.79$ & $>24.94$ & $>24.48$ & $>23.64$ & $>25.70$ & $>23.02$ & $>22.72$ & ... & $22.68\pm 0.12$ & $22.30\pm 0.12$ & $21.93\pm 0.22$ & $22.31\pm 0.17$\\
LESS\,27b & $>26.85$ & $>25.40$ & ... & $26.14\pm 0.18$ & $25.88\pm 0.16$ & $25.35\pm 0.21$ & $>24.94$ & $>24.48$ & $>23.64$ & ... & ... & $>22.72$ & ... & $22.01\pm 0.09$ & $21.68\pm 0.09$ & $21.45\pm 0.16$ & $21.95\pm 0.12$\\
LESS\,29 & $>26.85$ & ... & ... & $>26.81$ & $>26.68$ & $>25.79$ & $>24.94$ & $>24.48$ & $>23.64$ & ... & ... & $>22.72$ & ... & $22.22\pm 0.10$ & $21.70\pm 0.09$ & $21.27\pm 0.15$ & $21.19\pm 0.09$\\
LESS\,31 & $>26.85$ & $>25.40$ & ... & $>26.81$ & $>26.68$ & $>25.79$ & $>24.94$ & $>24.48$ & $>23.64$ & $26.49\pm 0.26$ & ... & $>22.72$ & $23.46\pm 0.05$ & $22.70\pm 0.12$ & $22.19\pm 0.11$ & $21.68\pm 0.18$ & $21.65\pm 0.10$\\
LESS\,34 & $24.98\pm 0.07$ & $24.69\pm 0.18$ & $24.39\pm 0.10$ & $24.09\pm 0.03$ & $23.67\pm 0.02$ & $23.06\pm 0.03$ & $21.87\pm 0.02$ & $21.57\pm 0.02$ & $20.94\pm 0.03$ & $20.78\pm 0.02$ & $20.29\pm 0.05$ & $20.04\pm 0.03$ & $19.90\pm 0.01$ & $19.51\pm 0.03$ & $19.71\pm 0.04$ & $20.37\pm 0.10$ & $20.73\pm 0.06$\\
LESS\,36 & $>26.85$ & $>25.40$ & ... & $>26.81$ & ... & $>25.79$ & $>24.94$ & $>24.48$ & $>23.64$ & $25.70\pm 0.16$ & ... & $22.67\pm 0.30$ & $22.76\pm 0.04$ & $21.53\pm 0.07$ & $21.06\pm 0.06$ & $20.65\pm 0.11$ & $20.98\pm 0.07$\\
LESS\,37 & $>26.85$ & ... & ... & $26.19\pm 0.19$ & $25.33\pm 0.10$ & $24.18\pm 0.08$ & $23.45\pm 0.08$ & ... & $22.86\pm 0.16$ & ... & $22.38\pm 0.18$ & $21.16\pm 0.08$ & ... & $20.55\pm 0.05$ & $20.41\pm 0.05$ & $20.82\pm 0.13$ & $20.94\pm 0.08$\\
LESS\,39 & $25.44\pm 0.10$ & $>25.40$ & ... & $25.17\pm 0.08$ & $24.77\pm 0.06$ & $24.15\pm 0.07$ & $23.91\pm 0.13$ & $23.41\pm 0.12$ & $23.59\pm 0.30$ & $24.32\pm 0.10$ & ... & $22.01\pm 0.17$ & ... & $21.62\pm 0.07$ & $21.28\pm 0.07$ & $21.06\pm 0.14$ & $21.07\pm 0.08$\\
LESS\,40 & $24.82\pm 0.06$ & $24.71\pm 0.19$ & $24.54\pm 0.10$ & $24.36\pm 0.04$ & $24.07\pm 0.03$ & $23.81\pm 0.05$ & $23.36\pm 0.08$ & $23.44\pm 0.12$ & $23.04\pm 0.19$ & $23.05\pm 0.05$ & $22.52\pm 0.21$ & $22.20\pm 0.20$ & $22.09\pm 0.03$ & $21.45\pm 0.07$ & $21.12\pm 0.07$ & $21.04\pm 0.13$ & $20.99\pm 0.07$\\
LESS\,41 & ... & ... & ... & ... & ... & ... & ... & ... & ... & ... & ... & ... & ... & $20.50\pm 0.04$ & $20.19\pm 0.04$ & $19.91\pm 0.08$ & $19.93\pm 0.05$\\
LESS\,43 & $>26.85$ & $>25.40$ & $28.17\pm 0.21$ & $>26.81$ & $>26.68$ & $>25.79$ & $>24.94$ & $>24.48$ & $>23.64$ & $23.93\pm 0.07$ & $23.11\pm 0.33$ & $>22.72$ & $22.58\pm 0.03$ & $21.46\pm 0.07$ & $21.04\pm 0.06$ & $21.09\pm 0.14$ & $21.73\pm 0.11$\\
LESS\,44 & $25.06\pm 0.07$ & $24.77\pm 0.20$ & ... & $24.42\pm 0.04$ & $24.32\pm 0.04$ & $24.34\pm 0.09$ & $24.58\pm 0.22$ & $24.32\pm 0.25$ & ... & ... & ... & ... & ... & $20.94\pm 0.05$ & $20.59\pm 0.05$ & $20.37\pm 0.10$ & $20.82\pm 0.07$\\
LESS\,47 & $>26.85$ & $>25.40$ & $27.05\pm 0.13$ & $26.31\pm 0.21$ & $25.58\pm 0.12$ & $25.26\pm 0.19$ & $>24.94$ & $>24.48$ & $>23.64$ & ... & ... & $>22.72$ & ... & $22.37\pm 0.10$ & $22.00\pm 0.10$ & $21.80\pm 0.19$ & $21.83\pm 0.12$\\
LESS\,48 & ... & ... & $>28.38$ & ... & ... & ... & ... & ... & ... & ... & ... & ... & ... & $20.27\pm 0.04$ & $20.15\pm 0.04$ & $20.14\pm 0.09$ & $20.80\pm 0.07$\\
LESS\,49a & $26.26\pm 0.21$ & $>25.40$ & ... & $25.30\pm 0.09$ & $25.22\pm 0.09$ & $24.59\pm 0.11$ & $24.07\pm 0.14$ & $24.25\pm 0.24$ & ... & $23.85\pm 0.07$ & ... & ... & $22.20\pm 0.03$ & $21.26\pm 0.06$ & $20.97\pm 0.06$ & $21.41\pm 0.16$ & $21.67\pm 0.11$\\
LESS\,49b & $>26.85$ & $>25.40$ & ... & $24.85\pm 0.06$ & $24.56\pm 0.05$ & $24.37\pm 0.09$ & $24.13\pm 0.15$ & $24.38\pm 0.26$ & ... & $23.85\pm 0.07$ & ... & ... & $22.49\pm 0.03$ & $22.16\pm 0.10$ & $21.88\pm 0.09$ & $21.65\pm 0.18$ & $21.67\pm 0.11$\\
LESS\,50a & $24.34\pm 0.04$ & $24.09\pm 0.11$ & $23.95\pm 0.10$ & $23.70\pm 0.02$ & $23.38\pm 0.02$ & $22.90\pm 0.02$ & $22.19\pm 0.03$ & $22.04\pm 0.03$ & $21.75\pm 0.06$ & $21.62\pm 0.02$ & $21.55\pm 0.09$ & $21.32\pm 0.09$ & $21.17\pm 0.02$ & $21.20\pm 0.06$ & $21.67\pm 0.09$ & $22.17\pm 0.25$ & $23.24\pm 0.32$\\
LESS\,50b & $>26.85$ & ... & $26.12\pm 0.12$ & $26.17\pm 0.18$ & $26.12\pm 0.20$ & $25.90\pm 0.33$ & $>24.94$ & $>24.48$ & $>23.64$ & $24.22\pm 0.08$ & $>23.02$ & $22.63\pm 0.29$ & $22.65\pm 0.03$ & $21.70\pm 0.08$ & $21.17\pm 0.07$ & $20.76\pm 0.12$ & $20.57\pm 0.06$\\
LESS\,54 & $>26.85$ & ... & $26.97\pm 0.13$ & $26.08\pm 0.17$ & $26.33\pm 0.24$ & $>25.79$ & $>24.94$ & $>24.48$ & $>23.64$ & ... & ... & $>22.72$ & ... & $21.66\pm 0.08$ & $21.26\pm 0.07$ & $21.31\pm 0.15$ & $21.97\pm 0.12$\\
LESS\,56 & $>26.85$ & $>25.40$ & $>28.38$ & $>26.81$ & $>26.68$ & $>25.79$ & $>24.94$ & $>24.48$ & $23.77\pm 0.35$ & $24.46\pm 0.09$ & $22.57\pm 0.22$ & $22.15\pm 0.20$ & $22.00\pm 0.03$ & $21.27\pm 0.06$ & $20.95\pm 0.06$ & $20.81\pm 0.12$ & $21.26\pm 0.09$\\
LESS\,57 & $>26.85$ & $>25.40$ & $26.57\pm 0.12$ & $25.14\pm 0.08$ & $25.38\pm 0.10$ & $24.96\pm 0.15$ & $>24.94$ & $>24.48$ & $>23.64$ & $24.25\pm 0.08$ & $23.05\pm 0.32$ & $>22.72$ & $22.55\pm 0.03$ & $22.02\pm 0.09$ & $21.63\pm 0.08$ & $21.35\pm 0.16$ & $20.54\pm 0.06$\\
LESS\,59 & $>26.85$ & $>25.40$ & $27.66\pm 0.14$ & $26.89\pm 0.33$ & $26.67\pm 0.31$ & $>25.79$ & $24.36\pm 0.19$ & $>24.48$ & $23.14\pm 0.21$ & $23.03\pm 0.04$ & $22.53\pm 0.21$ & $21.62\pm 0.12$ & $21.71\pm 0.02$ & $20.71\pm 0.05$ & $20.51\pm 0.05$ & $20.83\pm 0.12$ & $21.27\pm 0.09$\\
LESS\,60 & $25.50\pm 0.11$ & $25.12\pm 0.26$ & $24.98\pm 0.10$ & $24.95\pm 0.06$ & $24.59\pm 0.05$ & $24.08\pm 0.07$ & $23.18\pm 0.07$ & $22.87\pm 0.07$ & $22.54\pm 0.12$ & $22.23\pm 0.03$ & $22.01\pm 0.13$ & $21.49\pm 0.11$ & $21.31\pm 0.02$ & $20.59\pm 0.05$ & $20.39\pm 0.05$ & $20.54\pm 0.11$ & $20.65\pm 0.06$\\
LESS\,62 & $26.55\pm 0.26$ & $>25.40$ & $26.27\pm 0.12$ & $25.55\pm 0.11$ & $25.09\pm 0.08$ & $24.44\pm 0.10$ & $23.45\pm 0.08$ & $23.56\pm 0.13$ & $22.17\pm 0.09$ & $22.11\pm 0.03$ & $21.50\pm 0.09$ & $20.81\pm 0.06$ & $20.85\pm 0.01$ & $19.93\pm 0.03$ & $19.75\pm 0.04$ & $20.08\pm 0.09$ & $20.28\pm 0.05$\\
LESS\,63 & $25.39\pm 0.10$ & $>25.40$ & ... & $25.05\pm 0.07$ & $24.69\pm 0.06$ & $24.25\pm 0.08$ & $23.28\pm 0.07$ & $22.92\pm 0.08$ & $22.02\pm 0.08$ & $21.83\pm 0.03$ & $21.80\pm 0.11$ & $21.21\pm 0.09$ & $21.12\pm 0.02$ & $20.69\pm 0.05$ & $20.53\pm 0.05$ & $20.76\pm 0.12$ & $21.09\pm 0.08$\\
LESS\,64 & $>26.85$ & ... & $>28.38$ & $26.53\pm 0.25$ & $25.52\pm 0.12$ & $24.29\pm 0.08$ & $24.17\pm 0.16$ & $24.44\pm 0.28$ & $>23.64$ & $24.28\pm 0.09$ & ... & $>22.72$ & $23.06\pm 0.04$ & $22.51\pm 0.11$ & $22.41\pm 0.12$ & $22.27\pm 0.25$ & $21.87\pm 0.12$\\
LESS\,66 & $21.03\pm 0.00$ & $20.89\pm 0.01$ & ... & $21.18\pm 0.00$ & $21.13\pm 0.00$ & $20.79\pm 0.00$ & $20.64\pm 0.01$ & $20.23\pm 0.01$ & $21.21\pm 0.04$ & ... & $20.33\pm 0.05$ & $20.30\pm 0.04$ & ... & $19.61\pm 0.03$ & $19.44\pm 0.03$ & $19.36\pm 0.06$ & $19.32\pm 0.03$\\
LESS\,67 & $26.14\pm 0.19$ & $>25.40$ & $25.37\pm 0.10$ & $24.85\pm 0.06$ & $24.52\pm 0.05$ & $24.42\pm 0.09$ & $23.80\pm 0.11$ & $23.74\pm 0.16$ & $23.07\pm 0.20$ & $22.79\pm 0.04$ & $22.23\pm 0.16$ & $21.16\pm 0.08$ & $21.40\pm 0.02$ & $20.63\pm 0.05$ & $20.32\pm 0.05$ & $20.20\pm 0.09$ & $20.75\pm 0.06$\\
LESS\,70 & $25.19\pm 0.08$ & $24.72\pm 0.19$ & $24.56\pm 0.10$ & $23.87\pm 0.02$ & $23.78\pm 0.02$ & $23.69\pm 0.05$ & $23.61\pm 0.10$ & $23.53\pm 0.13$ & $22.56\pm 0.13$ & $22.56\pm 0.04$ & $22.51\pm 0.20$ & $21.70\pm 0.13$ & $21.60\pm 0.02$ & $20.96\pm 0.06$ & $20.78\pm 0.06$ & $20.58\pm 0.11$ & $20.89\pm 0.07$\\
LESS\,73 & $>26.85$ & $>25.40$ & $>28.38$ & $>26.81$ & $>26.68$ & $25.82\pm 0.31$ & $24.38\pm 0.19$ & $>24.48$ & $>23.64$ & $24.35\pm 0.09$ & $>23.02$ & $>22.72$ & $23.53\pm 0.07$ & $22.92\pm 0.13$ & $22.79\pm 0.14$ & $22.33\pm 0.25$ & $21.99\pm 0.12$\\
LESS\,74a & $>26.85$ & ... & $>28.38$ & $>26.81$ & $>26.68$ & $>25.79$ & $>24.94$ & $24.72\pm 0.35$ & $23.23\pm 0.22$ & $23.48\pm 0.05$ & $23.08\pm 0.33$ & $22.41\pm 0.24$ & $22.08\pm 0.03$ & $21.45\pm 0.07$ & $21.17\pm 0.07$ & $21.15\pm 0.14$ & $21.81\pm 0.12$\\
LESS\,74b & $>26.85$ & ... & $27.83\pm 0.15$ & $>26.81$ & $>26.68$ & $>25.79$ & $25.04\pm 0.32$ & $>24.48$ & $23.36\pm 0.25$ & $23.60\pm 0.06$ & $>23.02$ & $22.08\pm 0.18$ & $22.47\pm 0.03$ & $21.58\pm 0.07$ & $21.18\pm 0.07$ & $21.04\pm 0.14$ & $21.57\pm 0.10$\\
LESS\,75 & $25.14\pm 0.08$ & $24.80\pm 0.20$ & ... & $23.86\pm 0.02$ & $23.60\pm 0.02$ & $23.42\pm 0.04$ & $23.38\pm 0.08$ & $23.50\pm 0.13$ & $22.81\pm 0.16$ & $22.95\pm 0.04$ & ... & $22.10\pm 0.19$ & $22.17\pm 0.03$ & $21.07\pm 0.06$ & $20.43\pm 0.05$ & $19.79\pm 0.08$ & $19.07\pm 0.03$\\
LESS\,79 & $26.37\pm 0.23$ & $>25.40$ & $25.56\pm 0.11$ & $25.16\pm 0.08$ & $24.84\pm 0.06$ & $24.54\pm 0.10$ & $23.61\pm 0.10$ & $23.74\pm 0.16$ & $22.63\pm 0.13$ & $22.41\pm 0.03$ & $22.09\pm 0.14$ & $21.24\pm 0.09$ & $21.20\pm 0.02$ & $20.32\pm 0.04$ & $20.16\pm 0.04$ & $20.32\pm 0.10$ & $20.87\pm 0.07$\\
LESS\,81 & $25.94\pm 0.16$ & $>25.40$ & ... & $24.18\pm 0.03$ & $23.73\pm 0.02$ & $23.39\pm 0.04$ & $22.79\pm 0.05$ & ... & $21.38\pm 0.04$ & $22.45\pm 0.03$ & $21.49\pm 0.08$ & $20.90\pm 0.07$ & $21.70\pm 0.02$ & $20.18\pm 0.04$ & $19.91\pm 0.04$ & $20.11\pm 0.09$ & $20.61\pm 0.07$\\
LESS\,84 & $26.42\pm 0.24$ & $>25.40$ & $25.62\pm 0.11$ & $25.10\pm 0.07$ & $25.03\pm 0.08$ & $24.89\pm 0.14$ & $24.43\pm 0.20$ & $>24.48$ & $>23.64$ & $23.36\pm 0.05$ & $22.81\pm 0.26$ & $21.71\pm 0.13$ & $22.10\pm 0.03$ & $21.43\pm 0.07$ & $21.08\pm 0.06$ & $20.92\pm 0.13$ & $21.10\pm 0.08$\\
LESS\,87 & ... & ... & $25.52\pm 0.11$ & ... & ... & ... & ... & ... & ... & ... & ... & ... & ... & $21.32\pm 0.06$ & $21.07\pm 0.06$ & $21.03\pm 0.14$ & $20.90\pm 0.07$\\
LESS\,88 & $26.78\pm 0.32$ & $>25.40$ & $25.98\pm 0.11$ & $25.42\pm 0.10$ & $25.10\pm 0.08$ & $24.98\pm 0.15$ & $24.27\pm 0.17$ & $24.43\pm 0.28$ & $23.25\pm 0.23$ & $23.51\pm 0.06$ & $23.05\pm 0.32$ & $22.00\pm 0.17$ & $22.05\pm 0.04$ & $21.15\pm 0.06$ & $20.85\pm 0.06$ & $20.73\pm 0.12$ & $21.09\pm 0.08$\\
LESS\,96 & $21.92\pm 0.00$ & $21.73\pm 0.01$ & $21.32\pm 0.10$ & $20.82\pm 0.00$ & $20.62\pm 0.00$ & $20.40\pm 0.00$ & $19.99\pm 0.00$ & $19.72\pm 0.00$ & $19.74\pm 0.01$ & $19.50\pm 0.01$ & $19.77\pm 0.04$ & $19.76\pm 0.02$ & $19.59\pm 0.01$ & $19.63\pm 0.03$ & $19.43\pm 0.03$ & $19.07\pm 0.05$ & $18.59\pm 0.02$\\
LESS\,98 & $>26.85$ & ... & ... & $>26.81$ & $>26.68$ & $>25.79$ & $24.61\pm 0.23$ & $24.64\pm 0.33$ & $23.41\pm 0.26$ & $22.98\pm 0.04$ & ... & $21.49\pm 0.11$ & $21.40\pm 0.02$ & $20.22\pm 0.04$ & $19.84\pm 0.04$ & $19.99\pm 0.08$ & $20.31\pm 0.05$\\
LESS\,101 & $>26.85$ & $>25.40$ & $27.28\pm 0.13$ & $26.70\pm 0.29$ & $26.20\pm 0.21$ & $25.58\pm 0.25$ & $>24.94$ & $>24.48$ & $>23.64$ & $24.33\pm 0.10$ & $>23.02$ & $>22.72$ & $23.16\pm 0.05$ & $22.44\pm 0.11$ & $21.93\pm 0.10$ & $21.62\pm 0.18$ & $21.55\pm 0.10$\\
LESS\,102 & $>26.85$ & $>25.40$ & ... & $26.93\pm 0.34$ & $26.24\pm 0.22$ & $25.98\pm 0.35$ & $24.97\pm 0.31$ & $24.49\pm 0.29$ & $22.91\pm 0.17$ & ... & ... & $21.29\pm 0.09$ & ... & $20.43\pm 0.04$ & $20.13\pm 0.04$ & $20.17\pm 0.09$ & $20.92\pm 0.08$\\
LESS\,103 & $>26.85$ & ... & ... & $>26.81$ & $>26.68$ & $>25.79$ & $>24.94$ & $>24.48$ & $>23.64$ & ... & ... & $22.78\pm 0.33$ & ... & $21.51\pm 0.07$ & $21.25\pm 0.07$ & $21.12\pm 0.14$ & $21.77\pm 0.12$\\
LESS\,106 & $>26.85$ & $>25.40$ & ... & $>26.81$ & $>26.68$ & $>25.79$ & $24.51\pm 0.21$ & $24.38\pm 0.27$ & $23.07\pm 0.20$ & $22.71\pm 0.04$ & $21.76\pm 0.11$ & $21.33\pm 0.10$ & $21.22\pm 0.02$ & $20.22\pm 0.04$ & $19.83\pm 0.04$ & $19.91\pm 0.08$ & $20.06\pm 0.05$\\
LESS\,108 & $21.25\pm 0.00$ & $21.00\pm 0.01$ & $20.82\pm 0.10$ & $19.64\pm 0.00$ & $19.11\pm 0.00$ & $18.60\pm 0.00$ & $18.02\pm 0.00$ & $17.94\pm 0.00$ & $17.41\pm 0.00$ & $17.35\pm 0.00$ & $17.15\pm 0.04$ & $17.00\pm 0.00$ & $16.99\pm 0.00$ & ... & $17.79\pm 0.01$ & $17.84\pm 0.03$ & ...\\
LESS\,110 & ... & ... & ... & ... & ... & ... & ... & ... & ... & ... & ... & ... & ... & $23.13\pm 0.16$ & $22.44\pm 0.12$ & $21.82\pm 0.20$ & $21.72\pm 0.12$\\
LESS\,111 & $25.06\pm 0.07$ & $25.03\pm 0.25$ & ... & $23.82\pm 0.02$ & $23.31\pm 0.02$ & $22.79\pm 0.02$ & $22.38\pm 0.03$ & $22.27\pm 0.04$ & $22.75\pm 0.15$ & ... & ... & $21.57\pm 0.12$ & ... & $21.04\pm 0.06$ & $20.98\pm 0.06$ & $20.81\pm 0.12$ & $20.23\pm 0.05$\\
LESS\,112 & ... & ... & $26.98\pm 0.15$ & ... & ... & ... & ... & ... & ... & ... & ... & ... & ... & $20.93\pm 0.05$ & $20.59\pm 0.05$ & $20.43\pm 0.10$ & $21.06\pm 0.08$\\
LESS\,114 & $25.31\pm 0.09$ & $25.24\pm 0.29$ & $24.98\pm 0.10$ & $24.52\pm 0.04$ & $24.21\pm 0.04$ & $23.87\pm 0.06$ & $23.05\pm 0.06$ & $22.91\pm 0.07$ & $21.63\pm 0.06$ & $21.62\pm 0.02$ & $20.97\pm 0.05$ & $20.68\pm 0.05$ & $20.67\pm 0.01$ & $19.93\pm 0.03$ & $19.66\pm 0.03$ & $19.86\pm 0.08$ & $20.10\pm 0.05$\\
LESS\,117 & $26.42\pm 0.24$ & $>25.40$ & ... & $>26.81$ & $>26.68$ & $>25.79$ & $24.09\pm 0.15$ & ... & $22.58\pm 0.13$ & ... & $21.60\pm 0.09$ & $21.47\pm 0.11$ & ... & $20.56\pm 0.05$ & $20.31\pm 0.05$ & $20.53\pm 0.11$ & $20.93\pm 0.08$\\
LESS\,118 & ... & ... & ... & ... & ... & ... & ... & ... & ... & $24.30\pm 0.08$ & ... & ... & $23.98\pm 0.08$ & $23.39\pm 0.17$ & $23.81\pm 0.25$ & $22.77\pm 0.34$ & $>23.50$\\
LESS\,120 & $>26.85$ & $>25.40$ & ... & $26.77\pm 0.30$ & $25.99\pm 0.18$ & $>25.79$ & $24.60\pm 0.23$ & $23.79\pm 0.16$ & $23.34\pm 0.25$ & $23.23\pm 0.05$ & $22.79\pm 0.26$ & $21.77\pm 0.14$ & $21.87\pm 0.03$ & $20.99\pm 0.05$ & $20.70\pm 0.05$ & $20.74\pm 0.12$ & $21.33\pm 0.09$\\
LESS\,122 & $24.83\pm 0.06$ & $24.70\pm 0.19$ & $24.42\pm 0.10$ & $23.77\pm 0.02$ & $23.44\pm 0.02$ & $23.28\pm 0.03$ & $22.94\pm 0.05$ & $22.86\pm 0.07$ & $21.96\pm 0.07$ & $21.96\pm 0.03$ & $21.43\pm 0.08$ & $21.20\pm 0.09$ & $21.00\pm 0.02$ & $20.24\pm 0.04$ & $19.88\pm 0.04$ & $19.59\pm 0.07$ & $19.67\pm 0.04$\\
LESS\,126 & $>26.85$ & $>25.40$ & $26.87\pm 0.12$ & $26.32\pm 0.21$ & $26.61\pm 0.30$ & $25.87\pm 0.32$ & $>24.94$ & ... & $>23.64$ & $23.84\pm 0.07$ & $22.72\pm 0.24$ & $22.99\pm 0.38$ & $22.24\pm 0.03$ & $21.29\pm 0.06$ & $21.04\pm 0.06$ & $21.21\pm 0.14$ & $21.68\pm 0.10$\\
\hline
\end{tabular}
\end{tiny}

\label{tab:photo}
\end{minipage} 
\end{table}
\end{landscape}

\bibliographystyle{mn2e}
\bibliography{bibtex}

\appendix
\section{Discussion of unusual sources}
\label{sec:sources}

Most LESS SMGs have properties similar to previous
SMG populations. However, several robust counterparts have unusual properties, 
X-ray emission, or $8$\,\um\ excesses indicative of the presence of an
AGN. We discuss these galaxies on a case-by-case basis below. \\
{\bf LESS\,2a \& 2b}: LESS\,2 has both a robust $24$\,\um\ (LESS\,2a) and a robust 
radio counterpart (LESS\,2b), separated by $2.7''$, with $z=1.80^{+0.35}_{-0.14}$ 
and $2.27^{+0.16}_{-0.55}$ respectively. It is possible that the two counterparts 
are at the same redshift ($z\sim2$), separated by $\sim 20$ kpc and may in the
process of merging. \\
{\bf LESS\,6:} LESS\,6 is $\sim1^{\prime\prime}$ from the robust 24\um\ 
counterpart and $\sim2.5^{\prime\prime}$ from the robust 
radio counterpart and has a photometric redshift of $z=0.40^{+0.09}_{-0.03}$, 
therefore it is possible that the submillimetre source
is a background galaxy which is being gravitationally lensed by LESS\,6. \\
{\bf LESS\,9:} This SMG is X-ray luminous and has
$z_{phot}=4.63^{+0.10}_{-1.10}$ making it the highest redshift X-ray source in
our sample. Our best-fitting SED shows no $8$\,\um\ excess and thus
no suggestion of an AGN from the near-infrared and optical photometry.  \\ 
{\bf LESS\,10a:} LESS\,10a is one of two counterparts to LESS\,10 identified
from extended or blended radio emission. It has some $8$\,\um\ excess over
the best-fit SED, suggesting the presence of an AGN, but no detectable 
X-ray emission. \\
{\bf LESS\,11:} Similarly to LESS\,9 this galaxy shows no $8$\,\um\ excess
but is X-ray bright. \\
{\bf LESS\,19:} Based upon the SED fitting LESS\,19 has excess $5.8$ and 
$8$\,\um\ emission. However, the source is faint and the fit has 
$\chi^2 = 10.0$ so it is unclear whether the apparent excess is due to errors
in the fitting or the presence of an AGN. \\
{\bf LESS\,20:} This SMG is unusually radio bright with 
$S_{\rm 1.4GHz}=4.25$ mJy. There is no evidence of X-ray emission or an $8$\,\um\
excess, suggesting that LESS\,20 is a radio-bright AGN. If this is the 
case then the AGN contribution to the radio flux means that $T_D$ and 
$L_{FIR}$ are likely to be significantly over-estimated. Therefore, we 
exclude LESS\,20 from our analyses of the SFRs and luminosity function of SMGs.\\
{\bf LESS\,40:} LESS\,40 has X-ray emission and an $8$\,\um\ excess and is 
highly likely to contain an AGN. \\
{\bf LESS\,50b:} This galaxy is X-ray luminous but does not exhibit a 
compelling $8$\,\um\ excess. \\
{\bf LESS\,57:} LESS\,57 has a strong $8$\,\um\ excess and is X-ray bright, 
compelling evidence for the presence of an AGN in this SMG. \\
{\bf LESS\,66:} LESS\,66 lies near the diffraction spike 
of a bright star, so some of the photometry may be unreliable. However, 
it is an optically bright point source with an $8$\,\um\ excess and X-ray emission. 
There are broad emission lines in the spectra suggesting that LESS\,66 is a 
submillimetre-bright quasar. \\
{\bf LESS\,67:} This galaxy has coincident X-ray emission and may contain
an AGN. \\
{\bf LESS\,74a \& b:} The two counterparts to LESS\,74 have 
$z=1.84^{+0.32}_{-0.49}$ and $1.71^{+0.20}_{-0.17}$, are separated
by $2.7''$ ($\sim20$ kpc), and have some faint extended emission between them
in the optical images. Therefore, it is likely that LESS\,74a and LESS\,74b 
are undergoing an interaction at $z\sim1.8$ which triggered the 
submillimetre emission. \\
{\bf LESS\,75:} Although LESS\,75 is not X-ray detected there is strong $8$
and $5.8$\,\um\ excess above the best-fit SED, indicative of a highly obscured AGN. \\
{\bf LESS\,84:} LESS\,84 is X-ray detected, and has a small $8$\,\um\ 
excess suggesting it may contain an AGN.\\
{\bf LESS\,96:} LESS\,96 is similar to LESS\,66 -- it is X-ray luminous and has
a strong 8\,\um\ excess above the best-fit SED, and it is an bright optical
point source. We interpret this as evidence that LESS\,96 is a 
submillimetre-bright quasar. \\
{\bf LESS\,106:} This SMG is X-ray detected but no $8$\,\um\ excess is 
observed. \\ 
{\bf LESS\,108:} LESS\,108 is identified as a bright local ($z=0.086$) 
late-type spiral galaxy from its strong radio and $24$\,\um\ emission. 
The similarity of the radio, mid- and near-infrared, and optical morphologies 
suggests that this is not a case of gravitational lensing. \\
{\bf LESS\,111:} LESS\,111 is X-ray detected and has a $8$\,\um\ excess, 
indicating it may contain an AGN. It lies $\la3^{\prime\prime}$ from an extended foreground
galaxy and is likely to be gravitationally lensed. \\
{\bf LESS\,114:} This SMG is coincident with an X-ray source, but shows 
no evidence of an $8$\,\um\ excess above the best-fit SED. \\
{\bf LESS\,122:} LESS\,122 has excess flux at $8$\,\um\ compared to the best-fit SED
but is not X-ray detected.

\section{SED fits}

In Fig.~\ref{fig:seds} we show the measured photometry and best-fit
SED for each SMG counterpart. The calculated photometric redshifts and
errors are shown, and the probability distribution functions presented
for each galaxy. Most SMGs are well-fit by out template SEDs, although
nine (12\%) have excess 8\,\um\ flux above the best-fit SED.

\begin{figure*}
\begin{minipage}{14.5cm}
\includegraphics[width=14.5cm]{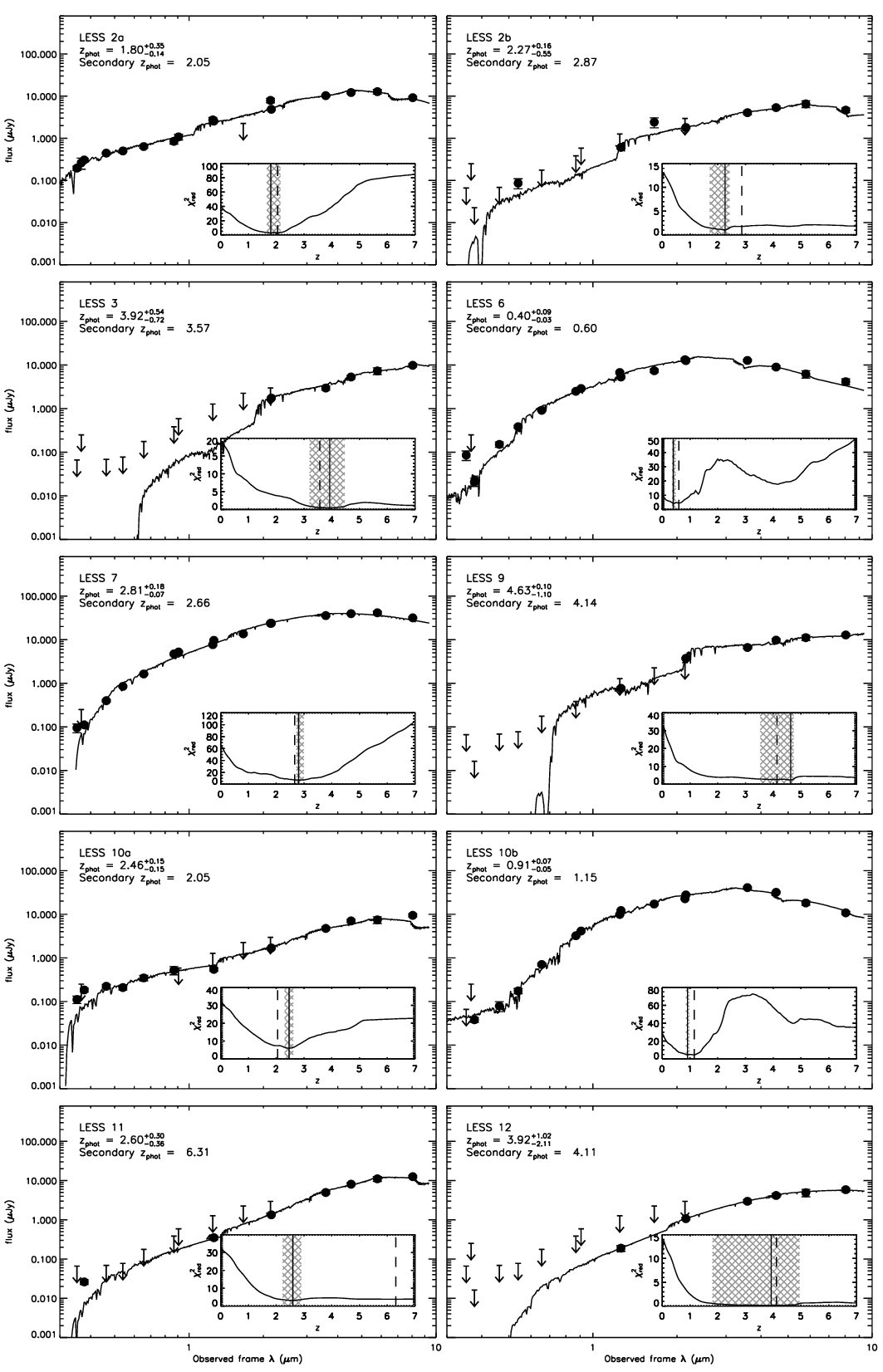}
\caption{Photometry and best-fit SEDs for robust SMG counterparts;
points and error-bars show measured photometry and arrows represent
$3\sigma$ detection limits. Error bars on the primary photometric
redshift are $99\%$ confidence limits. Inset panels show the minimum
reduced $\chi^2$ at each redshift step with the photometric redshift
primary and secondary solutions (where they exist) marked by solid and
dashed lines respectively; the photometric redshift error is
represented by the shaded region.  }
\label{fig:seds}
\end{minipage}
\end{figure*}

\begin{figure*}
\begin{minipage}{15cm}
\includegraphics[width=15cm]{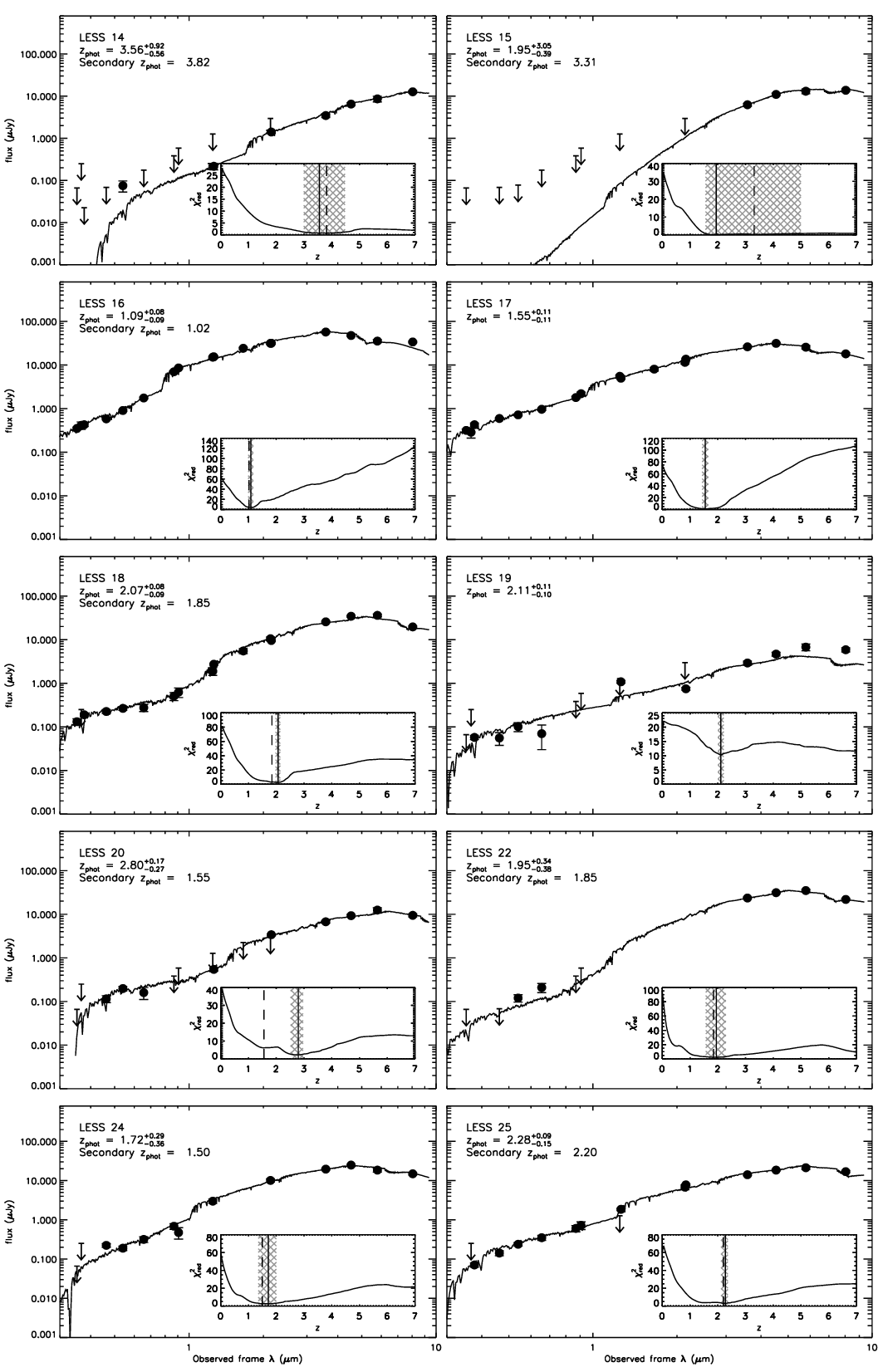}
\contcaption{}
\end{minipage}
\end{figure*}

\begin{figure*}
\begin{minipage}{15cm}
\includegraphics[width=15cm]{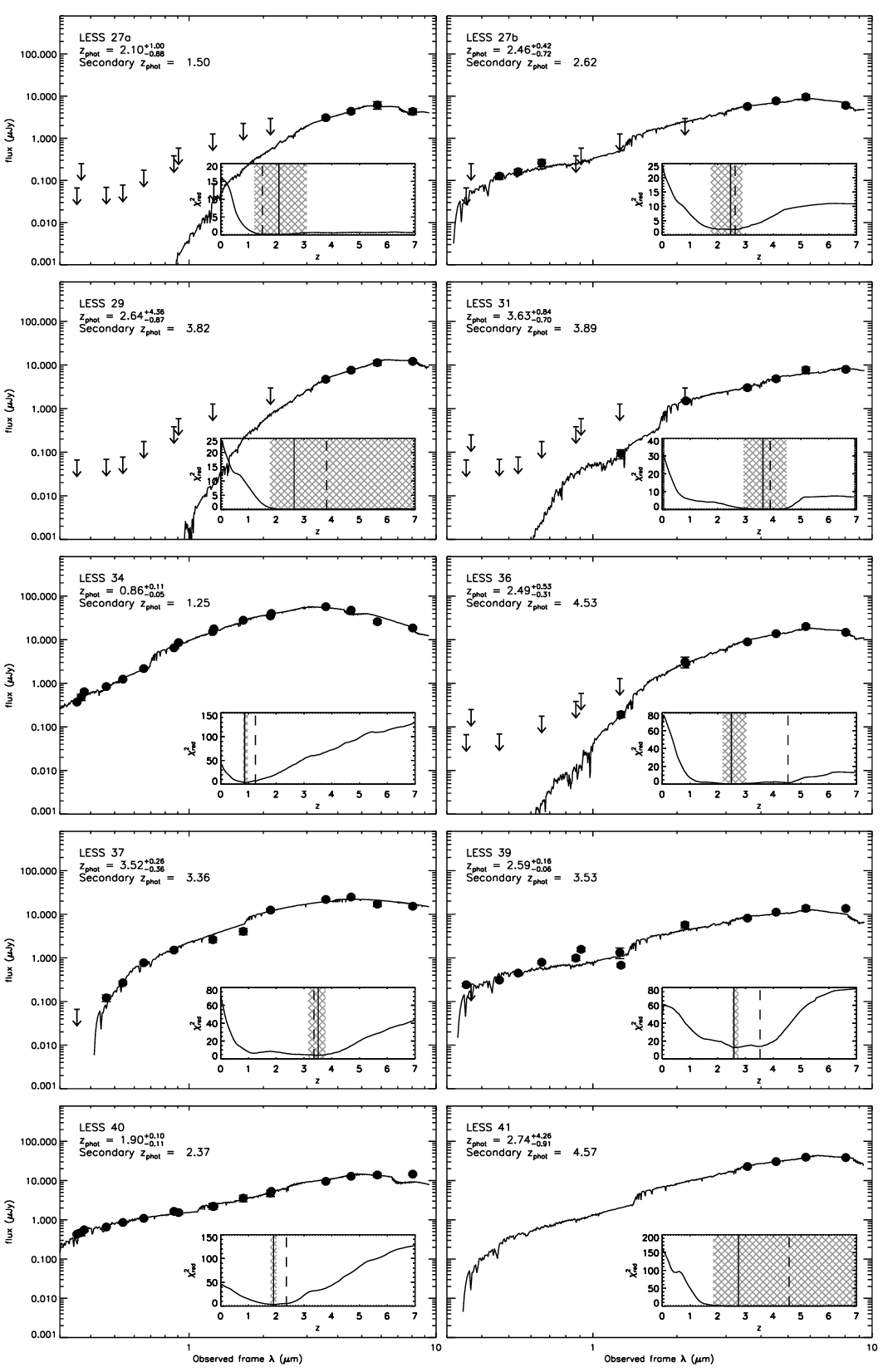}
\contcaption{}
\end{minipage}
\end{figure*}

\begin{figure*}
\begin{minipage}{15cm}
\includegraphics[width=15cm]{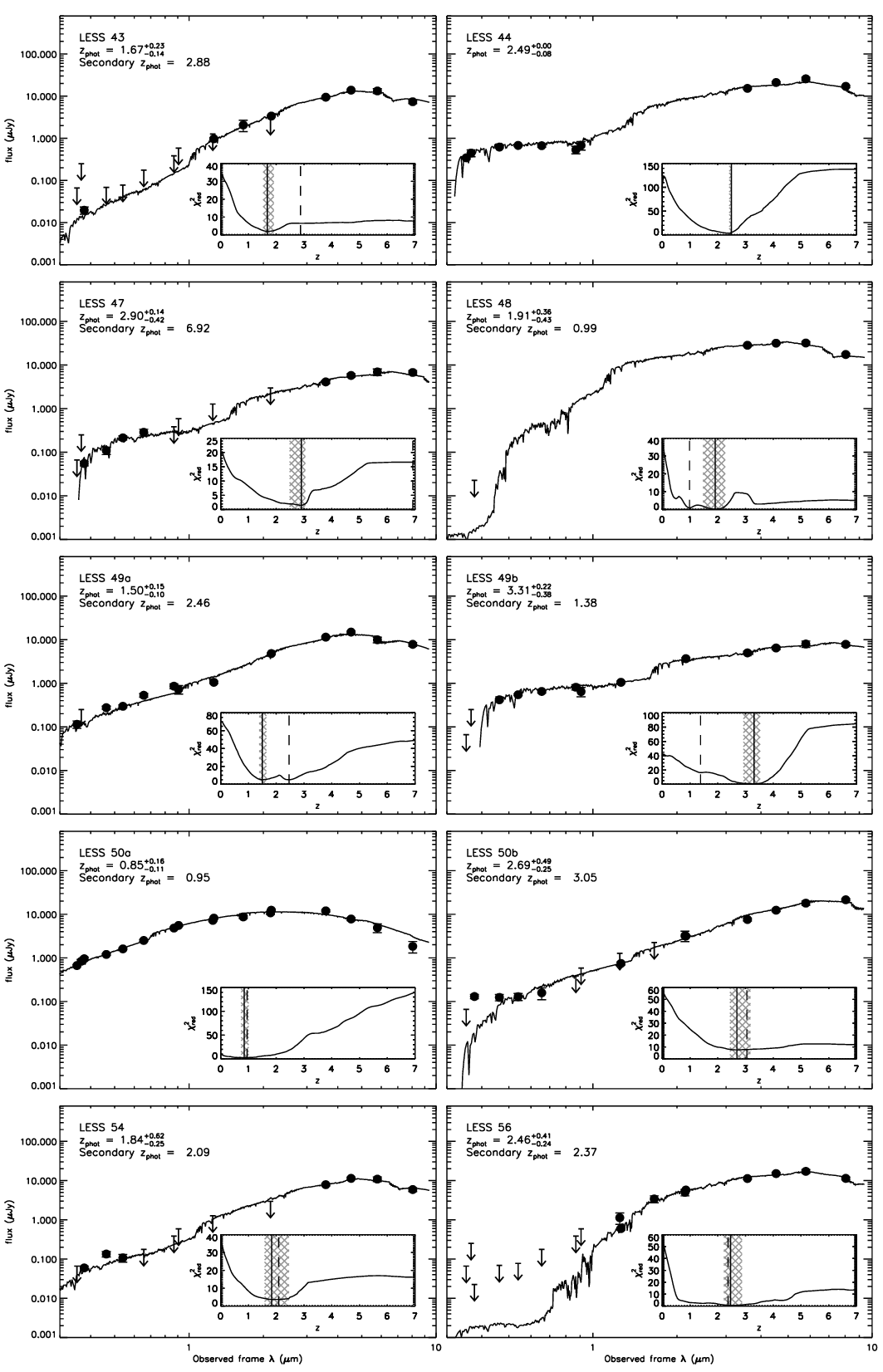}
\contcaption{}
\end{minipage}
\end{figure*}

\begin{figure*}
\begin{minipage}{15cm}
\includegraphics[width=15cm]{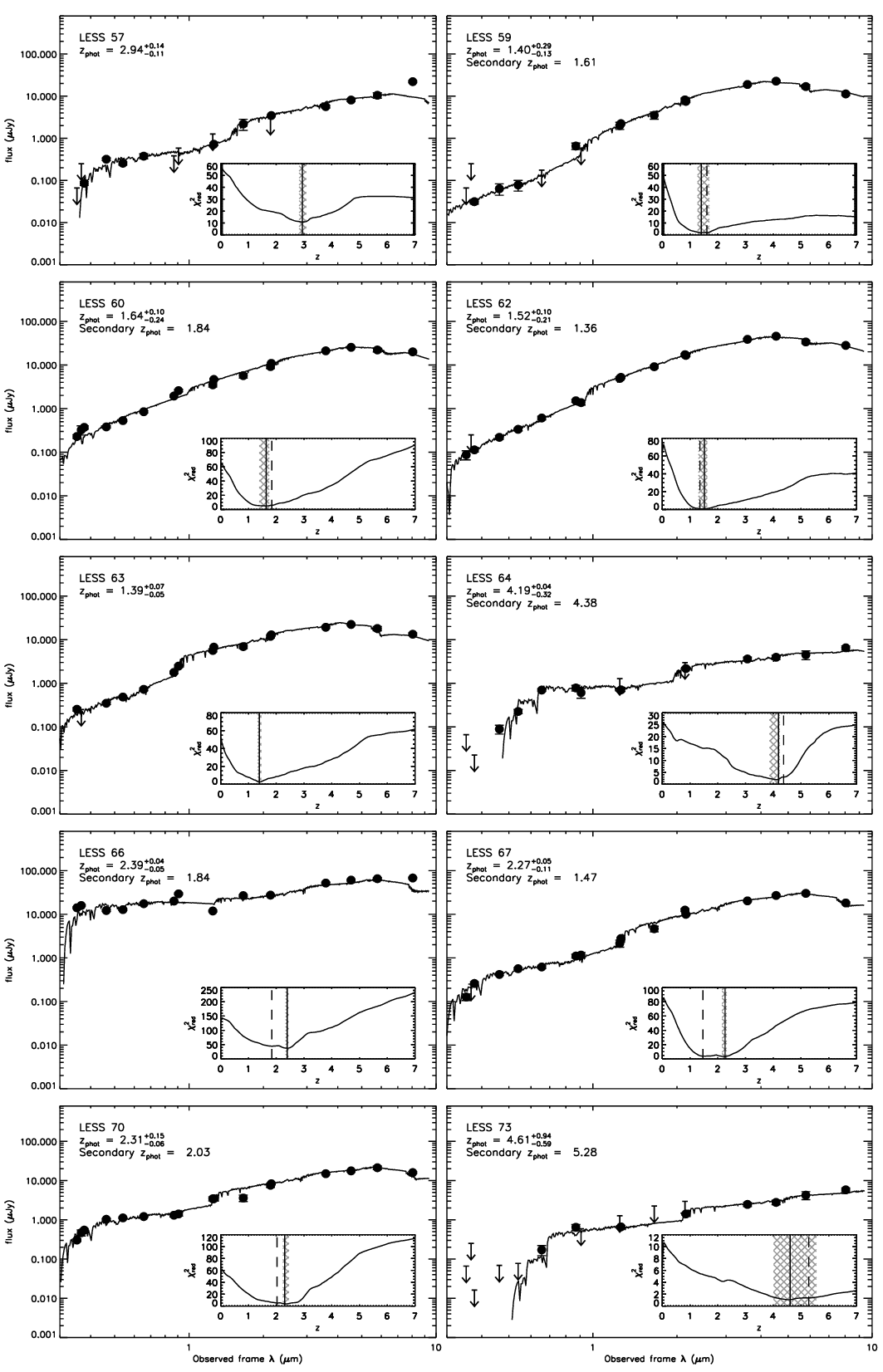}
\contcaption{}
\end{minipage}
\end{figure*}

\begin{figure*}
\begin{minipage}{15cm}
\includegraphics[width=15cm]{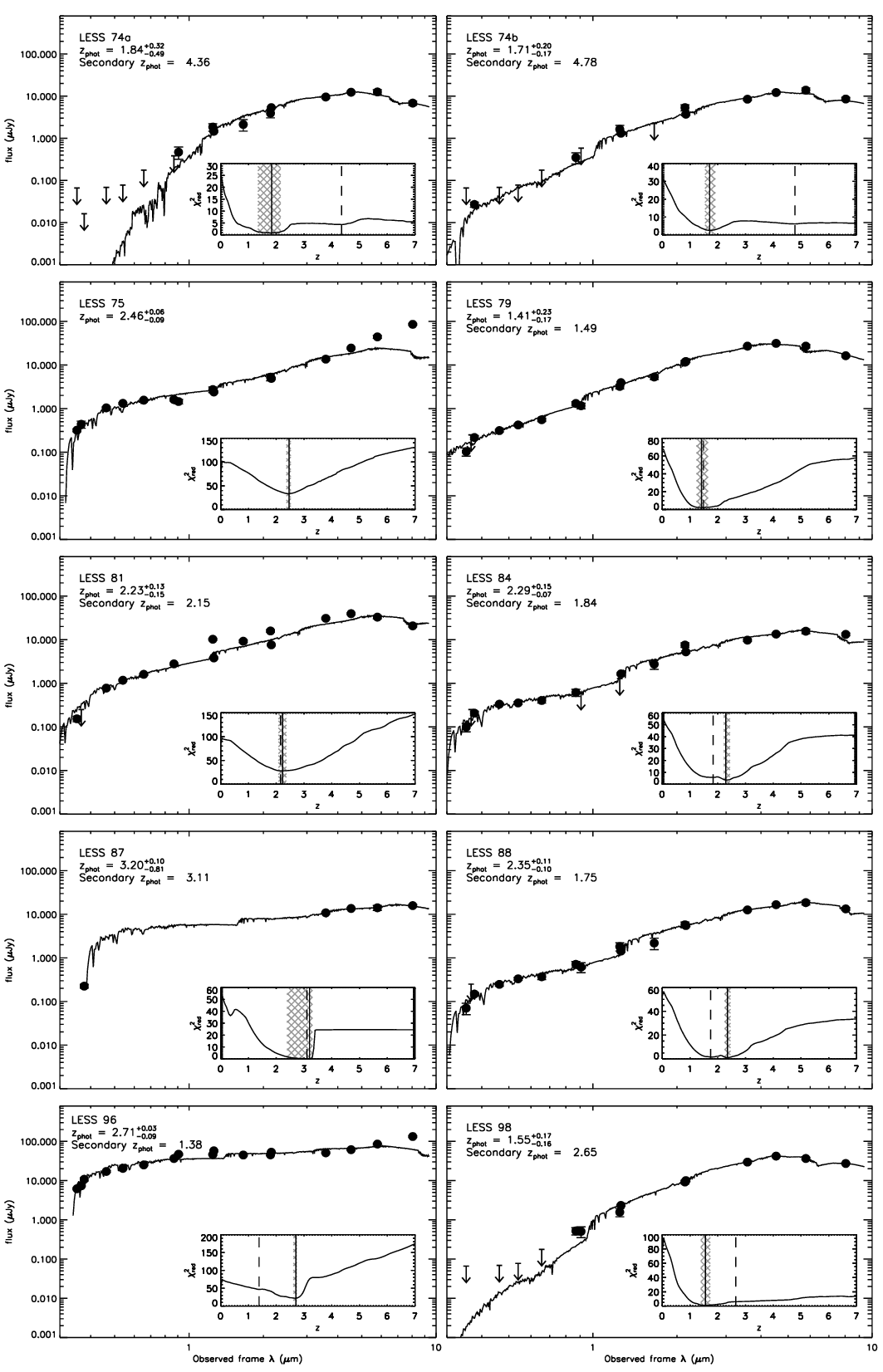}
\contcaption{}
\end{minipage}
\end{figure*}

\begin{figure*}
\begin{minipage}{15cm}
\includegraphics[width=15cm]{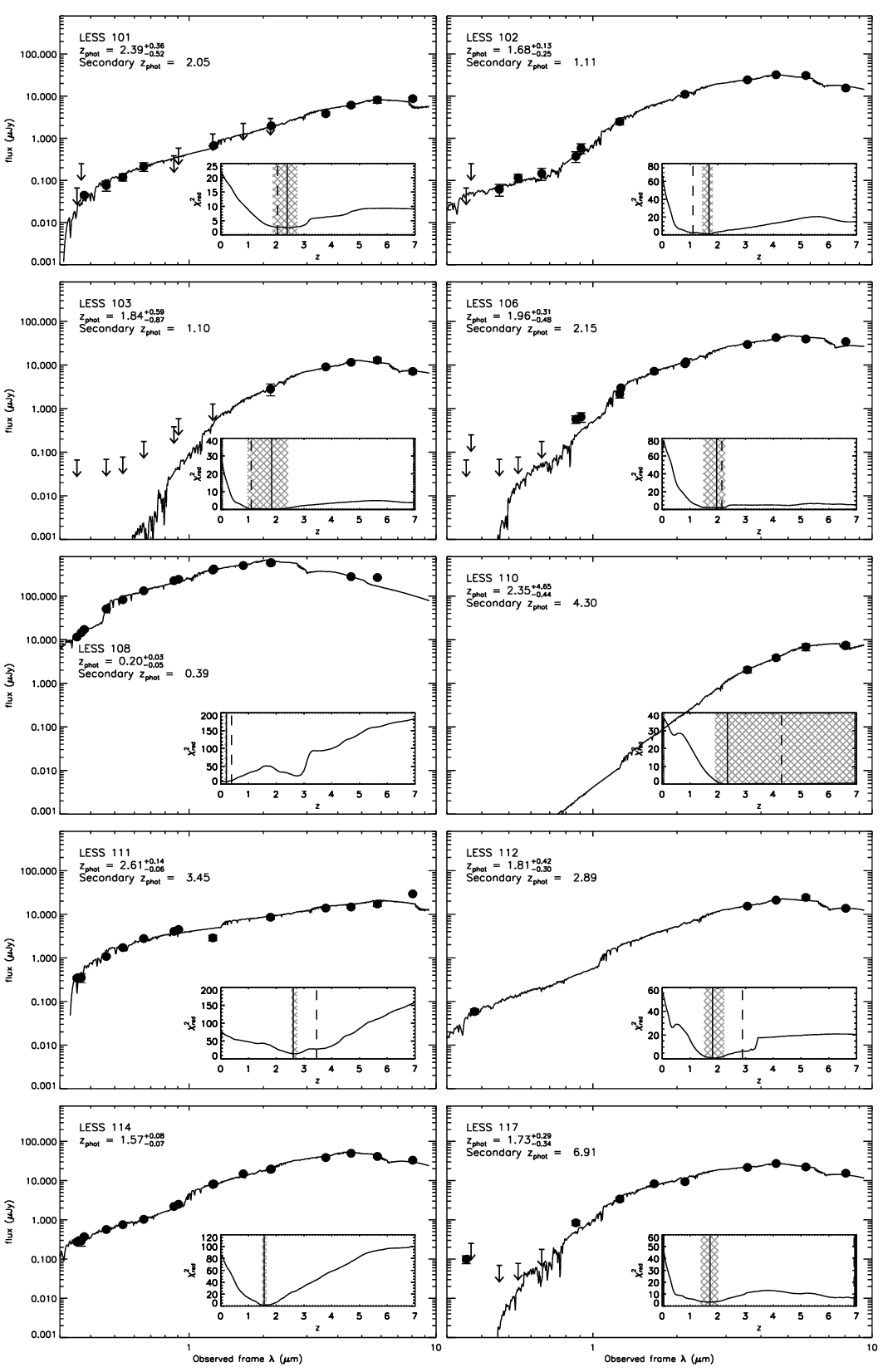}
\contcaption{}
\end{minipage}
\end{figure*}

\begin{figure*}
\begin{minipage}{15cm}
\includegraphics[width=15cm]{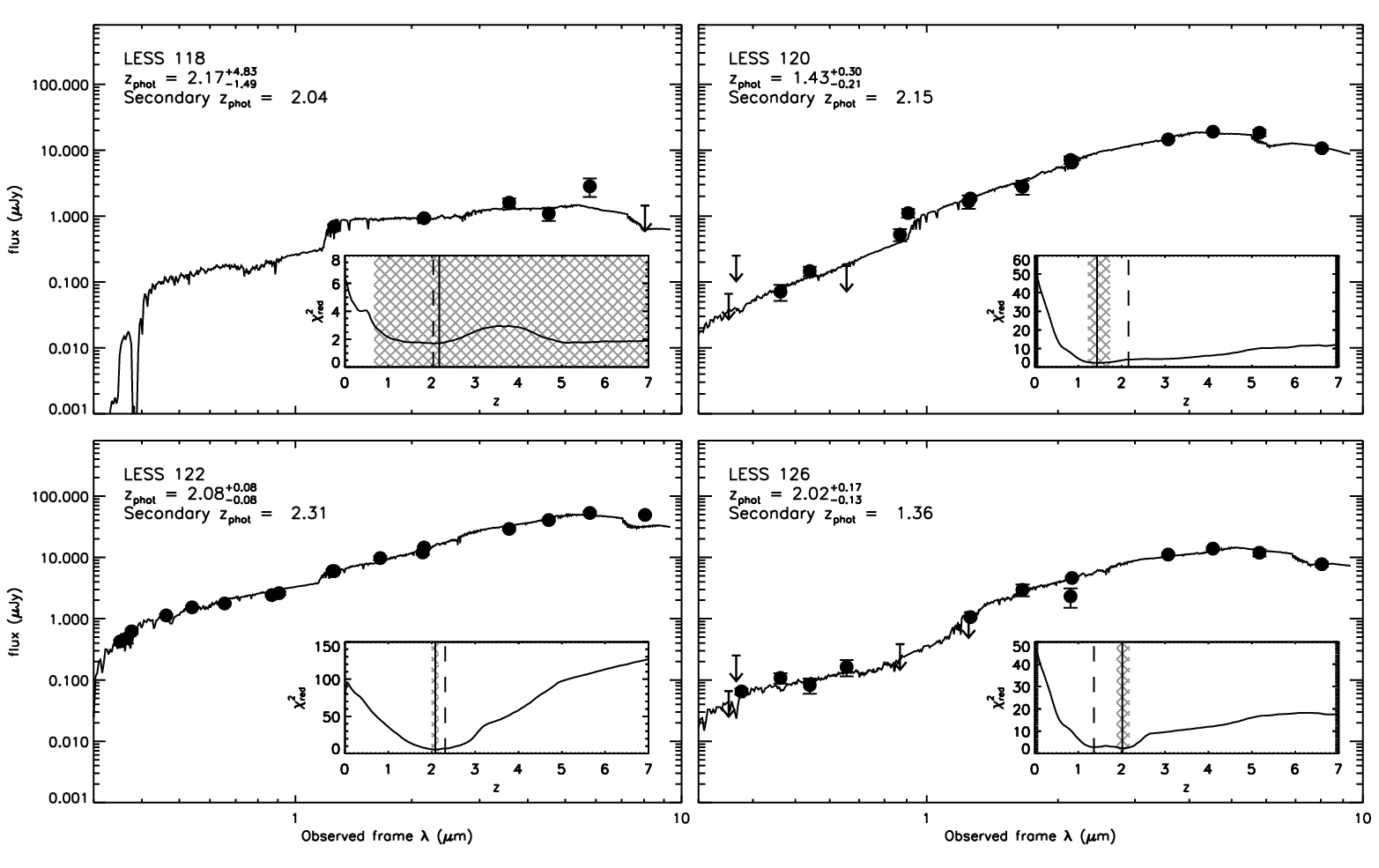}
\contcaption{}
\end{minipage}
\end{figure*}

\label{lastpage}
\end{document}